\documentclass[a4paper,fleqn,manuscript,usenatbib]{mnras}

\usepackage{amsmath}
\usepackage{amssymb}
\usepackage{multicol}
\usepackage{rotating}
\usepackage{lscape}
\usepackage{graphicx}
\usepackage{color}
\usepackage{placeins}
\usepackage{xcolor}
\usepackage{float}
\usepackage{url}
\usepackage{verbatim}
\usepackage{lipsum,adjustbox}

\usepackage{tikz}
\usetikzlibrary{shapes,snakes,calc,positioning}

\usetikzlibrary{arrows,positioning} 
\tikzset{
 myarrow/.style={->, >=latex', shorten >=1pt, thick},
    >=stealth',
    punkt/.style={
           rectangle,
           rounded corners,
           draw=black, very thick,
           text width=6.5em,
           minimum height=2em,
           text centered},
    pil/.style={
           ->,
           thick,
           shorten <=2pt,
           shorten >=2pt,}
}

\newcommand{\beq}{\begin{equation}}
\newcommand{\eeq}{\end{equation}}
\newcommand{\beqa}{\begin{eqnarray}}
\newcommand{\eeqa}{\end{eqnarray}}

\definecolor{gray}{gray}{0.55}

\newcommand{\kmsmpc}{\mbox{km\,s$^{-1}$\,Mpc$^{-1}$}}
\newcommand{\mbi}[1]{\mbox{\boldmath$#1$}}

\newcommand{\lsim}{\mbox{${\,\hbox{\hbox{$ < $}\kern -0.8em \lower 1.0ex\hbox{$\sim$}}\,}$}}
\newcommand{\gsim}{\mbox{${\,\hbox{\hbox{$ > $}\kern -0.8em \lower 1.0ex\hbox{$\sim$}}\,}$}}

\newcommand{\dd}{{\rm d}}

\def\beqn{\vspace{2mm}
\begin{eqnarray}} 
\def\eeqn{\vspace{2mm} 
\end{eqnarray}}

 \def\la{\langle}

\newcommand{\be}{\begin{equation}}
\newcommand{\ee}{\end{equation}}
\newcommand{\ba}{\begin{eqnarray}}
\newcommand{\ea}{\end{eqnarray}}
\newcommand{\brr}{\begin{array}}
 
\newcommand{\err}{\end{array}}
\newcommand{\bc}{\begin{center}}
\newcommand{\ec}{\end{center}}

\title[BOSS MultiDark PATCHY mocks]{{The clustering of galaxies in the SDSS-III Baryon Oscillation Spectroscopic Survey:}\\ mock galaxy catalogues for the BOSS Final Data Release} 

\author[Kitaura et al.]{Francisco-Shu Kitaura$^{1}$\thanks{email:kitaura@aip.de, Karl-Schwarzschild-fellow},  Sergio Rodr{\'i}guez-Torres$^{2,3,4}$\thanks{Campus de Excelencia Internacional UAM/CSIC Fellow}, Chia-Hsun Chuang$^{2}$\thanks{MultiDark Fellow}, \and  Cheng Zhao$^{5}$, Francisco Prada$^{2,3,6}$,  H{\'e}ctor Gil-Mar{\'i}n$^{7}$, Hong Guo$^{8,9}$, Gustavo Yepes$^{4}$, \and   Anatoly Klypin$^{10,11}$,  Claudia G.~Sc\'occola$^{2,12,13}$, Jeremy Tinker$^{14}$, Cameron McBride$^{15}$,  \and   Beth Reid$^{16,17}$,   Ariel G.~S{\'a}nchez$^{18}$, Salvador Salazar-Albornoz$^{18,19}$, Jan Niklas Grieb$^{18,19}$,  \and Mariana Vargas-Magana$^{20}$, Antonio J.~Cuesta$^{21}$, Mark Neyrinck$^{22}$, Florian Beutler$^{16}$,  \and  Johan Comparat$^2$, Will J.~Percival$^{7}$  \& Ashley Ross$^{23,7}$ \vspace{0.5cm} \\
{\it Affiliations are listed at the end of the paper}}

\date{\today.}

\begin{document}

\maketitle


\begin{abstract}
%
We reproduce the galaxy clustering catalogue from the SDSS-III Baryon Oscillation Spectroscopic Survey { Final} Data Release (BOSS { DR11\&DR12}) with high fidelity on all relevant scales in order to allow a robust analysis of baryon acoustic oscillations and redshift space distortions. We have generated { (6,000)} 12,288  MultiDark \textsc{patchy} BOSS { (DR11)} DR12  light-cones corresponding to an effective volume of ${\sim}192,000\,[h^{-1}\,{\rm Gpc}]^3$ (the largest ever simulated volume), including cosmic evolution in the { redshift} range from 0.15 to 0.75. The mocks have been calibrated using a reference galaxy catalogue based on the halo abundance matching modelling of the BOSS { DR11\&DR12} galaxy clustering data and on the data themselves. The production follows three steps. First, we apply the \textsc{patchy} code to generate a dark matter field and an object distribution including nonlinear stochastic galaxy bias. Secondly, we run the halo/stellar distribution reconstruction \textsc{hadron} code to assign masses to the various objects. This step uses the mass distribution as a function of local density and non-local indicators (i.e., tidal field tensor eigenvalues and relative halo exclusion separation for massive objects) from the reference simulation applied to the corresponding patchy dark matter and galaxy distribution. Finally, we apply the \textsc{sugar} code to build the light cones. { The resulting MultiDark \textsc{patchy} mock light cones reproduce the number density, selection function, survey geometry, and  in general within 1-$\sigma$, for arbitrary stellar mass bins, the power spectrum up to $k=0.3\,h\,{\rm Mpc}^{-1}$, the two-point correlation functions down to a few Mpc scales, and the three-point statistics of the BOSS DR11\&DR12 galaxy samples. }
\vspace{-0.cm}
 \end{abstract}

\begin{keywords}
cosmology:  methods: numerical - galaxies: haloes - galaxies: statistics - large-scale structure of Universe

\vspace{-1.cm}
\end{keywords}

\section{Introduction}

The observable Universe represents a unique realization of an underlying physical cosmological process. Large galaxy redshift surveys like the Baryon Oscillation Spectroscopic Survey \citep[BOSS; e.g.,][]{BSA12,DSA13,AAA15}, 
a branch of the ongoing Sloan Digital Sky Survey \citep[SDSS-III][]{EWA11}  scan the sky with unprecedented accuracy trying to unveil  structure formation in an expanding Universe.  One important question arises in the analysis of the data provided by such surveys: if the Universe is comparable to a huge unique experiment, how can we determine the uncertainties in the measurement of quantities derived from observing it? 
One strategy consists of dividing the observations into subvolumes, treating each of the subsamples as independent measurements, and computing the errors with jackknife or bootstrap estimates.
While this approach continues being relevant as a way to obtain error estimates directly from the data \citep[see e.g.][]{Norberg09}, it also implies several disadvantages. First, it does not include systematic errors present in all subvolumes, secondly it does not lead to a physical understanding of the data by itself, and thirdly it introduces variance beyond the one already present in the data on scales larger than the subvolumes.  The last point is especially critical when the signal sought has a large characteristic scale and its detection significance crucially depends on the volume, as  is the case for baryon acoustic oscillations \citep[BAOs; see e.g.][]{SE05,WSP09}.
During the past decades, there has been a huge effort  to encode our physical knowledge of structure formation in computational algorithms, and compare the theoretical models to the actual observations. 
Pioneering works started with qualitative comparisons \citep[see e.g.][]{Klypin1983,Blumenthal1984,Davis1985}.
Since then simulations have grown and such comparisons have turned increasingly more quantitative \citep[see e.g.][]{klypin,Springel2005,Boylan2009,Klypin2011}.
These efforts are essential to understand structure formation and yet they suffer from a strong limitation: as simulations always push the computational limits they are not suited for massive production. In fact the number of  current state-of-the-art large volume $N$-body simulations is of order 10 \citep[][]{KimHorizon2009,AnguloXXL2012,Prada2012,DeusSimulation2012,jubilee2013,MICE2013,Skillman2014,Klypin2014,Ishiyama2014}. 
However, an ideal approach to determine the uncertainties from current and upcoming surveys scanning large sky areas, and hence covering huge volumes, such as 
BOSS\footnote{\url{http://www.sdss3.org/surveys/boss.php}} \citep[][]{White11}, DESI\footnote{\url{http://desi.lbl.gov/}}/BigBOSS \citep[][]{bigboss2011}, DES\footnote{\url{http://www.darkenergysurvey.org}} \citep[][]{des2013}, LSST \footnote{\url{http://www.lsst.org/lsst/}} \citep[][]{lsst2012}, J-PAS\footnote{\url{http://j-pas.org/}} \citep[][]{jpas2014}, 4MOST\footnote{\url{http://www.aip.de/en/research/research-area-ea/research-groups-and-projects/4most}} \citep[][]{4most}, or Euclid\footnote{\url{http://www.euclid-ec.org}} \citep[][]{2009ExA....23...39C,euclid2009},  requires thousands of such simulations if the simplest error determination methods are used \citep[][]{Dodelson13,Taylor13,2014MNRAS.439.2531P}. 
Alternative more efficient methods need to be considered to face this challenge.
A few pioneering works explored a viable strategy  more than a decade ago relying on simplified fast gravity solvers using perturbation theory: \textsc{pinocchio} \citep[][]{2002ApJ...564....8M,Monaco2013} and \textsc{pthalos} \citep[][]{scocci}.
Nevertheless, these methods are not trivial, need calibration with $N$-body simulations, and still demand high computational efforts.
For this reason, some of the first  analysis of large surveys \citep[][]{Percival01,CPP05} was done based on lognormal realizations \citep[see also][]{PVP04,BBC11}, which match the two-point statistics by construction \citep[][]{CJ91}, although their three-point statistics is very different from the true one \citep[see e.g.][]{WTM14,ChuangComp15}. It is also not clear that their four-point statistics will be accurate \citep[][]{Cooray01,Takada13}.

 The analysis of past data releases of the BOSS collaboration utilized 1,000 mocks, created based on an improved version of \textsc{pthalos} \citep[][]{manera12,Manera15}. The use of approximate gravity solvers in these methods came at the expense of only matching  clustering statistics on a wide range of scales to $\sim10\%$ precision (and strongly deviating towards small scales $\lsim20\,h^{-1}$ Mpc).  

This sets the agenda for the current BOSS data release  { DR11{\&}DR12} and the requirements for a new generation of mock galaxy catalogues. 
 Ideally one would like to base these on efficient solvers that are trained on exact solutions and deliver a comparable precision.
A new generation of methods that can meet these high requirements have been developed during the past two years, in particular,
\textsc{patchy} \citep[][]{KitauraPatchy}, \textsc{qpm} \citep[][]{WTM14}, and  \textsc{ezmocks} \citep[][]{ezmocks}.
The key concept exploited by these methods is to rely only on the large-scale density field obtained from approximate gravity solvers and use biasing prescriptions to populate it with mock galaxies, in a similar way to the methods proposed to augment the resolution of $N$-body simulations \citep[][]{delaTorre2012,dlTGP2013,Angulo2013,Ahn15}.   One should however be careful, as computing an accurate dark matter field is a necessary, but not sufficient condition to reproduce the correct halo/galaxy three-point statistics. The bias parameters are degenerate in the two-point statistics and need to be additionally constrained to reproduce higher order statistics \citep[][]{KGS15}.
We will rely in this work on the \textsc{patchy} method due to its verified accuracy in the two and three-point statistics for different populations of objects \citep[see application of the \textsc{hadron} code to \textsc{patchy} and \textsc{ezmock};][]{Zhao15}. { An additional set of galaxy mocks fitting the BOSS { DR11{\&}DR12} (CMASS and LOWZ) data at two  mean redshifts (respectively) based on \textsc{qpm} have been produced in an unprecedented effort. These are constructed with a different structure formation model based on low resolution particle mesh solvers, and a different galaxy bias, based on a rank-ordering scheme assigning most massive objects to the highest density peaks, \citep[for a comparison of both sets of catalogues see \S \ref{sec:results} and][]{2015arXiv150906386G}.}

Another approach uses approximate PT based solutions to speed up $N$-body solvers \citep[see \textsc{Cola} method,][]{cola2013,HMP15,KBB15}. This method is very promising to generate ensembles of reference mock catalogues; however, it has the drawback of requiring large computational memory for the force calculation and large number of particles to resolve the haloes \citep[see][]{ChuangComp15}, and is therefore not suitable for the massive production aimed in this work. The speed of the method over $N$-body simulations comes at the expense of not resolving the sub-structures required to produce a realistic galaxy catalogue. This problem can be circumvented by, e.g., augmenting the missing objects with the halo occupation distribution model, hereby losing some of the advantage of having a higher precise description of the nonlinear clustering over the above mentioned methods which rely only on the large scale dark matter field, as shown in a comparison study \citep[see][and references therein]{ChuangComp15}. One may need an approach like \textsc{cola}, to model the large-scale structure, combined with the galaxy bias presented in this work for future emission line galaxy-based surveys. We will, however, demonstrate here that this is not necessary to model the distribution of luminous red galaxies (LRGs) aimed in this work.

One could argue whether mock catalogues are required at all, as  analytical models may deliver an almost direct computation of error bars and covariance matrices \citep[][]{Hartlap07,Hamaus10,Dodelson13,Taylor13,KPS15}. It still remains to be shown that these methods making simple assumptions, such as that the density field is Gaussian distributed, yield the same accuracy as covariance matrices based on large sets of mock catalogues.

Nevertheless, the purpose of mock catalogues is manifold, as they not only serve to provide error estimates, but also to provide an understanding of the systematics of the survey and of the methodology.
 Any analytical prediction or data analysis method should be cross-checked with  large ensembles of mock galaxy catalogues for which the products of this work could be useful. One clear example is the case of BAO reconstruction techniques \citep[see e.g.][]{ESS07,PXE12,AAB14,RSH15}.

We exploit the efficiency and accuracy of the \textsc{patchy} code to produce 12,288 galaxy mock catalogues\footnote{This corresponds to an effective volume of ${\sim}192,000\,[h^{-1}\,{\rm Gpc}]^3$, a factor of $\sim20$ times larger than the volume of the DEUS FUR simulation \citep[][]{DeusSimulation2012}, and a factor of $\sim375$ times larger than the DarkSky  ds14 simulation \citep[][]{Skillman2014}.}  including the lightcone evolution of galaxy bias based on the halo abundance matching technique applied to the reference {BigMultiDark} $N$-body simulation \citep[see][companion paper]{2015arXiv150906404R}, and to the peculiar motions based on the observational data, matching the two, three-point statistics, in real and redshift space of the BOSS { DR11{\&}DR12} galaxy clustering data at different redshifts and for arbitrary stellar mass bins. Special care has been taken to include all relevant observational effects including selection functions and masking. 
The \textsc{MultiDark patchy} BOSS DR11 mock catalogues presented in this work are publicly available\footnote{\url{http://data.sdss3.org/sas/dr11/boss/lss/dr11_patchy_mocks/}\\The BOSS DR12 mock catalogues will be made publicly available together with the data catalogue: \url{http://data.sdss3.org/sas/dr12/boss/lss/dr12_patchy_mocks/}.}.

This paper is structured as follows: in section \S \ref{sec:method} we describe the methodology. This section starts with  the generation of the reference catalogue using $N$-body simulations and the halo abundance matching technique. Subsequently the scheme to massively generate mock catalogues is described.  Then we show in \S \ref{sec:results} the statistical comparison between the mock catalogues  and the BOSS DR12 data. Subsequently we discuss future work (\S \ref{sec:future}). Finally, in section \S \ref{sec:conc} we present the conclusions.
The reader interested only in the results may skip \S \ref{sec:method} and directly go to \S \ref{sec:results}.

\section{Methodology}

\label{sec:method}

To construct high-fidelity mock light cones for interpreting the BOSS  { DR11{\&}DR12} galaxy clustering, we adopt an iterative training procedure in which a reference catalogue is statistically reproduced with approximate gravity solvers and analytical-statistical biasing models. The whole algorithm involves several steps and is summarized in the flow chart in Fig.~\ref{fig:flow}. 
\begin{enumerate}

\item The first step consists of  the  generation of  an accurate reference catalogue. Here we rely on a large $N$-body simulation capable of resolving distinct haloes and the corresponding substructures. This permits us to apply the HAM technique to reproduce the clustering of the observations with only one parameter: the scatter in the stellar mass-to-halo mass relation \citep[see][companion paper; and \S \ref{sec:ham}]{2015arXiv150906404R}. This technique is applied at different redshift bins to obtain a detailed galaxy bias evolution spanning the  redshift range covered by BOSS { DR11{\&}DR12} galaxies.
In this way we obtain mock galaxy catalogues in full cubical volumes of 2.5 $h^{-1}$ Gpc side at different redshifts. 

\item In the second step we train the \textsc{patchy} code \citep[][]{KitauraPatchy,KGS15} to match the two- and three-point clustering of the full mock galaxy catalogues for each redshift bin. Here we consider all the mock galaxies together in a single bin  irrespectively of their stellar mass.

\item In the third step we apply the \textsc{hadron} code \citep[][]{Zhao15} to assign stellar masses to the individual objects.

\item In the fourth step we apply the \textsc{sugar} code  \citep[see][companion paper]{2015arXiv150906404R} which includes selection effects, masking, and combines different boxes at different redshifts into a light-cone. 

\item In the fifth step the resulting  \textsc{MultiDark patchy} mock catalogues are compared to the observations. The process is iterated until the desired accuracy for different statistical measures is reached.

\end{enumerate}

In the next sections we will describe in detail these steps described above for the massive generation of accurate mock galaxy catalogues.
The reader interested only in the results may directly go to \S \ref{sec:results}.

\begin {figure*}
\centering
\begin{adjustbox}{width=1\textwidth}
\begin{tikzpicture}[node distance=1cm, auto]
\hspace{1.cm}
 \node[ellipse,draw] (start) {Cosmological parameters};
 \node[above=of start] (dummy) {};
 \node[right=of dummy] (title) {\hspace{-15cm}\textsc{MultiDark patchy Mocks for BOSS { DR11{\&}DR12}: training mock catalogs from observed and simulated data sets}};
 \node[rectangle,draw, inner sep=5pt,below=0.5cm of start] (IC) {Initial conditions};
  \node[rectangle,draw,text width=4.2cm, inner sep=3pt,left=.7cm of start] (BOSS) {\hspace{0.1cm}\Large\bf\textsc{BOSS { DR11{\&}DR12}}};
\node[left=of BOSS] (dummy) {};
\node[rectangle,draw,text width=2.95cm, inner sep=5pt,left=2.cm of BOSS] (BMD) {\hspace{0.4cm}{{BigMultiDark}}\hspace{1cm} $N$-body simulation*};
\node[above=0.3cm of start] (dummy0) {};
\node[left=11.cm of dummy0] (dummy1) {};
\draw [-] (start)--(dummy0.center);
\draw [-] (dummy0.center)--(dummy1.center);
\draw [->] (dummy1.center)--(BMD);
 \node[ellipse,draw,below=.7cm of dummy] (HAM) {\textsc{sugar*}};
\node[rectangle,draw,text width=3.7cm, inner sep=5pt,below=.4cm of HAM] (BMDM) {\hspace{0.2cm}{{i) BigMultiDark}} mock{*}};
\draw [->] (HAM)--(BMDM);
\draw [->] (BOSS.south) to  [out=270,in=0] (HAM.east);
\draw [->] (BMD.south) to  [out=270,in=180] (HAM.west);
\draw [->] (start)--(IC);
 \node[below=of IC] (dummy) {};
 \node[ellipse,draw,right=of dummy] (ALPT) { {ALPT}}; 
 \draw [->] (IC.south) to [out=-90,in=180] (ALPT); 
 \node[below=of ALPT] (dummy) {};
 \node[rectangle,draw,text width=2cm,right=of dummy] (DF) {Density + \\velocity field};
\draw [->] (ALPT.east) to  [out=-0,in=90] (DF.north);
 \node[ellipse,draw,below=2.2cm of IC] (PA0) {{\textsc{ii) patchy}}-bias{*}};
\draw [->] (BMDM.south) to  [out=270,in=90] (PA0.north);
\draw [->] (DF)--(PA0);
 \node[rectangle,draw,text width=2.9cm,left=of PA0] (HD) {Halo distribution + peculiar velocities};
\draw [->] (PA0)--(HD);
 \node[ellipse,draw,below=1.cm of PA0] (HA) {\textsc{iii) hadron*}};
\draw [->] (BMDM.south) to  [out=270,in=180] (HA.west);
\draw [->] (DF.south) to  [out=270,in=-0] (HA.east);
\draw [->] (HD.south) to  [out=-90,in=180] (HA.west);
 \node[rectangle,draw,text width=2.8cm,below=0.5cm of HA] (HM) {Halo distribution + halo masses + \\peculiar velocities};
\draw [->] (HA.south) to (HM.north);
 \node[ellipse,draw,below=0.5cm of HM] (SG) {\textsc{iv) sugar*}};
\draw [->] (BMDM.south) to  [out=-90,in=180] (SG.west);
\draw [->] (HM.south) to (SG.north);
\node[rectangle,draw,text width=5.4cm,below=5.4cm of BMDM] (GD) {{\textsc{v) MultiDark patchy Mocks}} \hspace{1cm} Galaxy distribution + stellar masses + peculiar velocities + completeness};

\node[left=.5cm of BMD] (md1) {};
\node[above=1.cm of md1] (md2) {};
\node[above=.775cm of BOSS] (md3) {};
\node[left=2.0cm of GD] (md4) {};
\draw [dashed] (md2.south) to [out=0,in=180] (md3.south);
\draw [<-,dashed] (BOSS.north) to [out=90,in=90] (md3.south);
\draw [dashed] (md2.south) to [out=-90,in=90] (md4.east);
\draw [->,dashed] (md4.east) to [out=0,in=180] (GD.west);


 \node[below=-3.cm of GD] (dummy) {};
 \node[rectangle,text width=5.cm,right=-4.3cm of dummy] (CR9) {dashed lines: \\self-consistent/iterative steps};
 \node[below=-0.4cm of GD] (dummy) {};
 \node[rectangle,text width=15.cm,below=9.cm of BOSS] (CR8) {*: the asterisc indicates the steps in which calibration  with observations and simulations is required};
 \node[below=1.cm of GD] (dummy) {};
 \node[rectangle,text width=9.5cm,below=1.cm of GD] (CR) {ALPT: Augmented Lagrangian Perturbation Theory:\\ \hspace{0cm}\citet{KitauraHess2013}};
 \node[rectangle,text width=9.5cm,below=0.1cm of CR] (CR2) {{BigMultiDark}: $N$-body Planck simulation \hspace{3.5cm} (2.5 $h^{-1}$\,Gpc)$^3$ with 3,840$^3$ particles: \citet{Klypin2014}};
 \node[rectangle,text width=9.5cm,below=0.1cm of CR2] (CR3) {\textsc{hadron}: Halo mAss Distribution ReconstructiON: \\\citet{Zhao15}};
 \node[below=1.6cm of SG] (dummy) {};
 \node[rectangle,text width=9.5cm,below=1.2cm of SG] (CR4) {\textsc{ham}: Halo Abundance Matching,  \citet[][companion paper]{2015arXiv150906404R}};
 \node[rectangle,text width=9.5cm,below=.1cm of CR4] (CR5) {\textsc{MockFactory}: \citet{WTM14}};
 \node[rectangle,text width=9.5cm,below=.1cm of CR5] (CR6) {\textsc{patchy}: PerturbAtion Theory Catalog generator \\of Halo and galaxY distributions: \citet{KitauraPatchy,KGS15}};
 \node[rectangle,text width=9.5cm,below=0.1cm of CR6] (CR7) {\textsc{sugar}:  SUrvey GenerAtoR,  \citet[see][companion paper]{2015arXiv150906404R} \\this code contains \textsc{ham} and \textsc{MockFactory}};
\draw [->] (SG.west) to (GD.east);
\draw [<->,dashed] (GD.west) to  [out=180,in=180] (BMDM.west);
\draw [<->,dashed] (BMDM.east) to  [out=0,in=-90] (BOSS.south);
\end{tikzpicture}
\end{adjustbox}
\caption{\label{fig:flow} Flowchart of the methodology applied in this work for the generation of high fidelity BOSS { DR11{\&}DR12} mock galaxy catalogues: i) starting from a reference mock catalogue calibrated with the observations, ii) followed by the reproduction of the whole catalogue, iii) with the subsequent mass assignment, iv) and survey generation. v) The final catalogues are compared with the observations and the simulation, and the previous steps are repeated until the mock catalogues are compatible with the observations within 1-$\sigma$ for the monopole and quadrupole up to $k\sim0.3\,h\,{\rm Mpc}^{-1}$.}
\end{figure*}

\subsection{Reference mock catalogues}
\label{sec:ham}

The reference catalogues are extracted from one of the {BigMultiDark} simulations\footnote{\url{http://www.multidark.org/MultiDark/}} \citep[][]{Klypin2014}, which was performed using \textsc{gadget-2} \citep{Springel2005} 
with $3,840^3$ particles on a volume of $(2.5\,h^{-1}{\rm Mpc}$ $)^3$ assuming $\Lambda$ cold dark matter Planck cosmology with \{$\Omega_{\rm M}=0.307115, \Omega_{\rm b}=0.048206,\sigma_8=0.8288,n_s=0.9611$\}, and a Hubble constant ($H_0=100\,h\kmsmpc$) given by  $h=0.6777$.  Haloes were defined based on  the Bound Density Maximum halo finder \citep{Klypin1997}.

We rely here on the HAM technique to connect haloes to galaxies \citep{Kravtsov04,Neyrinck04,Tasitsiomi04,Vale04,Conroy06,Kim08,Guo10,Wetzel10,Trujillo11,Nuza13}.

We note that there are alternative methods connecting haloes to galaxies like the HOD
model, which we are not going to consider here \citep[e.g.,][]{Berlind02,Kravtsov04,Zentner05,Zehavi05,Zheng07,Skibba09,RossBrunner09,Zheng09,White11}. These methods are based on a statistical relation  describing the probability that a 
halo of virial mass $M$ hosts $N$ galaxies with some specified properties. 
In general, theoretical HODs require the fitting of a function with several parameters, 
which we want to avoid here.

At first order HAM assumes a one-to-one correspondence between the 
luminosity and 
stellar or dynamical masses: 
 galaxies with more stars are assigned
to more massive haloes or subhaloes. 
The luminosity in a red-band is sometimes used instead of stellar mass.
 There should be some degree of stochasticity in the relation between stellar and dynamical
masses due to deviations in the merger history, angular momentum,
halo concentration, and even observational errors \citep{Tasitsiomi04,Behroozi10,Leauthaud11,Trujillo11}. 
 Therefore, we include a scatter in that relation necessary to accurately  fit the clustering of the BOSS data  \citep[see][companion paper]{2015arXiv150906404R}.
To do this, we  modify the maximum circular velocity ($V_{\rm max}$) of each object adding a Gaussian noise: $V_{\rm max}^{\rm scat}=V_{\rm max}(1+\mathcal{N}(0,\sigma))$, 
where $\mathcal{N}(0,\sigma)$ is a Gaussian random number with mean 0, and standard deviation $\sigma$. Then, we sort all objects by $V_{\rm max}^{\rm scat}$ and then, we selected objects starting from the one with larger $V_{\rm max}^{\rm scat}$ and we continue until we get the proper number density at different redshifts bins. 

By construction, the method reproduces the observed
luminosity function (or stellar mass function). It also reproduces the
scale dependence of galaxy clustering over a large range of epochs
\citep{Conroy06,Guo10}. When abundance matching is used for the
observed stellar mass function \citep{Li09}, it gives also a reasonable 
fit to lensing results \citep{Mandelbaum06} and to 
the relation between stellar and virial mass
\citep{Guo10}.


\subsection{Generation of mock galaxy catalogues}
\label{sec:massprod}

All covariance matrix estimates based on a finite number of mock catalogues, $N_{\rm s}$, are affected by noise, which must be propagated into the final constraints. The impact of the uncertainties in the covariance matrix on the derived cosmological constraints has been subject of several recent analyses \citep[][]{Dodelson13,Taylor13,2014MNRAS.439.2531P}. In particular, \citet[][]{Dodelson13} showed that this additional uncertainty can be described by a rescaling of the parameter covariances derived from the distribution of measurements from a set of mocks with a factor given by
\begin{equation}
m=1+ \frac{(N_{\rm s}-N_{\rm b}-2)\left(N_{\rm b}-N_{\rm p}\right)}{(N_{\rm s}-N_{\rm b}-1)(N_{\rm s}-N_{\rm b}-4)},
\end{equation}
where $N_{\rm b}$ is the number of bins in the corresponding clustering measurements and $N_{\rm p}$ is
the number of parameters measured. This implies that a large number of mock catalogues are necessary for a robust analysis of the galaxy clustering data.

{For the anisotropic BAO measurements of \citet[][]{Cuesta15} the estimation of the full covariance matrix of the monopole and quadrupole of the two-dimensional correlation function from the ensemble of 1,000 \textsc{qpm} corresponds to an additional uncertainty of 2\% on the constraints on $H(z)r_{\rm d}$ and $D_{\rm A}(z)/r_{\rm d}$. Using the  2,048 \textsc{MultiDark patchy}  mock catalogues, the effect is reduced to the order of 1\%. Large sets of catalogues are even more important for full-shape fits of anisotropic clustering measurements, where the inclusion of information from smaller scales can significantly improve the constraints based on redshift space distortions  (RSD; requiring a larger number of bins). For example, in the analysis of S{\'a}nchez et al.~(in prep.), based on measurements of the clustering wedges statistic \citep[][]{Kazin12}, the use of  mock catalogues corresponds to a rescaling of the parameter covariances by $m=1.04$ and 1.085 when using 1,000 or 2,048 catalogues, respectively.
This additional uncertainty corresponds to a degradation of the true constraining power of the clustering measurements, which should be minimized by using a larger number of mock catalogues. For this reason we have made the effort in the BOSS collaboration of producing at least 1,000 mocks for each BOSS  { DR11{\&}DR12} sub-sample.}

The strategy for the massive production of mock galaxy catalogues relies on generating dark matter fields with approximate gravity solvers on a mesh. We use grids of $960^3$ cells with volumes of (2.5 $h^{-1}$ Gpc)$^3$ and resolutions of $2.6 \,h^{-1}$ Mpc for which the structure formation model can be  considered to be accurate   \S \ref{sec:approx}. Then the galaxies are populated on the mesh according to a combined nonlinear deterministic \S \ref{sec:bias} and stochastic bias model \S \ref{sec:stoc}.
In a post-processing step we assign halo/stellar masses to each object  \S \ref{sec:mass}.
Finally we apply the survey geometry and selection functions \S \ref{sec:survey}.

Let us start describing the \textsc{patchy} code ({{\bf P}erturb{\bf A}tion {\bf T}heory {\bf C}atalog generator of {\bf H}alo and galax{\bf Y} distributions}).

\subsubsection{Approximate fast structure formation model}
\label{sec:approx}

We rely on  augmented Lagrangian Perturbation Theory (ALPT) to simulate structure formation. Let us recap the basics of this method and refer for details to \citet[][]{KitauraHess2013}.
In this approximation the displacement field  $\mbi\Psi(\mbi q,z)$, mapping a distribution of dark matter particles at initial Lagrangian positions $\mbi q$ to the final Eulerian positions $\mbi x(z)$ at redshift $z$ ($\mbi x(z)=\mbi q+\mbi\Psi(\mbi q,z)$), is split into a long-range $\mbi\Psi_{\rm L}(\mbi q,z)$ and a short-range component $\mbi\Psi_{\rm S}(\mbi q,z)$, i.e. 
$\mbi\Psi(\mbi q,z)=\mbi\Psi_{\rm L}(\mbi q,z)+\mbi\Psi_{\rm S}(\mbi q,z)$.

We rely on second order  Lagrangian Perturbation Theory (2LPT) for the long-range component $\mbi\Psi_{\rm 2LPT}$
 \citep[for details on 2LPT see][]{1994MNRAS.267..811B,bouchet1995,catelan}. 

 The resulting displacement field is filtered  with a kernel $\cal K$: $\mbi\Psi_{\rm L}(\mbi q,z)={\cal K}(\mbi q,r_{\rm S}) \circ \mbi\Psi_{\rm 2LPT}(\mbi q,z)$.
We apply a Gaussian filter ${\cal K}(\mbi q,r_{\rm S})$$=\exp{(-|\mbi q|^2/(2r_{\rm S}^2))}$, with $r_{\rm S}$ being the smoothing radius.
 We use the  spherical collapse approximation to model the short-range component $\mbi\Psi_{\rm SC}(\mbi q,z)$ \citep[see][]{1994ApJ...427...51B,2006MNRAS.365..939M,Neyrinck2013}. 
The combined ALPT displacement field
\be
\label{eq:disp}
\mbi\Psi_{\rm ALPT}(\mbi q,z)={\cal K}(\mbi q,r_{\rm S}) \circ \mbi\Psi_{\rm 2LPT}(\mbi q,z)+\left(1-{\cal K}(\mbi q,r_{\rm S}) \right)\circ \mbi\Psi_{\rm SC}(\mbi q,z)
\ee
 is used to move a set of homogenously distributed particles from Lagrangian initial conditions to the Eulerian final ones. We then grid the particles following a clouds-in-cell scheme to produce a smooth density field $\delta^{\rm ALPT}$.
One may get some improvements preventing voids within larger collapsing regions, which essentially extends the collapsing  region towards moderate underdensities \citep[see \textsc{muscle} method in][]{Neyrinck15}.
This approach requires about eight additional convolutions being about twice as expensive, as the approached used here.
Moreover, we have checked that the improvement provided by including \textsc{muscle} is not perceptible when using grids with cell sizes of 2.6 $h^{-1}$ Mpc.

\subsubsection{Deterministic bias relations}
\label{sec:bias}

In this section we describe the deterministic part of our bias model. This is combined with a stochastic element, described in \S \ref{sec:stoc}, and a nonlocal element, described in \S \ref{sec:mass}, to produced the full model. 
{ The deterministic bias relates the expected number counts of haloes or galaxies $\rho_{\rm g}\equiv\langle N_{\rm g}\rangle_{\partial V}$ at a given finite volume to the underlying dark matter field $\rho_{\rm M}$,   with $\langle[\cdots]\rangle_{\partial V}$ being the ensemble average over the differential volume element ${\partial V}$ (in our case the cell of a regular mesh)}.
This relation  is known to be nonlinear, nonlocal and stochastic \citep[][]{1974ApJ...187..425P,1985MNRAS.217..805P,bbks1986,1993ApJ...413..447F,1996MNRAS.282..347M,1999ApJ...520...24D,1999MNRAS.304..767S,2000MNRAS.318..203S,MoWhite02,Berlind02,2007PhRvD..75f3512S,2010PhRvD..82j3529D,2011PhRvD..83j3509B,2011A&A...527A..87V,2012MNRAS.421.3472E,Chan12,2012PhRvD..86h3540B,baldauf2013,Ahn15}.
In general this bias relation will be arbitrarily complex:
\be
\rho_{\rm g}=f_{\rm g}\,{ B}(\rho_{\rm M}){,}
\ee
with ${B}(\rho_{\rm M})$ being a general bias function, { $f_{\rm g}=\frac{\langle\rho_{\rm g}\rangle_V}{\langle{B}(\rho_{\rm M})\rangle_V}$, $\langle\rho_{\rm g}\rangle_V$ being the number density $\bar{N}_{\rm g}$,   and $\langle[\cdots]\rangle_V$ being the ensemble average over the whole considered volume $V$ (in our case the volume of the  mesh)}.

The deterministic bias model we consider in this work has the following form: 
\be
\rho_{\rm g}=f_{\rm g}\,\theta(\rho_{\rm M}-\rho_{\rm th})\,\exp\left[-\left(\frac{\rho_{\rm M}}{\rho_\epsilon}\right)^{\epsilon}\right]\,\rho_{\rm M}^\alpha\,(\rho_{\rm M}-\rho_{\rm th})^\tau{,}
\ee
with 
\be
\label{eq:numden}
{ f_{\rm g}=\bar{N}_{\rm g}/\langle\theta(\rho_{\rm M}-\rho_{\rm th})\,\exp\left[-\left(\frac{\rho_{\rm M}}{\rho_\epsilon}\right)^{\epsilon}\right]\,\rho_{\rm M}^\alpha\,(\rho_{\rm M}-\rho_{\rm th})^\tau\rangle_V}{,}
\ee
 and \{$\rho_{\rm th},\alpha,\epsilon,\rho_\epsilon,\tau$\} the parameters of the model. 
We have modelled threshold bias \citep[][]{Kaiser84,bbks1986,Cole89,Sheth01,MoWhite02} as a combination of a step function $\theta(\rho_{\rm M}-\rho_{\rm th})$ \citep[][]{KitauraPatchy}  and an exponential cut-off $\exp\left[-\left(\frac{\rho_{\rm M}}{\rho_\epsilon}\right)^{\epsilon}\right]$ \citep[][]{Neyrinck2014}.
The local bias expansion \citep[][]{Cen1993,1993ApJ...413..447F} is summarized by a power-law \citep[][]{delaTorre2012,KitauraPatchy}.
In addition we consider a  bias $(\rho_{\rm M}-\rho_{\rm th})^\tau$ which compensates for the missing power of PT based methods.

Nonlocal bias has been recently found to be relevant \citep[][]{Mcdonald09,2012PhRvD..86h3540B,Chan12,Sheth13,Saito2014}.
A non-local bias introduces a scatter in the local deterministic bias relations described above. In this work, the scatter is first described by a stochastic bias relation (see \S \ref{sec:stoc}). 
We have investigated second order nonlocal bias with \textsc{patchy} without finding that this can have a relevant effect on the mock catalogues considering stochastic bias and the full (one single mass bin) catalogue (see Autefage et al.~in prep.).
In fact once one considers different populations of halo or stellar mass objects, then nonlocal bias plays an important role. We solve this in a post-processing step when assigning the masses to each galaxy \citep[see \S \ref{sec:mass} and ][]{Zhao15}.

\subsubsection{Stochastic biasing}
\label{sec:stoc}

The halo distribution is a discrete sample $N_{{\rm g},i}$ of the continuous underlying dark matter distribution $\rho_{{\rm g},i}$: 
\be
N_{{\rm g},i}\curvearrowleft P(N_{{\rm g},i}\mid\rho_{{\rm g},i},\{p_{\rm SB}\})\,,
\ee 
for each cell $i$ and $\{p_{\rm SB}\}$ being the set of stochastic bias parameters. To account for the shot noise one could do Poissonian realizations of the halo density field as given by the deterministic bias and the dark matter field \citep[see e.g.][]{delaTorre2012}. However, it is known that the excess probability of finding haloes  in high density regions generates over-dispersion \citep[][]{2001MNRAS.320..289S,2002MNRAS.333..730C}.

The strategy up to now has been to generate a mock catalogue which reproduces the clustering of the whole population of galaxies for a given redshift. This has the advantage that by mixing massive and low mass galaxies we will always be dominated by overdispersion, which is much easier to model than underdispersion.
In particular we consider the negative binomial (NB) probability distribution function \citep[for non-Poissonian distributions see][]{1984ApJ...276...13S,1995MNRAS.274..213S}  including an additional parameter $\beta$ to model over-dispersion (tends towards the Poisson probability distribution function for $\beta\rightarrow\infty$ and for low  $\lambda$ values).

We note that a proper treatment of the deviation from Poissonity is also crucial to get accurate density reconstructions \citep[see][and Ata et al in prep.]{Ata15}.

We will need, however, to take care of the different statistical nature of each population of galaxies when we assign masses to each object (see \S \ref{sec:mass}).

\subsubsection{Redshift space distortions}

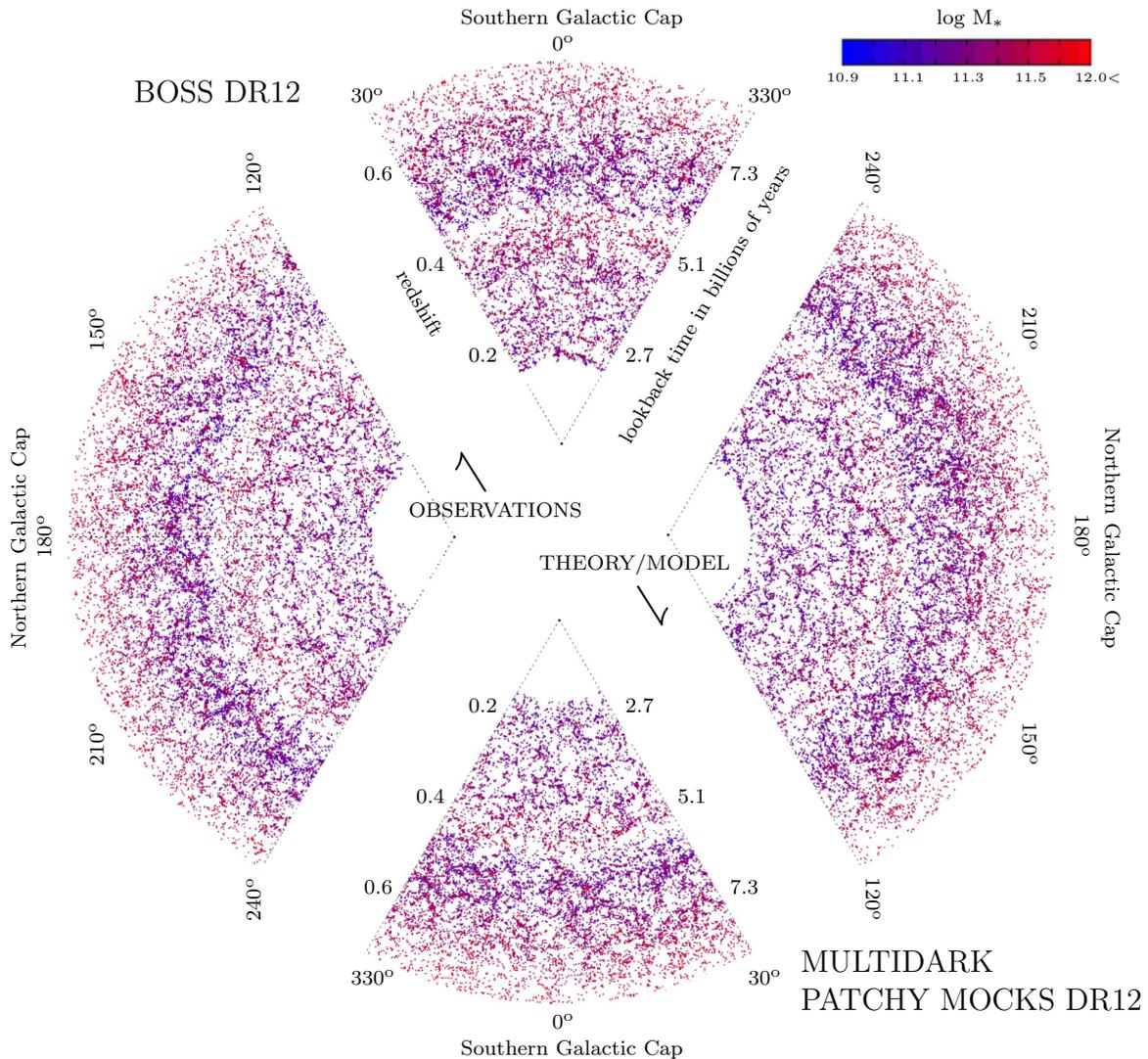
\begin{figure*}
\includegraphics[width=15.cm]{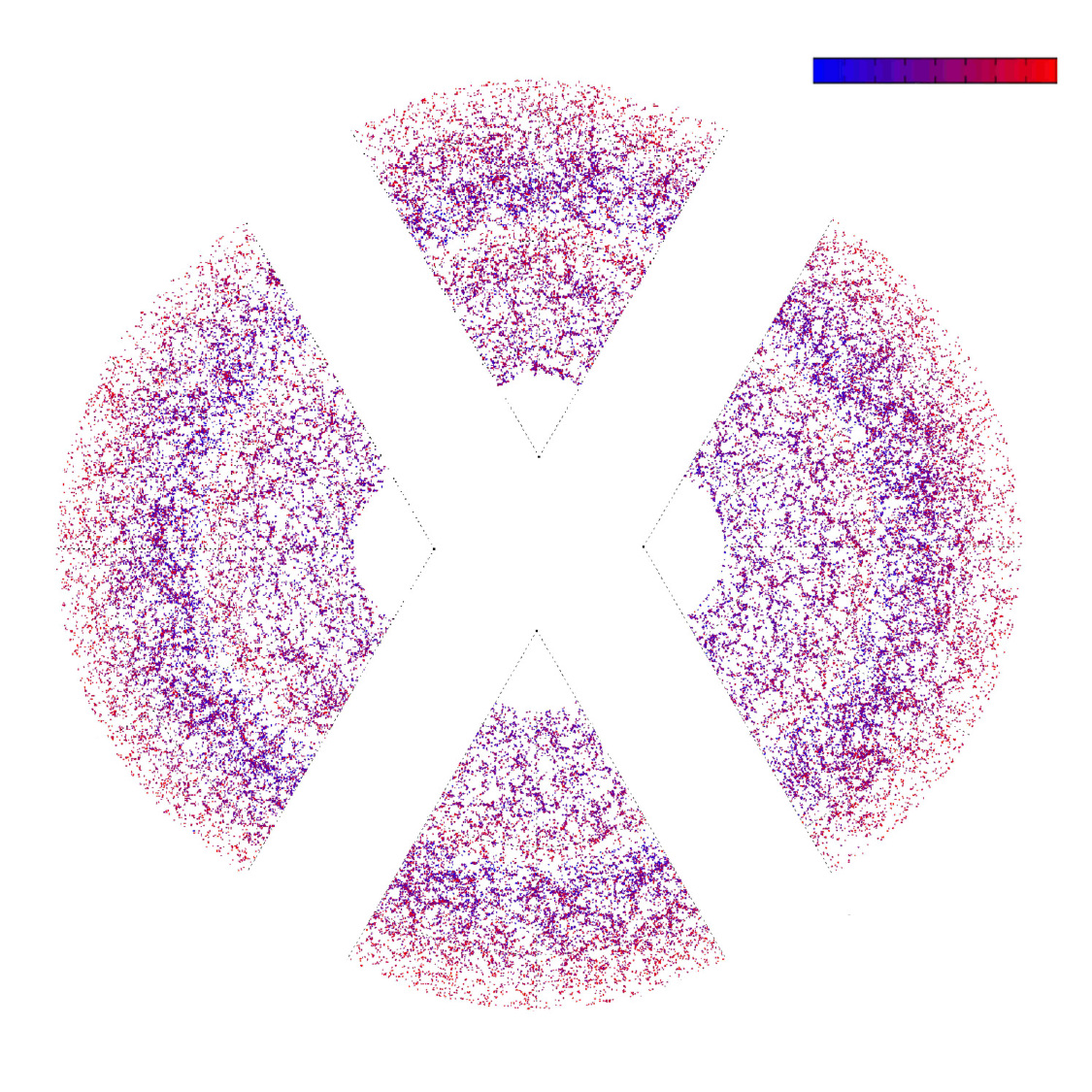}
\put(-380,380){\rotatebox[]{0}{\Large\text{BOSS DR12}}}
\put(-125,45){\rotatebox[]{0}{\Large\text{MULTIDARK}}}
\put(-125,30){\rotatebox[]{0}{\Large\text{PATCHY MOCKS DR12}}}
\put(-20,210){\rotatebox[]{-90}{180$^{\rm o}$}}
\put(-417,210){\rotatebox[]{90}{180$^{\rm o}$}}
\put(-10,210){\rotatebox[]{-90}{\text{Northern Galactic Cap}}}
\put(-427,210){\rotatebox[]{90}{\text{Northern Galactic Cap}}}
\put(-40,290){\rotatebox[]{-90}{210$^{\rm o}$}}
\put(-397,290){\rotatebox[]{90}{150$^{\rm o}$}}
\put(-100,350){\rotatebox[]{-90}{240$^{\rm o}$}}
\put(-337,350){\rotatebox[]{90}{120$^{\rm o}$}}
\put(-40,130){\rotatebox[]{-90}{150$^{\rm o}$}}
\put(-397,130){\rotatebox[]{90}{210$^{\rm o}$}}
\put(-100,70){\rotatebox[]{-90}{120$^{\rm o}$}}
\put(-337,70){\rotatebox[]{90}{240$^{\rm o}$}}
\put(-145,380){\rotatebox[]{0}{330$^{\rm o}$}}
\put(-297,380){\rotatebox[]{0}{30$^{\rm o}$}}
\put(-145,40){\rotatebox[]{0}{30$^{\rm o}$}}
\put(-297,40){\rotatebox[]{0}{330$^{\rm o}$}}
\put(-255,410){\rotatebox[]{0}{\text{Southern Galactic Cap}}}
\put(-220,400){\rotatebox[]{0}{0$^{\rm o}$}}
\put(-220,23){\rotatebox[]{0}{0$^{\rm o}$}}
\put(-255,13){\rotatebox[]{0}{\text{Southern Galactic Cap}}}
\put(-192,280){\rotatebox[]{0}{2.7}}
\put(-252,280){\rotatebox[]{0}{0.2}}
\put(-172,315){\rotatebox[]{0}{5.1}}
\put(-272,315){\rotatebox[]{0}{0.4}}
\put(-152,350){\rotatebox[]{0}{7.3}}
\put(-292,350){\rotatebox[]{0}{0.6}}
\put(-195,300){\rotatebox[]{60}{\text{lookback time in billions of years}}}
\put(-280,300){\rotatebox[]{-60}{\text{redshift}}}
\put(-260,233){\rotatebox[]{-58}{\Huge$\leftharpoondown$}} 
\put(-190,183){\rotatebox[]{-238}{\Huge$\leftharpoondown$}} 
\put(-275,220){\rotatebox[]{0}{\text{OBSERVATIONS}}}
\put(-225,200){\rotatebox[]{0}{\text{THEORY/MODEL}}}
\put(-192,145){\rotatebox[]{0}{2.7}}
\put(-252,145){\rotatebox[]{0}{0.2}}
\put(-172,110){\rotatebox[]{0}{5.1}}
\put(-272,110){\rotatebox[]{0}{0.4}}
\put(-152,75){\rotatebox[]{0}{7.3}}
\put(-292,75){\rotatebox[]{0}{0.6}}
\put(-73,410){\rotatebox[]{0}{\text{log M$_*$}}}
\put(-115,388){\rotatebox[]{0}{\tiny\text{10.9}}}
\put(-90,388){\rotatebox[]{0}{\tiny\text{11.1}}}
\put(-67,388){\rotatebox[]{0}{\tiny\text{11.3}}}
\put(-43,388){\rotatebox[]{0}{\tiny\text{11.5}}}
\put(-20,388){\rotatebox[]{0}{\tiny\text{12.0$<$}}}
\caption{\label{fig:pie}   Pie plot of the BOSS DR12 observations (upper left region), and one \textsc{MultiDark patchy} mock realization (lower right region).}
\end{figure*}
 
\label{sec:rsd}

Let us recap here the way in which RSDs are treated in the \textsc{patchy}-code \citep[see][]{KitauraPatchy}. 

The mapping between Eulerian real space $\mbi x(z)$ and redshift space $\mbi s(z)$ is given by: $\mbi s(z)=\mbi x(z)+\mbi v_r(z)$, with $\mbi v_r\equiv(\mbi v\cdot\hat{\mbi r})\hat{\mbi r}/(Ha)$; where  $\hat{\mbi r}$ is the unit sight line vector, $H$ the Hubble constant, $a$ the scale factor, and $\mbi v=\mbi v(\mbi x)$ the 3-d velocity field interpolated at the position of each halo in Eulerian-space $\mbi x$ using the displacement field $\mbi\Psi_{\rm ALPT}(\mbi q,z)$.
We split the peculiar velocity field into a coherent $\mbi v^{\rm coh}$ and a (quasi) virialized component $\mbi v_{\sigma}$: $\mbi v=\mbi v^{\rm coh}+\mbi v^{\sigma}$. 
The coherent peculiar velocity field is computed in Lagrangian-space from the linear Gaussian field $\delta^{(1)}(\mbi q)$ using the ALPT formulation consistently with the displacement field (see Eq.~\ref{eq:disp}):
\be
\mbi v_{\rm ALPT}^{\rm coh}(\mbi q,z)={\cal K}(\mbi q,r_{\rm S}) \circ \mbi v_{\rm 2LPT}(\mbi q,z)+\left(1-{\cal K}(\mbi q,r_{\rm S}) \right)\circ \mbi v_{\rm SC}(\mbi q,z)\,,
\ee
with $\mbi v_{\rm 2LPT}(\mbi q,z)$ being the second order and $\mbi v_{\rm SC}(\mbi q,z)$ being the spherical collapse component \citep[for details see][]{KitauraPatchy}.

We use the high correlation between the local density field and the velocity dispersion to model the displacement due to (quasi) virialized motions. Effectively, we sample a Gaussian distribution function ($\mathcal G$) with a dispersion given by $\sigma_v\propto\left(1+b^{\rm ALPT}\delta^{\rm ALPT}\left(\mbi x\right)\right)^\gamma$. Consequently, 
\be
\mbi v^{\sigma}_r\equiv(\mbi v^{\sigma}\cdot\hat{\mbi r})\hat{\mbi r}/(Ha)={\mathcal G}\left(g\times\left(1+\delta^{\rm ALPT}\left(\mbi x\right)\right)^\gamma\right)\hat{\mbi r}\,.
\ee 
 
For the Gaussian streaming model see \citet[][]{ReidWhite11}, for non-Gaussian models see e.g.~\citet[][]{Tinker07}.
  In closely virialized systems the kinetic energy approximately equals the gravitational energy and a Keplerian law predicts $\gamma$ close to $0.5$, leaving only the proportionality constant $g$ as a free parameter in the model  \citep[see also][]{hesscs}.
We assign larger dispersion velocities to low mass objects considered to be satellites.
The parameters $g$ and $\gamma$ have been adjusted to fit the damping effect in the monopole and quadrupole as found in the {BigMultiDark} $N$-body simulation first and later further constrained with the BOSS DR12 data for different redshift bins (see discussion in \S \ref{sec:results}).

\subsubsection{Halo/stellar mass distribution reconstruction}
\label{sec:mass}

Once we have a spatial distribution of objects $\{\mbi{r}_{\rm g}\}$ which accurately reproduce the clustering of the whole galaxy sample at a given redshift, we assign the halo/stellar masses $M_{\rm g}^l$ to each object $l$ according to the statistical information extracted from the {BigMultiDark} simulation using the {\bf H}alo m{\bf A}ss {\bf D}istribution {\bf R}econstructi{\bf ON}  (\textsc{hadron}) code \citep[for technical details see][]{Zhao15}. In particular we sample the following conditional probability distribution function 
\be
M_{\rm g}^l\curvearrowleft P(M_{\rm g}^l|\{\mbi{r}_{\rm g}\},\rho_{\rm M},T,\Delta r_{\rm min}^{\rm M},\{p_{\rm c}\},z)\,,
\ee 
where  $\rho_{\rm M}$ is the local density, $T$ the tidal field tensor (in particular the eigenvalues), $\Delta r_{\rm min}^{\rm M}$ a minimum separation between massive objects due to exclusion effects, $\{p_{\rm c}\}$ a set of cosmological parameters, and $z$ the redshift at which we want to apply the mass reconstruction. We note that at this stage we consider nonlocal biasing through the tidal field tensor and the minimum separation of objects.
Using all this information it has been proven that one can recover compatible  clustering for arbitrary halo mass cuts with the $N$-body simulation up to scales of about $k=0.3\,h^{-1}$ Mpc \citep[][]{Zhao15}. We extend the algorithm to stellar masses including the rank ordering relation and scatter described in \S \ref{sec:ham}.

\subsubsection{Survey generator}
\label{sec:survey}

The {\bf SU}rvey {\bf G}ener{\bf A}to{\bf R}  (\textsc{sugar}) code is an openMP code which constructs light-cones from mock galaxy catalogues  \citep[see][companion paper]{2015arXiv150906404R}.
This code applies geometrical features of the survey, including the geometry (using the publicly available \textsc{mangle} mask; \citet[][]{Swanson08}), sector completeness, veto masks and radial selection functions.

The \textsc{sugar} code can construct light-cones using a single box or multiples boxes at different redshifts, in order to include the redshift evolution in the final catalogue.  The first step in the construction of the lightcone is to locate the observer ($z=0$) and to transform from comoving Cartesian coordinates to equatorial coordinates (RA,Dec) and redshift. To compute the observed redshift (redshift space) of an object, first we compute the comoving distance from the observer to the object, and then we transform it to redshift space following ${s} ={r}_{\rm c} + ({\mbi{v}\cdot\hat{\mbi{r}}})/{aH(z_{\rm real})}$ (see \S \ref{sec:rsd}), where $r_{\rm c}(z)$ is computed from
$r_{\rm c}(z)=\int\limits_0^{z_{\rm real}}\frac{c\dd z'}{H_0\sqrt{\Omega_{\rm M}(1+z')^3+\Omega_\Lambda}}$.

Once we compute the redshift of each galaxy,  we consider two options to select objects in the radial direction:
\begin{enumerate}
\item downsampling: this option preserves the clustering of the input box selecting objects randomly to have the desired number density. 
\item selecting by  halo property: this consists of rank ordering objects by a halo property and selecting them sequentially until the correct number density is obtained.
\end{enumerate}

\section{Results: statistical comparison between  the MultiDark patchy mocks and the BOSS DR12 data}
\label{sec:results}

Following the method described in \S \ref{sec:method}  we generate  12,288 mock light-cone galaxy catalogues for BOSS DR12\footnote{We have produced half the amount of mock catalogues for DR11, i.e., 1,024 for each LOWZ, CMASS, combined, southern and northern galactic cap.}  (2,048 for each LOWZ, CMASS, combined, southern and northern galactic cap). We call these catalogues  \textsc{MultiDark patchy} mocks,  \textsc{md patchy} mocks in short. {The corresponding computations required about 500,000 CPU hours (30-50 min for each box on 16 cores and a total of 40,960 boxes). Since each \textsc{patchy}+\textsc{hadron}-run requires less than 24 Gb shared memory for a grid with 960$^3$ cells, we were able to make use of 128 nodes with 32 Gb each  in parallel from the BSC Marenostrum facilities, taking about one week wall clock time for all 40,960 catalogues. The light-cone generation with \textsc{sugar} required an additional $\sim$1,000 CPU hours.  The equivalent computations based on  $N$-body simulations would have required about 9,000 million CPU hours ($\sim$2.3 million CPU hours for each light-cone). The effective number of particles is $\sim$(61,440)$^3$ (given that the reference catalogue required $3,840^3$ particles to resolve the objects we reproduce in the \textsc{md patchy} catalogues).}

\begin{figure*}
\begin{tabular}{cc}
\hspace{-1cm}
\includegraphics[width=8.cm]{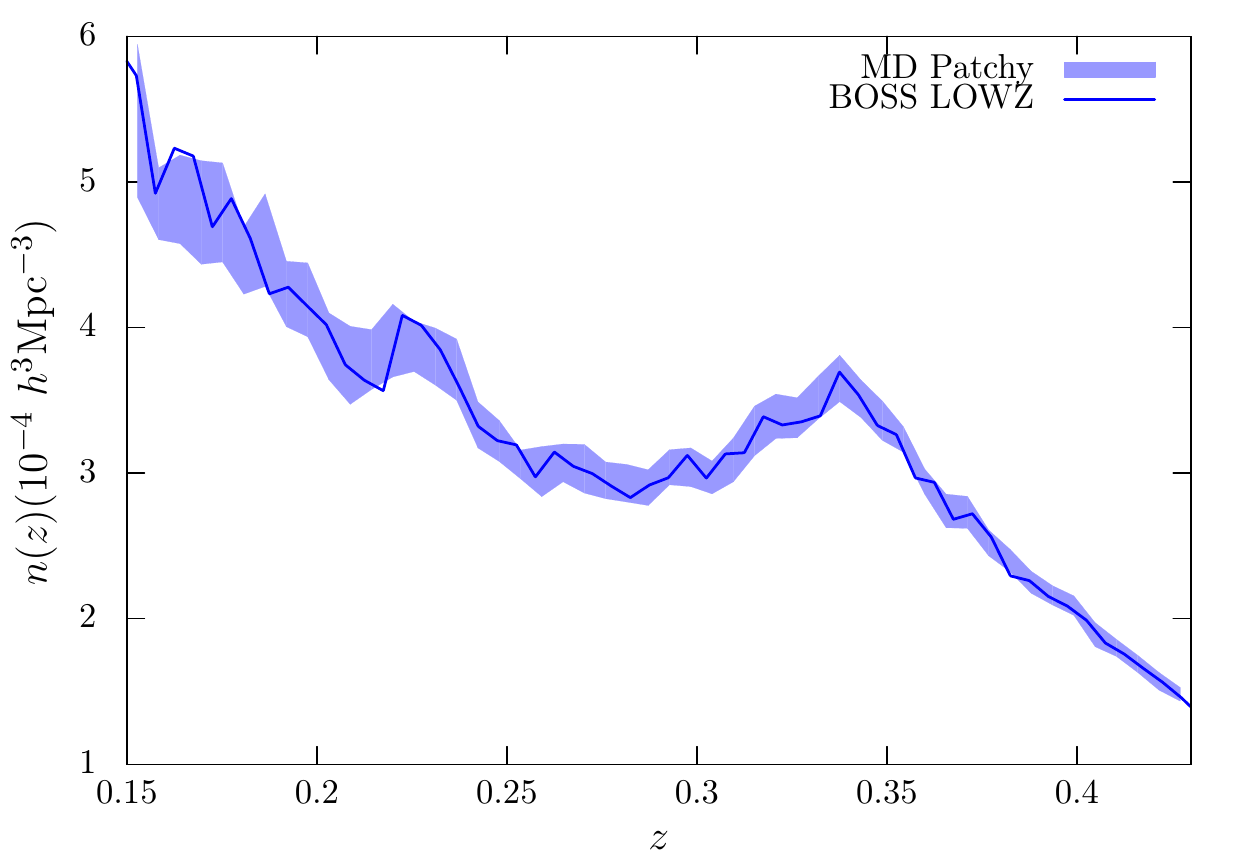}
\hspace{-0.45cm}
\includegraphics[width=7.5cm]{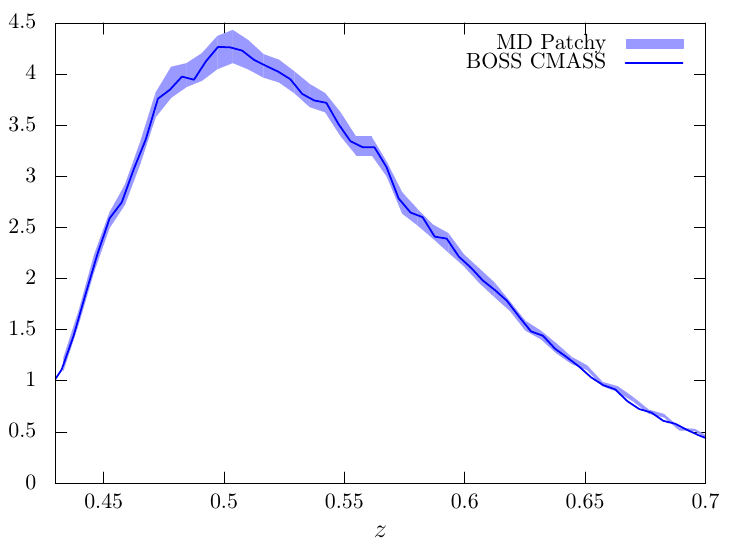}
\end{tabular}
\caption{\label{fig:ND} Number density for the LOWZ (left) and CMASS (right) samples. The observations are given by the blue solid lines. The shaded contours represent the 1-$\sigma$ regions according to the \textsc{md patchy} mocks.   }
\end{figure*}

\begin{figure*}
\begin{tabular}{cc}
\hspace{-1.cm}
\includegraphics[width=8.cm]{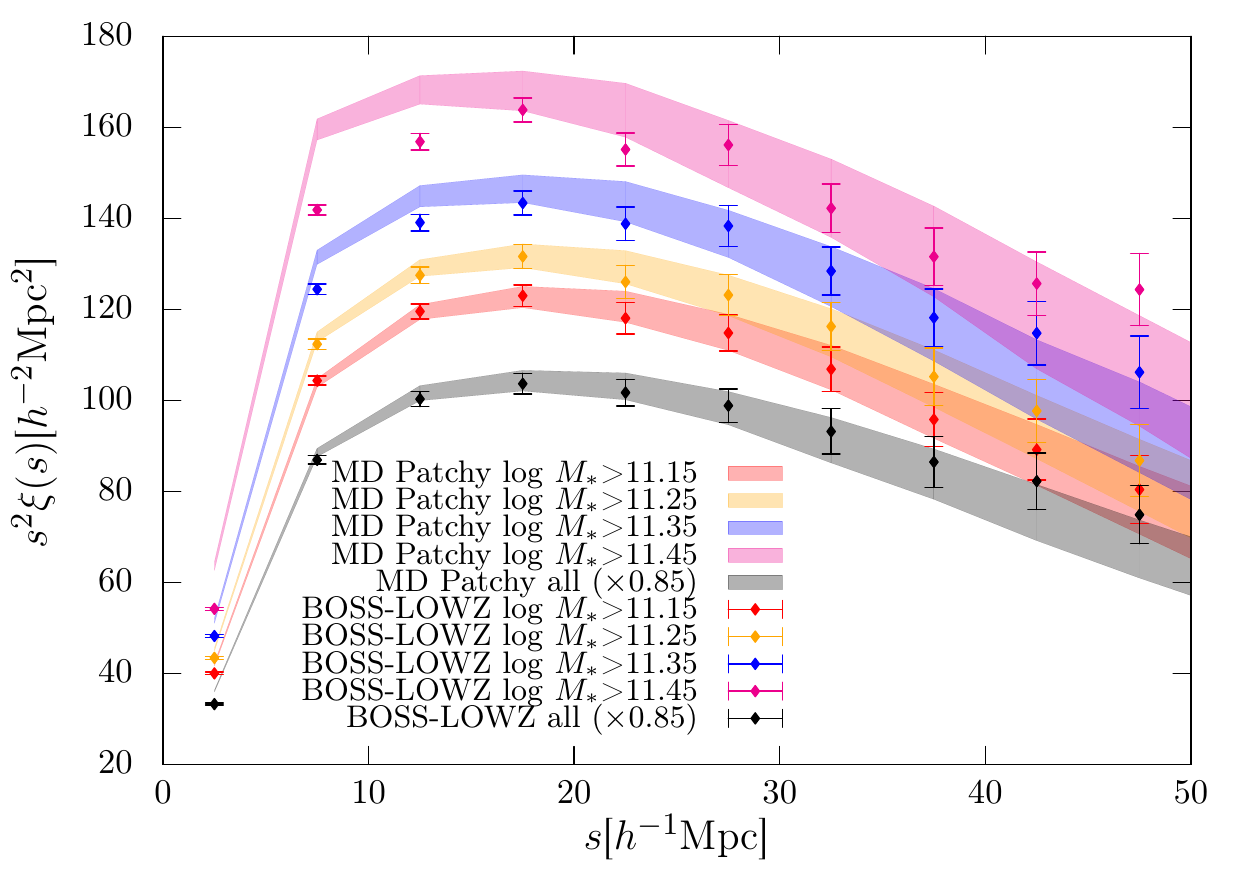}
\includegraphics[width=7.5cm]{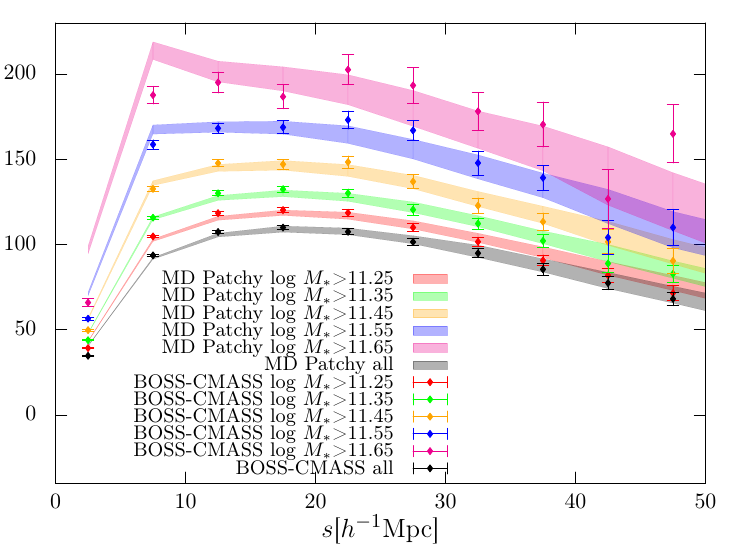}
\end{tabular}
\caption{\label{fig:stellar}  Monopole for different stellar mass bins as indicated in the legend with the corresponding colour code. The error bars represent the BOSS DR12 data.  The shaded contours represent the 1-$\sigma$ regions according to the \textsc{md patchy} mocks.}
\end{figure*}

We used 10 redshift bins to construct the light-cones. This permits us to obtain the  galaxy bias, the growth, and the peculiar motion evolution as a function of redshift.  A visualization of the BOSS DR12 and one \textsc{md patchy} mock realization is shown in Fig.~\ref{fig:pie}. We can clearly see from this plot that both the data and the mocks follow the same selection criteria including the survey mask (the colour code stands for the stellar mass), and there are no 
 obvious visual differences beyond cosmic variance. The empty regions seem to be similarly distributed for both cases, indicating that the three-point statistics should be close, and the statistical comparison between the \textsc{md patchy} mock galaxy catalogues and the observations of BOSS DR12 yield good agreement. The number densities for LOWZ and CMASS galaxy samples are recovered by construction (see Fig.~\ref{fig:ND}).
We investigate the performance of the mock galaxy catalogues in detail in the following subsections.

{  To avoid redundancy we show only the results for BOSS DR12, as the only difference with respect to the BOSS DR11 mocks is the applied mask and selection function.}


\begin{figure*}
\begin{tabular}{cc}
\includegraphics[width=9.cm]{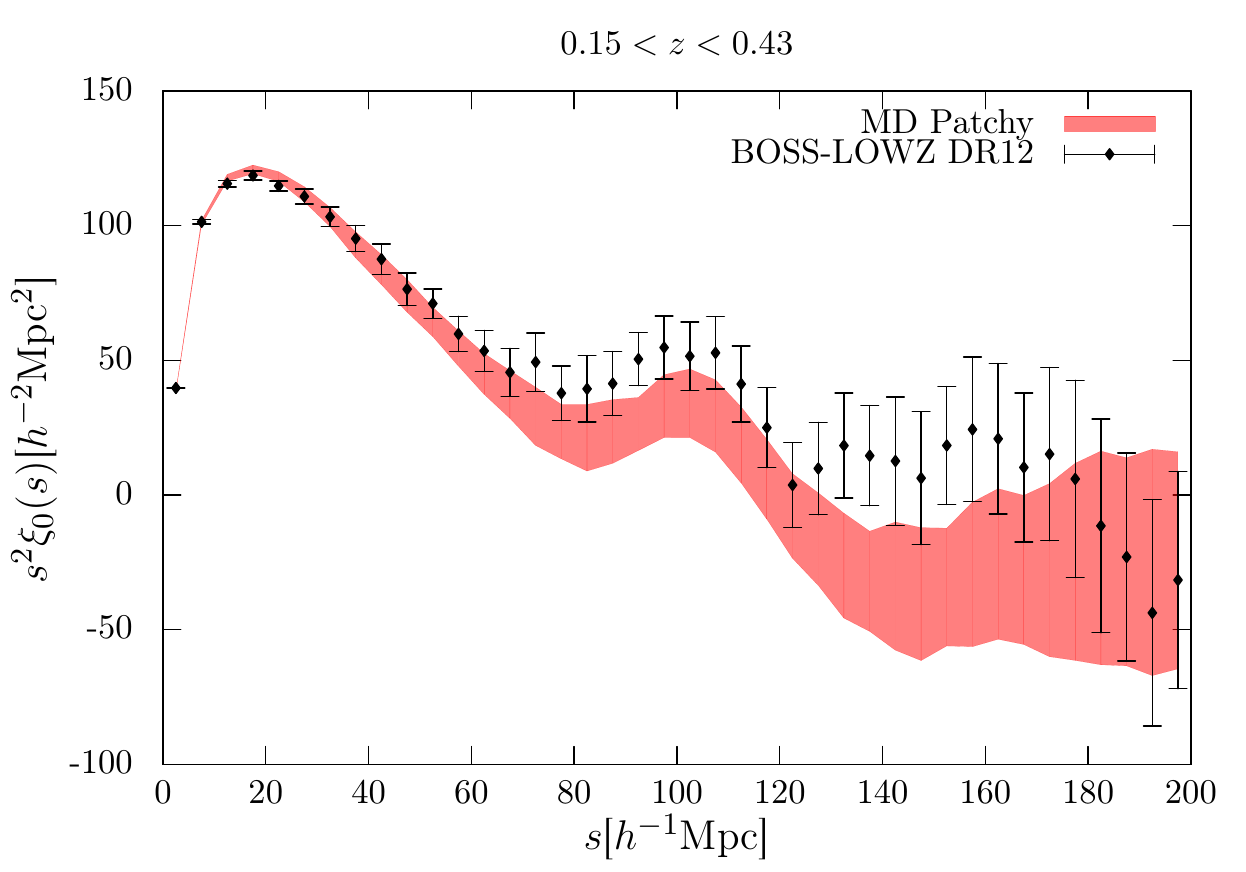}
\includegraphics[width=9.cm]{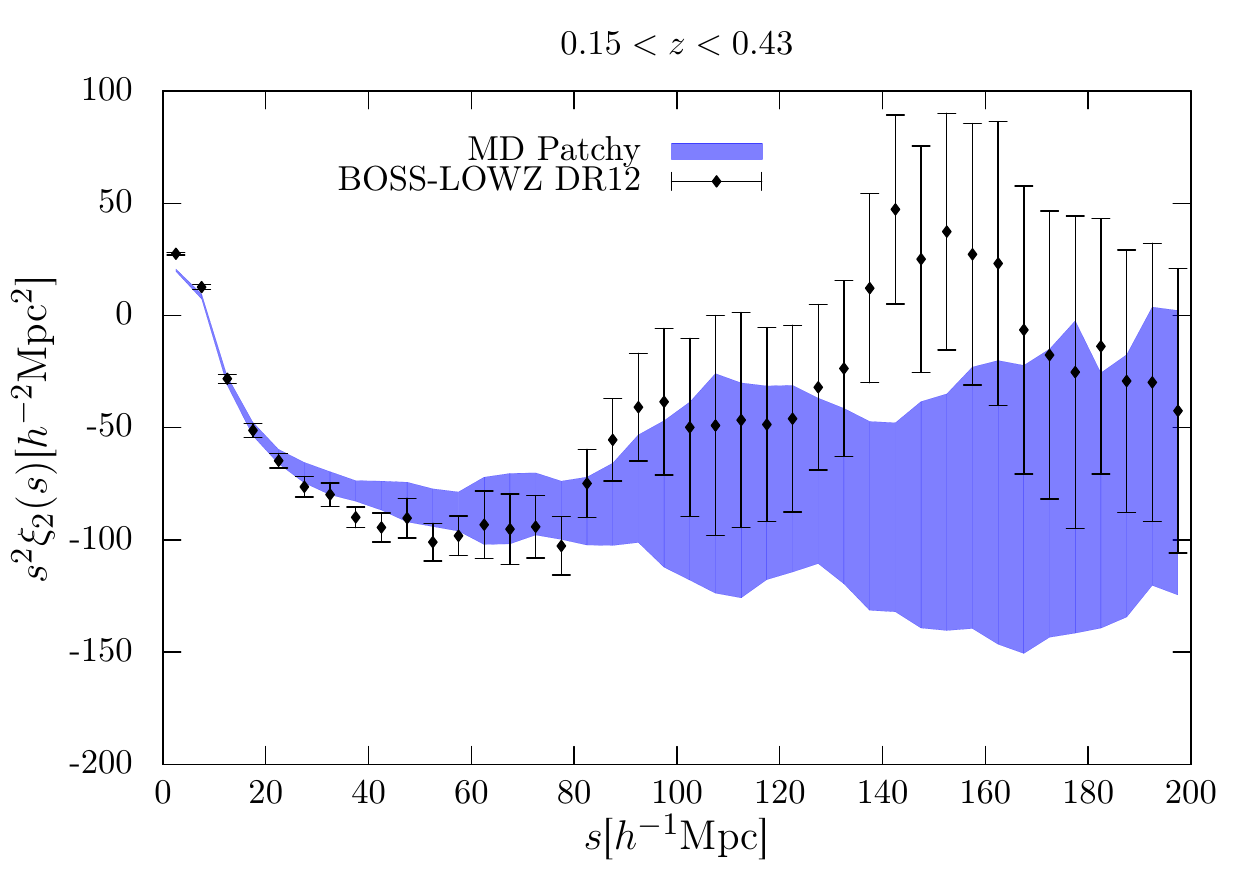}
\\
\includegraphics[width=9.cm]{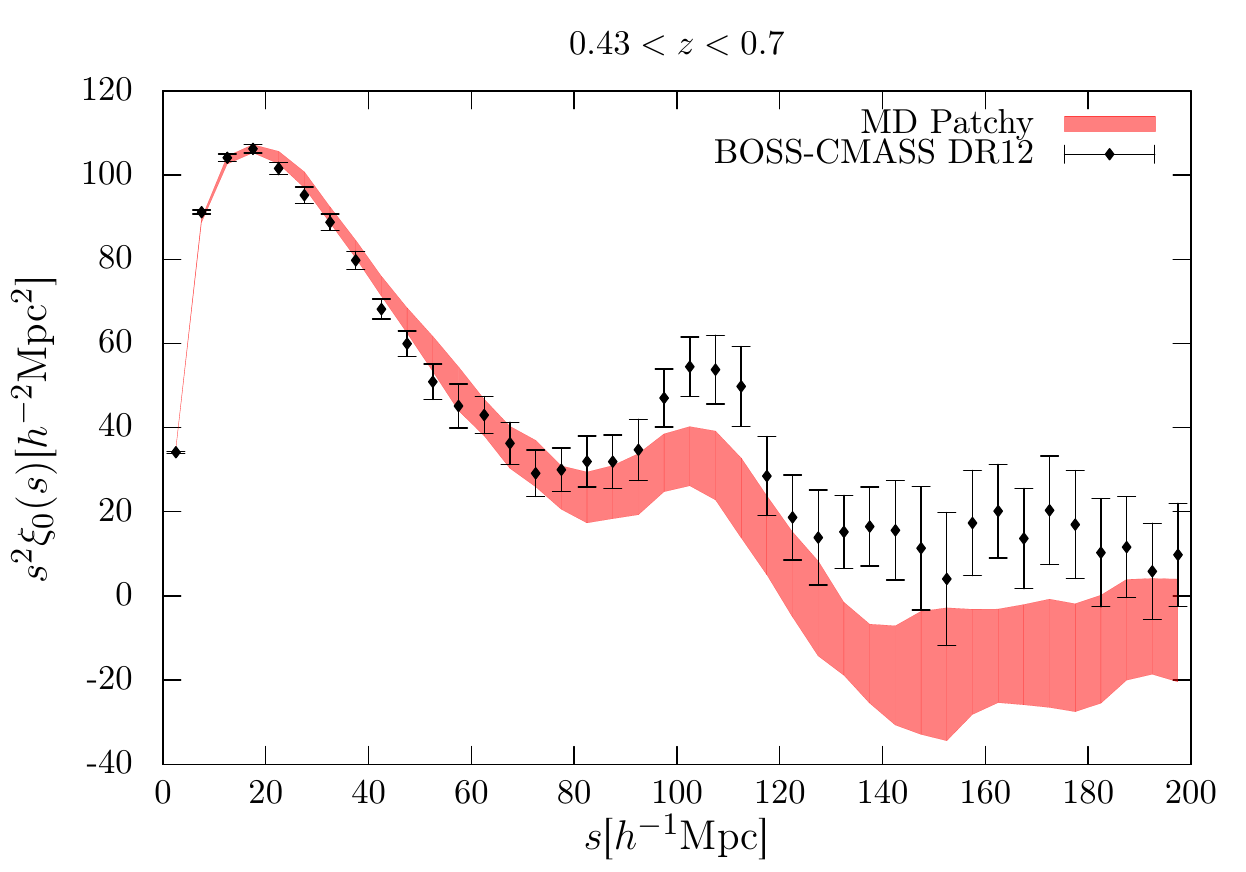}
\includegraphics[width=9.cm]{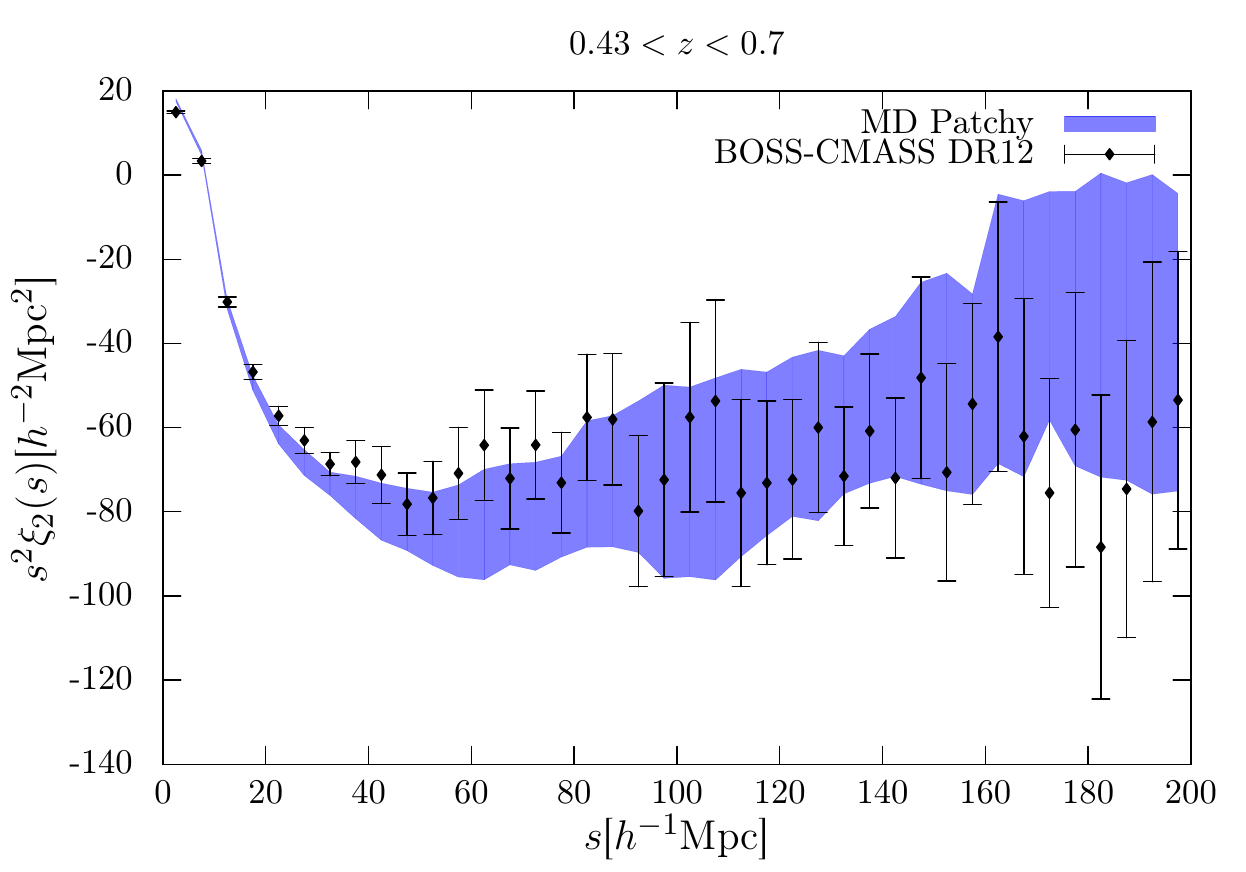}
\end{tabular}
\caption{\label{fig:CFQLOWZCMASS}  Monopole (on the left) and quadrupole  (on the right) for LOWZ and CMASS in the first and second rows, respectively. 
The shaded contours represent the 1-$\sigma$ regions according to the \textsc{md patchy} mocks, correlation function in red, quadrupole in blue.
}
\end{figure*}

\begin{figure*}
\begin{tabular}{cc}
\includegraphics[width=6.cm]{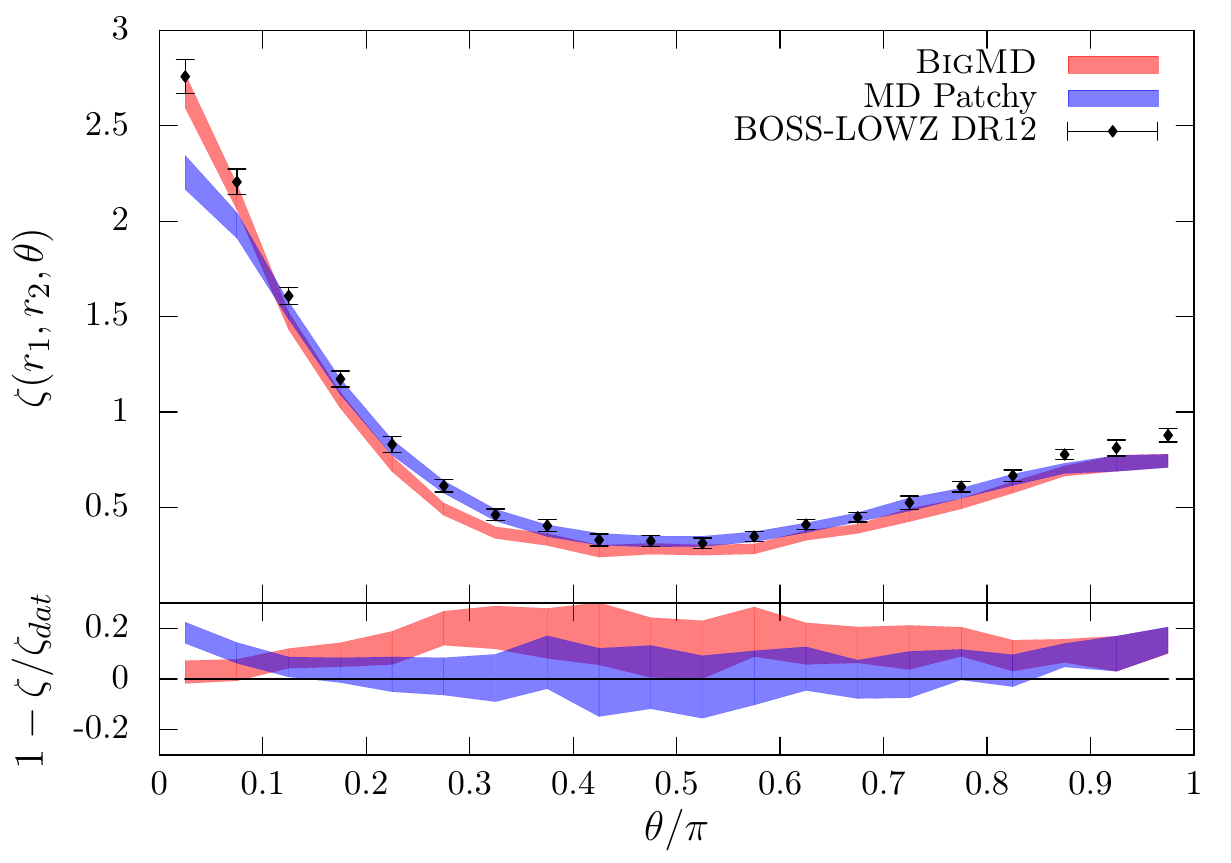}
\includegraphics[width=5.75cm]{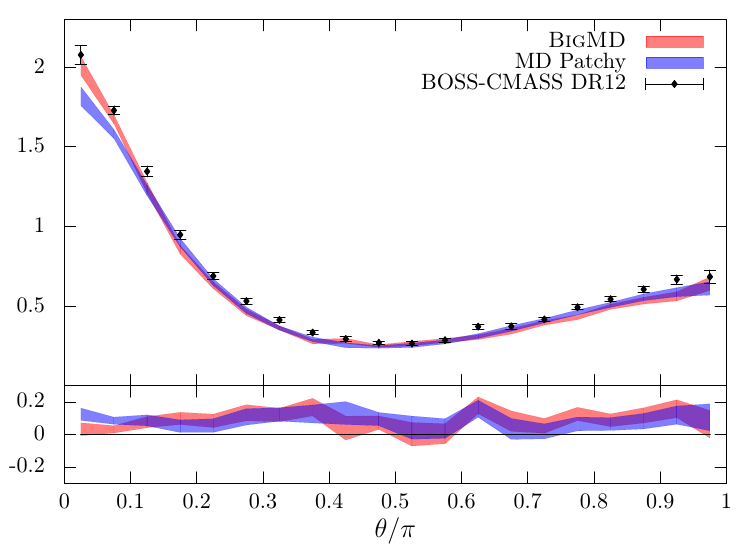}
\includegraphics[width=5.75cm]{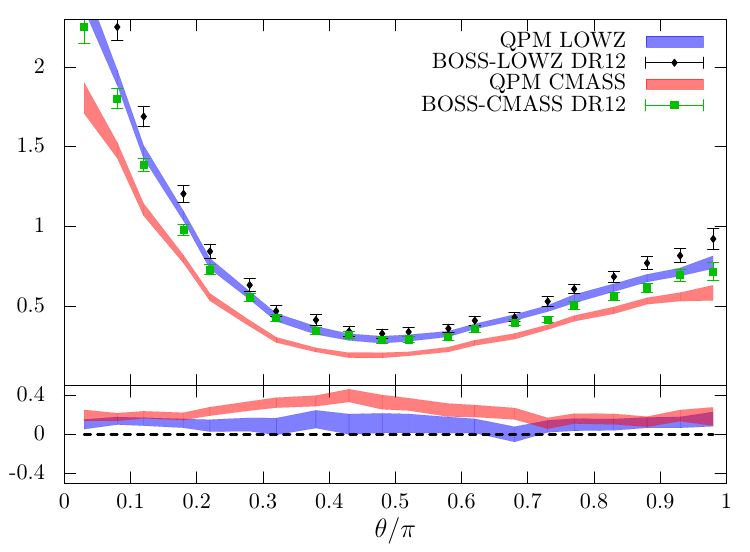}
\end{tabular}
\caption{\label{fig:3pt} Left-hand  and central panels: three-point statistics comparing the \textsc{md patchy} mocks (blue shaded region) with the {BigMultiDark} mocks of the $N$-body simulation (red shaded region) and the observations (black error bars) for LOWZ (left) and CMASS galaxies (central). Right-hand panel: three-point statistics comparing the \textsc{qpm} mocks (LOWZ: blue shaded region, CMASS: red shaded region) to the  observations (LOWZ: black error bars, CMASS: green error bars). Corresponding ratios are shown in the bottom panels. 
Shaded area shows 1-$\sigma$ uncertainties, $r_1 = 10$ and $r_2 = 20$ $h^{-1}$ Mpc and $\theta$ is the angle between $r_1$ and $r_2$
 $h^{-1}$ Mpc.}
\end{figure*}

\subsection{Two-point and three-point correlation functions}
\label{sec:xi}

We perform first an analysis in configuration space computing the two- and three-point correlation functions.
To compute the clustering signal in the correlation function for the \textsc{md patchy} mock lightcones and the observed data we rely on the \citet[][]{LandySzalay93} estimator. We will follow their notation
 referring to the data sample (either simulation or observed data) as $D$ and to the random catalogue as $R$. 
The correlation function is then constructed in the following way:
\be
\xi(s) = \frac{DD-2DR+RR}{RR}\,,
\ee
as a function of separation between pairs of galaxies in redshift space $s$.

\begin{figure*}
\begin{tabular}{cc}
\includegraphics[width=8.5cm]{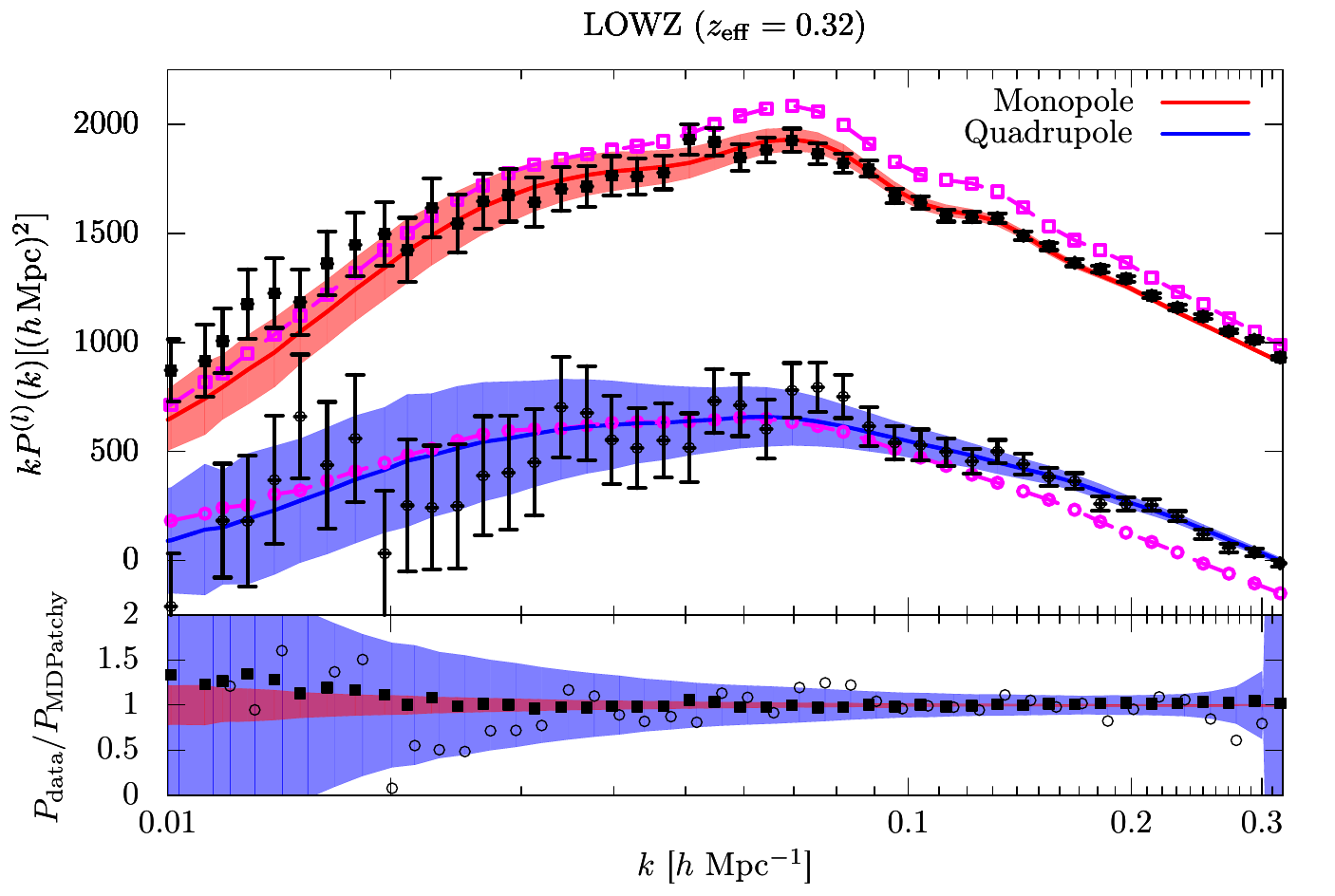}
\includegraphics[width=8.5cm]{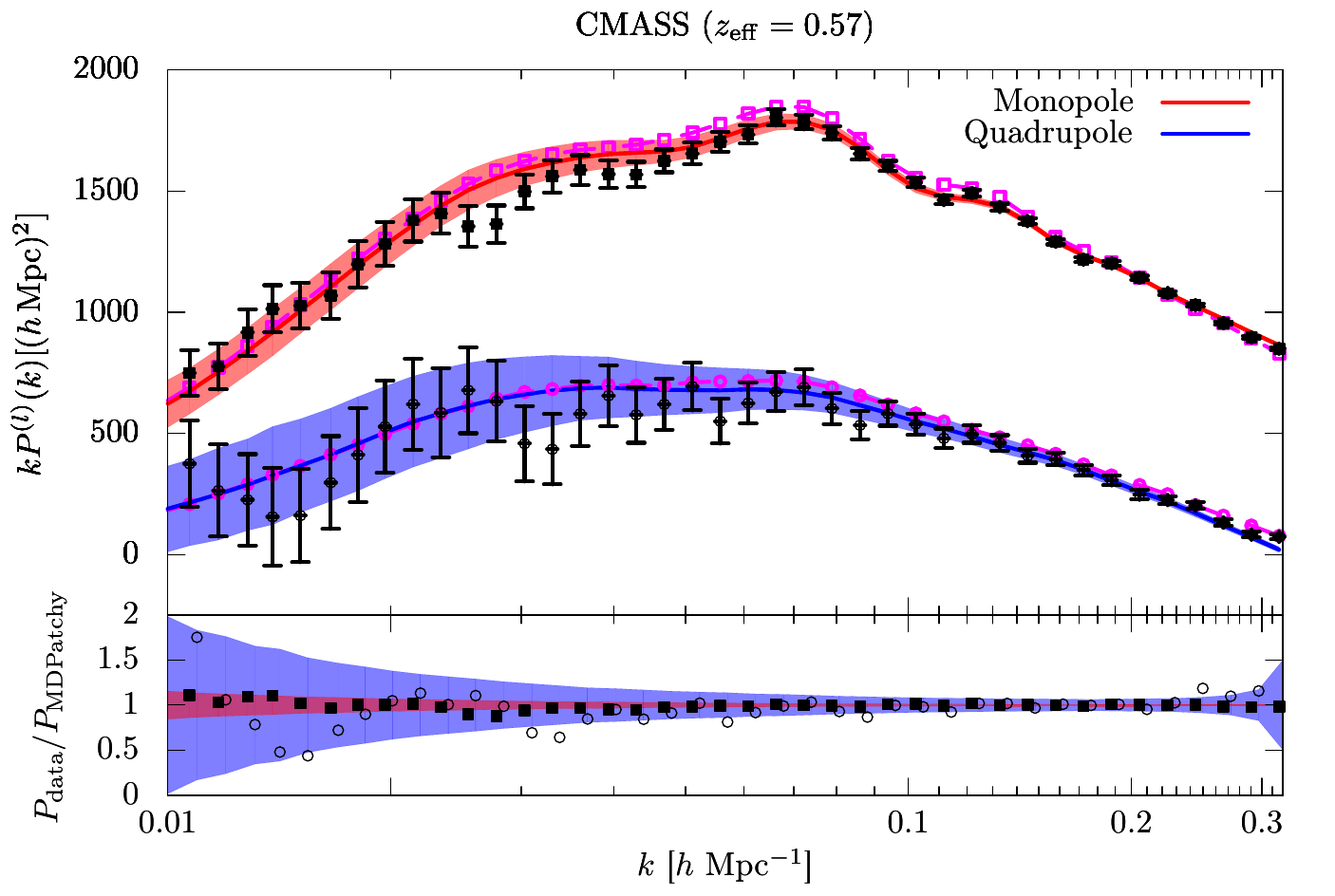}
\end{tabular}
\caption{\label{fig:MQFourier} Monopole (red) and quadrupole (blue) in Fourier space for the LOWZ (left) and CMASS galaxies (right) for the mean over 2,048 \textsc{md patchy} mocks for both southern and northern galactic caps, the average and 1-$\sigma$ uncertainties are shown. The results for \textsc{qpm} (1,000 mocks for each LOWZ/CMASS, and north/south) are shown with dashed magenta lines.   The error bars assigned to the data points have been computed based on 2,048 \textsc{md patchy} mocks. The ratio plots in the bottom panels have been only done for the \textsc{md patchy} mocks.
 }
\end{figure*}

The  three-point correlation function gives a description of the probability of finding three objects in three different volumes, 
and can be computed following \citet[][]{SzapudiSzalay98}
\be
\zeta(s_{12}, s_{23}, s_{13}) = \frac{DDD-3DDR+3DRR-RRR}{RRR}\,,
\ee
as a function of separation between the vertices of triangles spanned by triplets of galaxies in redshift space $s_{12}, s_{23}, s_{13}$.

Fig.~\ref{fig:stellar} shows that we accurately recover the clustering (monopole) for arbitrary stellar mass bins showing an almost perfect agreement with observations. Only for the two largest stellar mass bin, we find deviations larger than 1-$\sigma$. This is mainly due to the ``halo exclusion effect'', which is only approximately modelled, assuming a minimum separation for massive galaxies, and not the full separation distribution function \citep[][]{Zhao15}.  We find, however, that these differences are not critical, as they are restricted to small scales ($\lsim20\,h^{-1}$ Mpc) and only a low number of objects are affected.
 We further compute the monopole and quadrupole for LOWZ and CMASS (see Fig.~\ref{fig:CFQLOWZCMASS} and \S \ref{sec:evol}).
The monopole agrees towards small scales down to a few Mpc within 1-$\sigma$.

\begin{figure*}
\begin{tabular}{cc}
\includegraphics[width=8.cm]{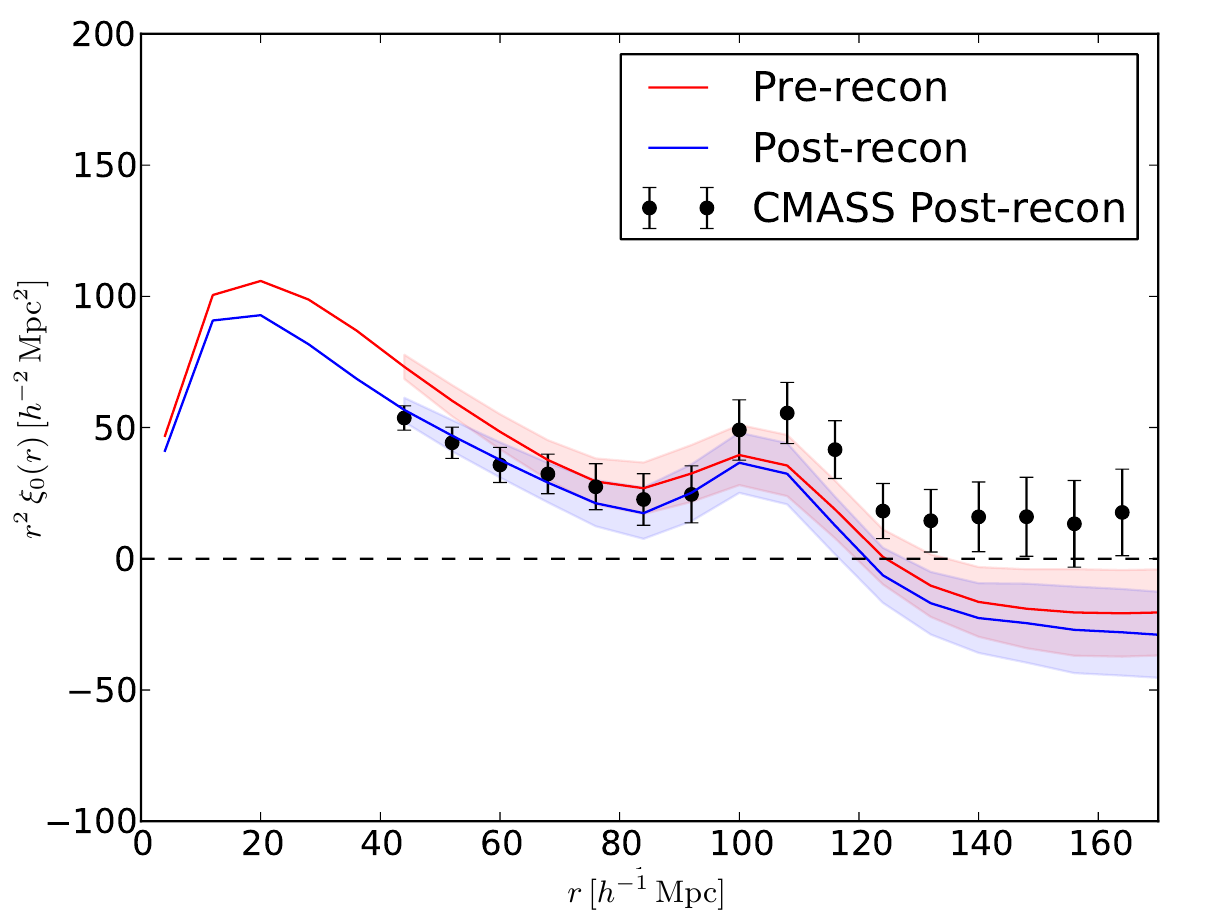}
\includegraphics[width=8.cm]{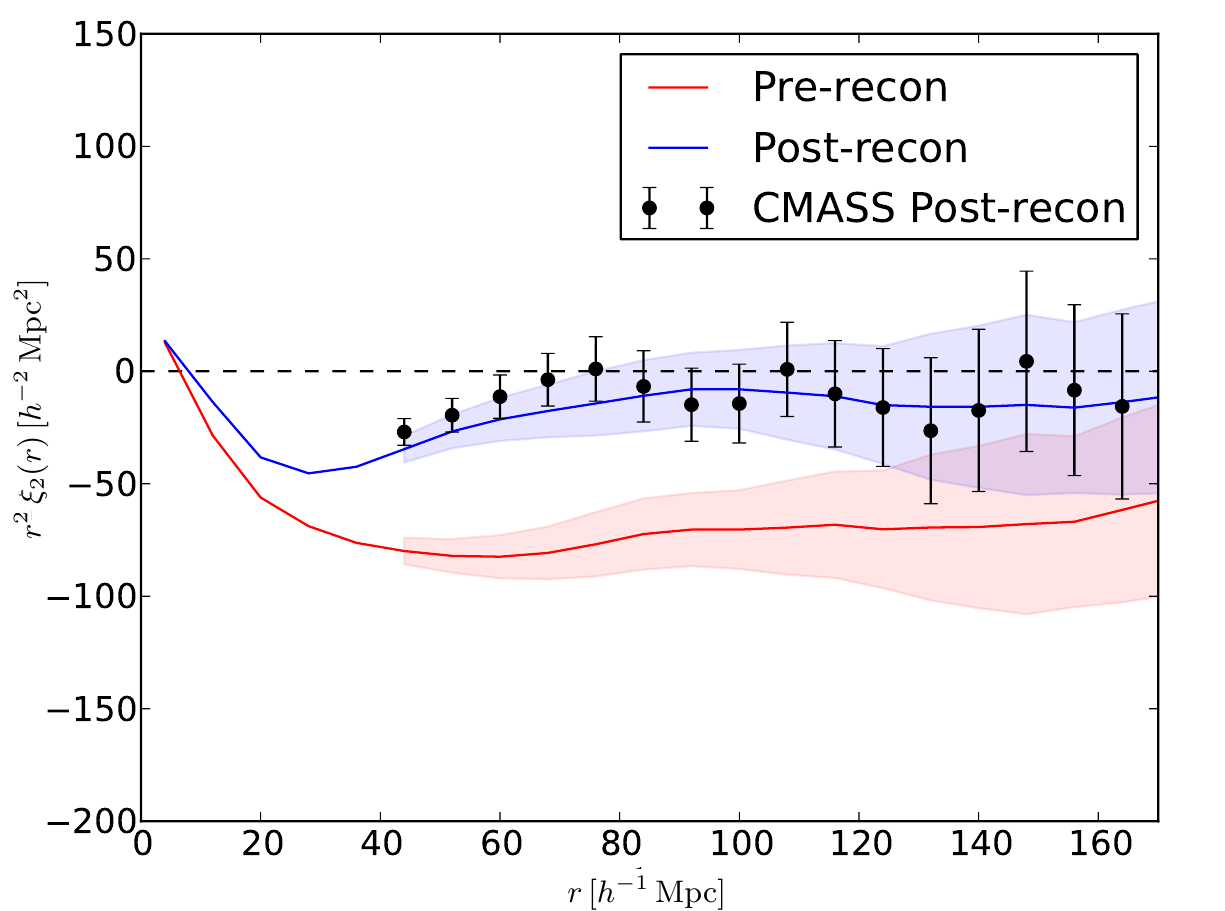}
\end{tabular}
\caption{\label{fig:baorec}  Monopole (on the left) and quadrupole (on the right) before and after BAO reconstruction (see Vargas-Magana et al.~in prep.). The error bars represent the BOSS DR12 data.  The solid lines correspond to the mean, and the shaded contours represent the 1-$\sigma$ regions, according to the \textsc{md patchy} mocks (red pre-, and blue post-reconstruction).}
\end{figure*}

There is a deviation of the monopole around the BAO peak and towards larger scales. While the galaxy mock catalogues cross zero right after the BAO peak, the observations do not. In this study, we have applied all of the systematic weights, such as the stellar density contamination, detailed in \citet[][]{Reid15} and Ross et al. (in prep.). The correlation function measurements are quite covariant between s bins at these scales, making the deviations less significant than one would expect by the visual impression. The significance and potential causes of the large-scale excess are studied in Ross et al. (in prep.), where it is also shown that it has no significant impact on BAO measurements. { This is even more so, as the overall shape of $\xi(s)$ in BAO measurements is marginalized over with a polynomial  \citep[see e.g.][]{AAB14}. See also \citet[][]{Ross12,Chuang13} for similar studies on an earlier BOSS data set and \citet[][]{Huterer13} for potential photometric calibration systematics, which have not been accounted for in this analysis.}

In the case of RSD measurements one has to make sure that the analysis is performed on scales which are not affected by systematics  \citep[][companion paper]{2015arXiv150906386G}. 
 The quadrupole is in very good agreement on all scales,  further supporting that RSD analysis should be safe, even in case there are some remnant systematics in the data.

An investigation of the three-point function demonstrates that the \textsc{md patchy} mocks have a quality very similar to those based on $N$-body simulations after calibration (see left-hand and central panels in Fig.~\ref{fig:3pt}).
We have constrained the galaxy bias parameters (see \S \ref{sec:bias}, \ref{sec:stoc}, and \ref{sec:mass}) based on the reference catalogues from the {BigMultiDark} simulation on cubical full volumes at each of the 10 redshift bins, matching the two- and the three-point statistics. To fit the latter we focused on matching the higher order correlation functions through the probability distribution function of galaxies in the reference catalogues following the approach presented in \citet[][]{KGS15}. Using the observations to constrain the three point statistics is not trivial, due to incompleteness effects. This explains why the \textsc{md patchy} mock catalogues better fit  the reference catalogue than the data, especially for the CMASS galaxies.  The three-point statistics performs worse for the \textsc{qpm} mocks, possibly because they do not include an iterative validation step fitting higher order statistics (beyond the two-point correlation function).
The nonlinear RSD parameter (see \S \ref{sec:rsd}) was iteratively constrained based on the observations, as we explain in the next section.

\begin{figure*}
\begin{tabular}{ccc}
\hspace{-1.3cm}
\includegraphics[width=6.8cm]{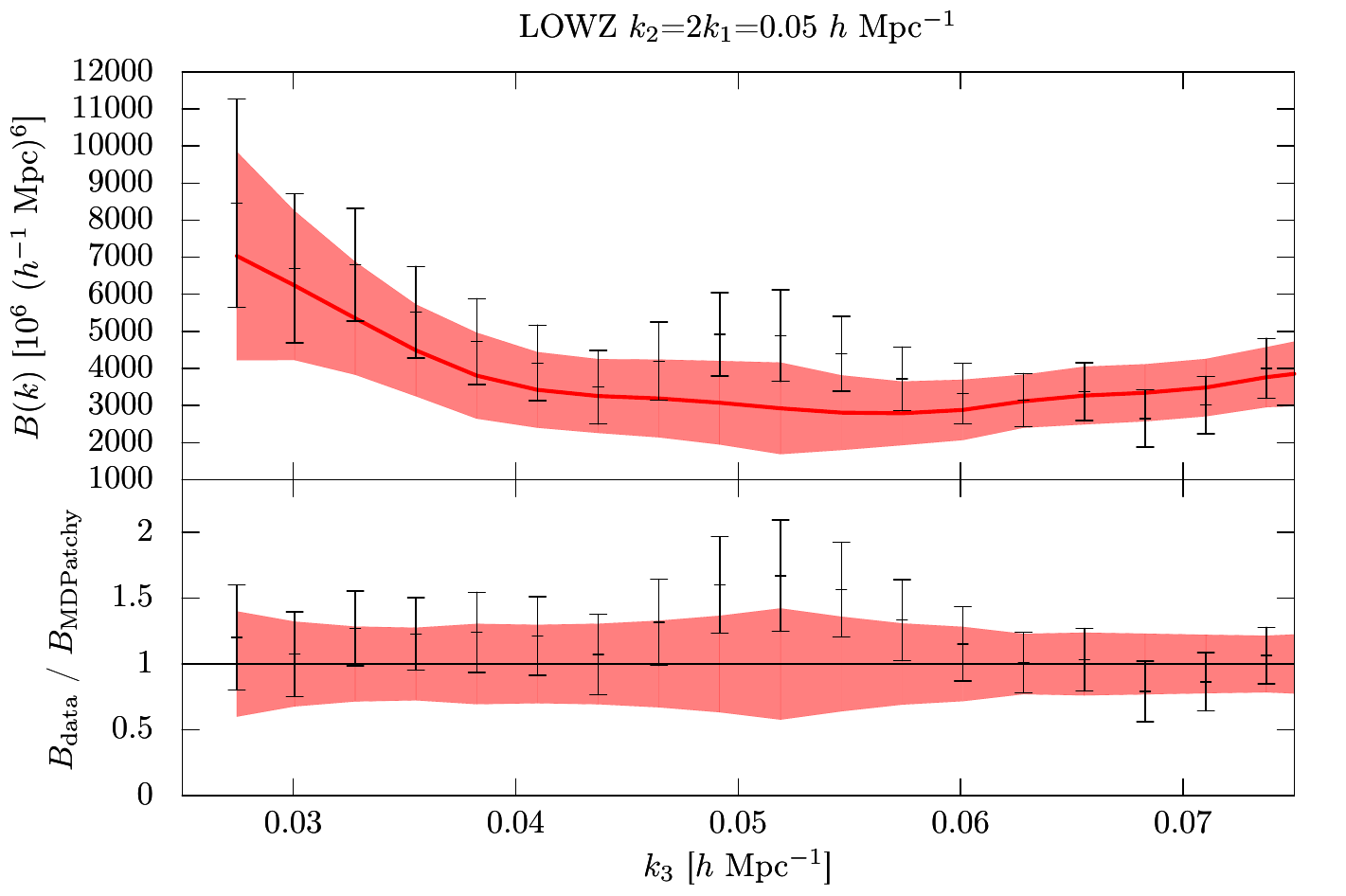}
\hspace{-0.2cm}
\includegraphics[width=6.73cm]{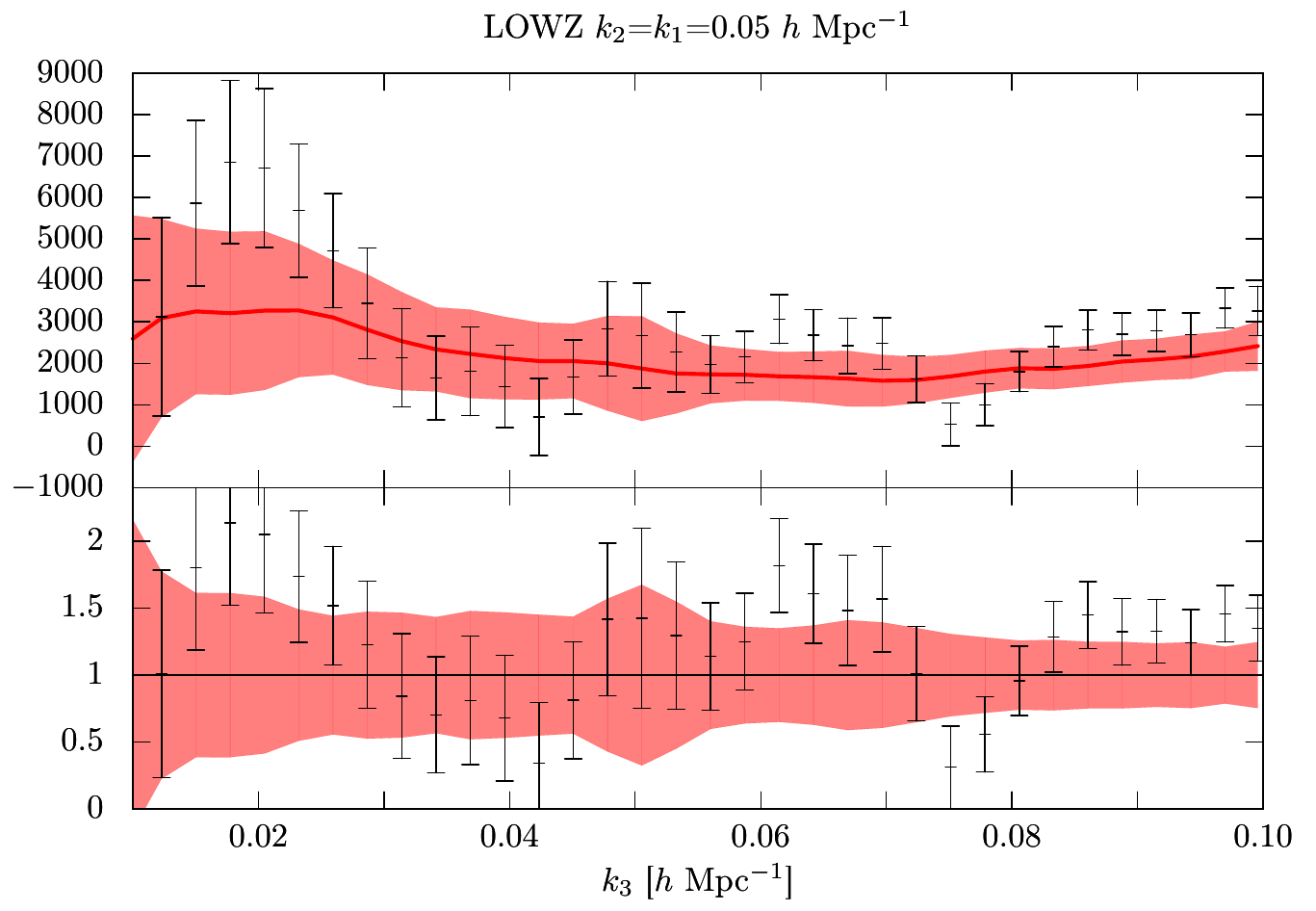}
\hspace{-0.2cm}
\includegraphics[width=6.6cm]{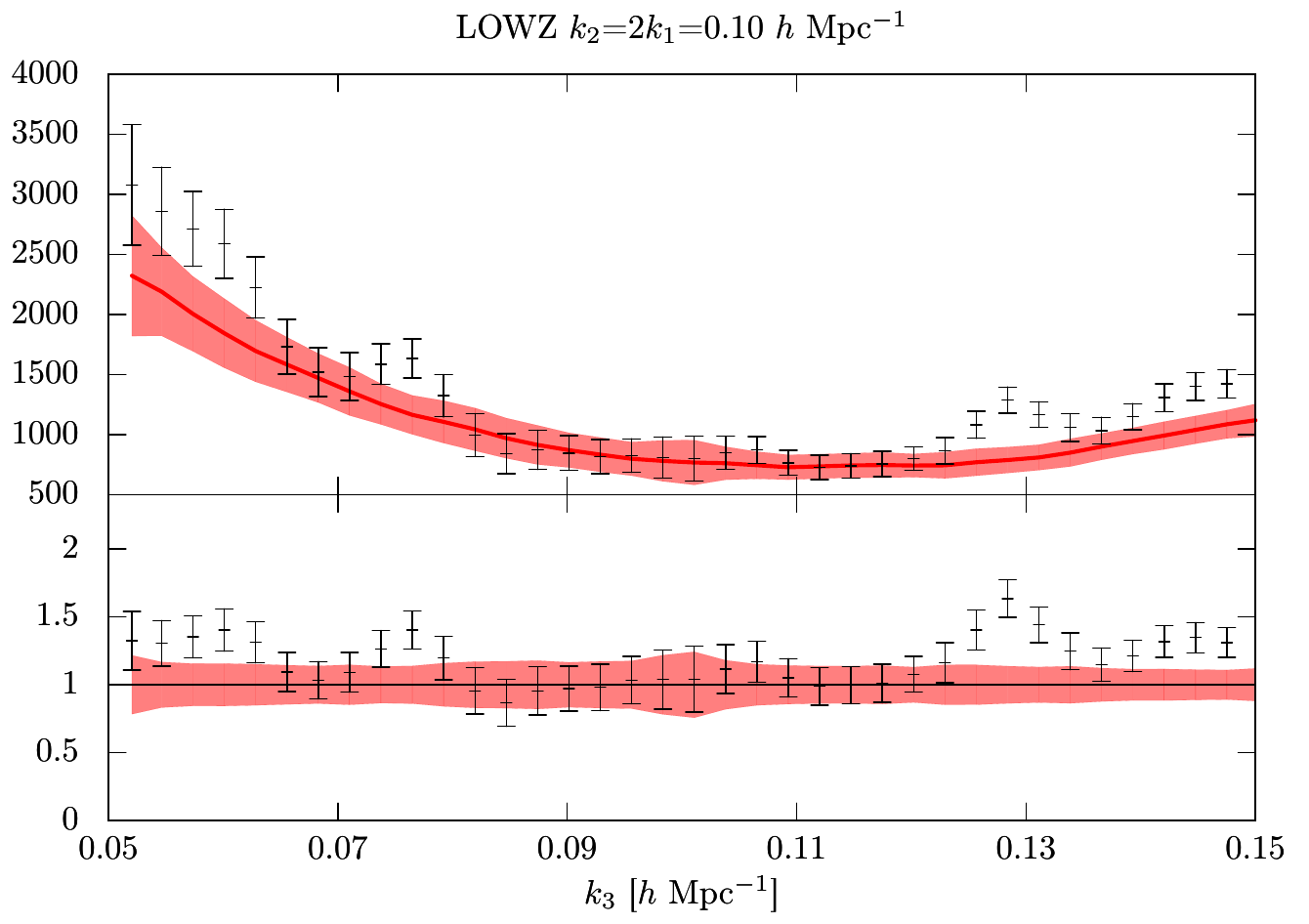}
\\
\hspace{-1.3cm}
\includegraphics[width=6.9cm]{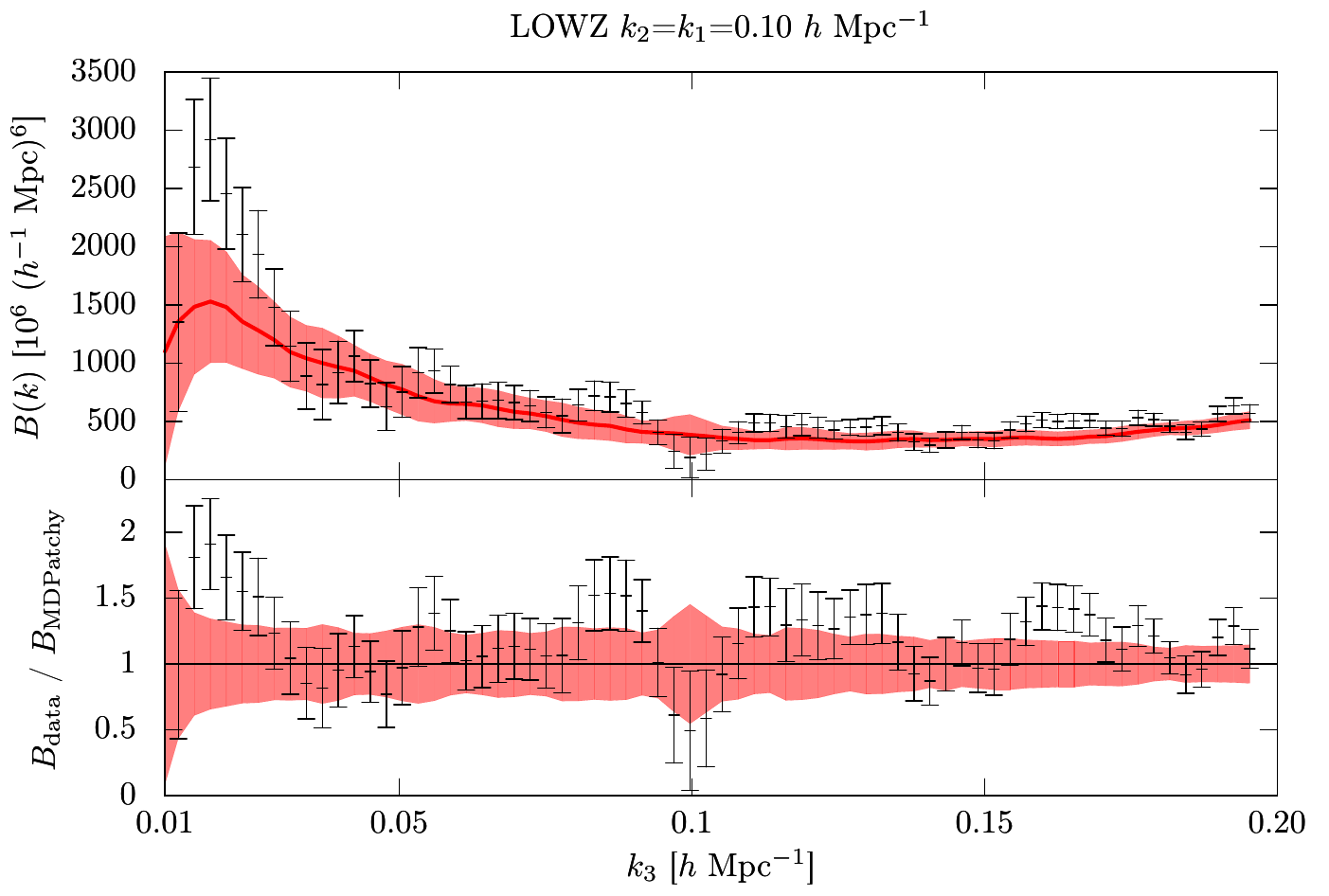}
\hspace{-0.2cm}
\includegraphics[width=6.6cm]{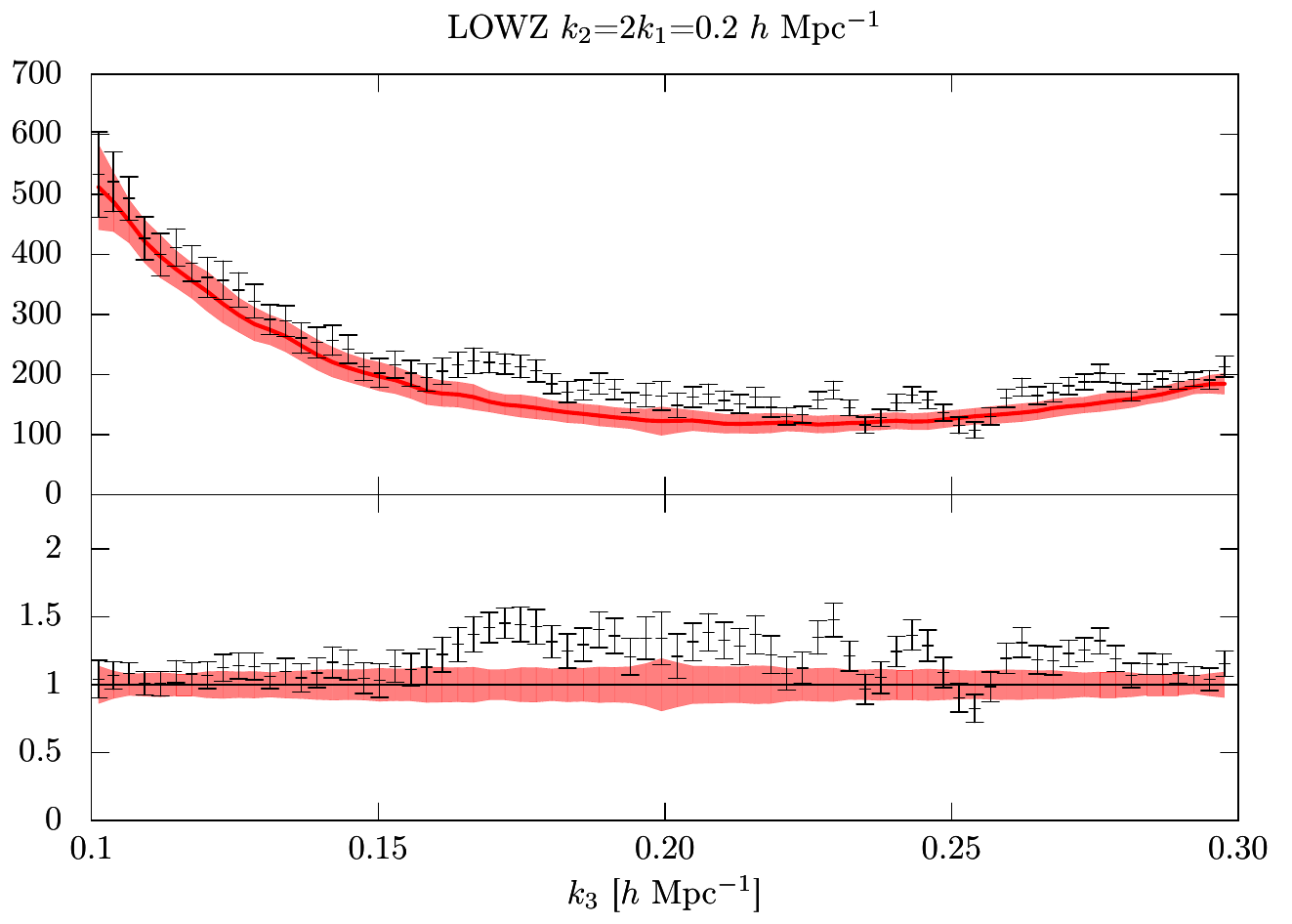}
\hspace{-0.2cm}
\includegraphics[width=6.7cm]{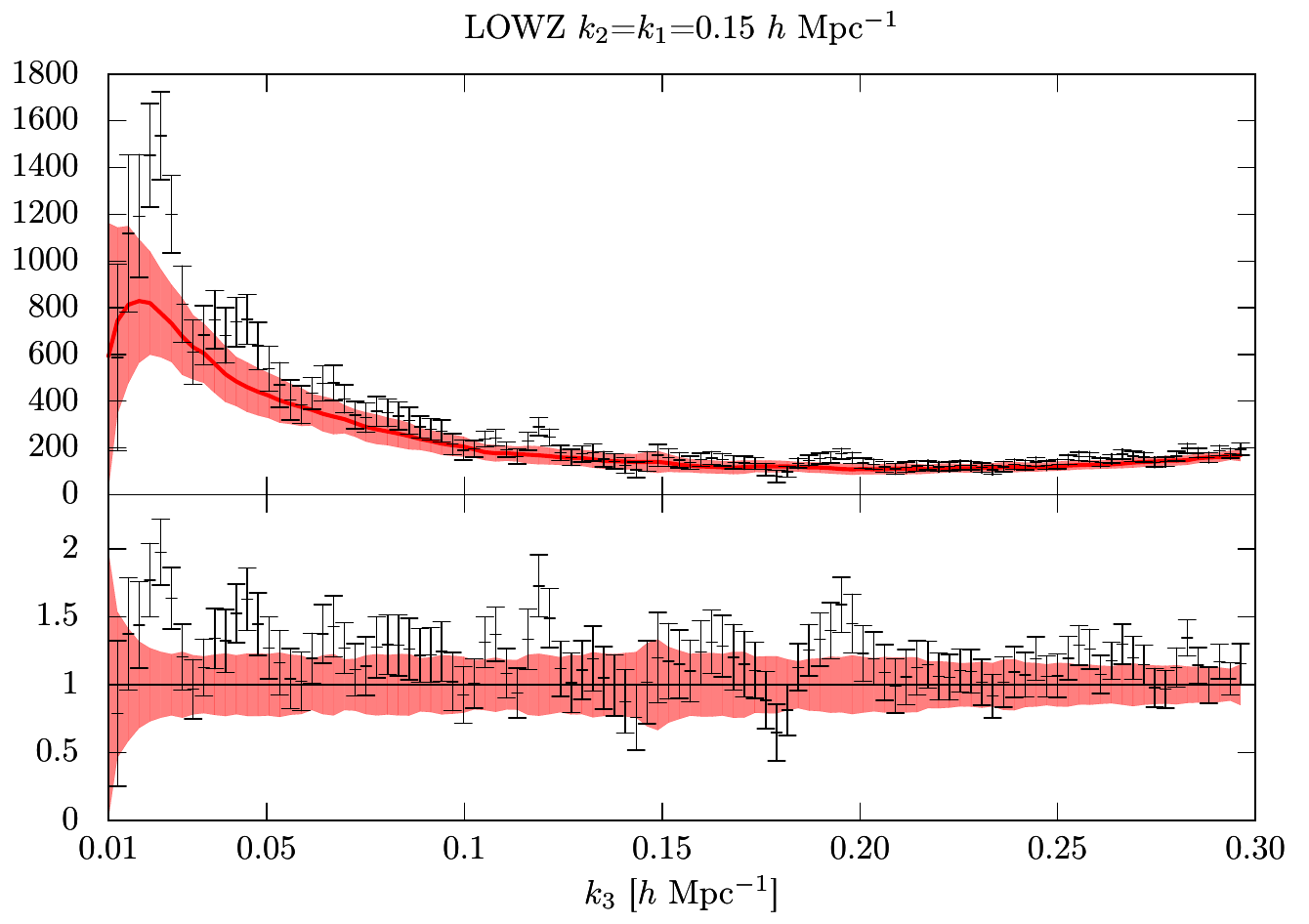}
\\
\hspace{-1.3cm}
\includegraphics[width=6.95cm]{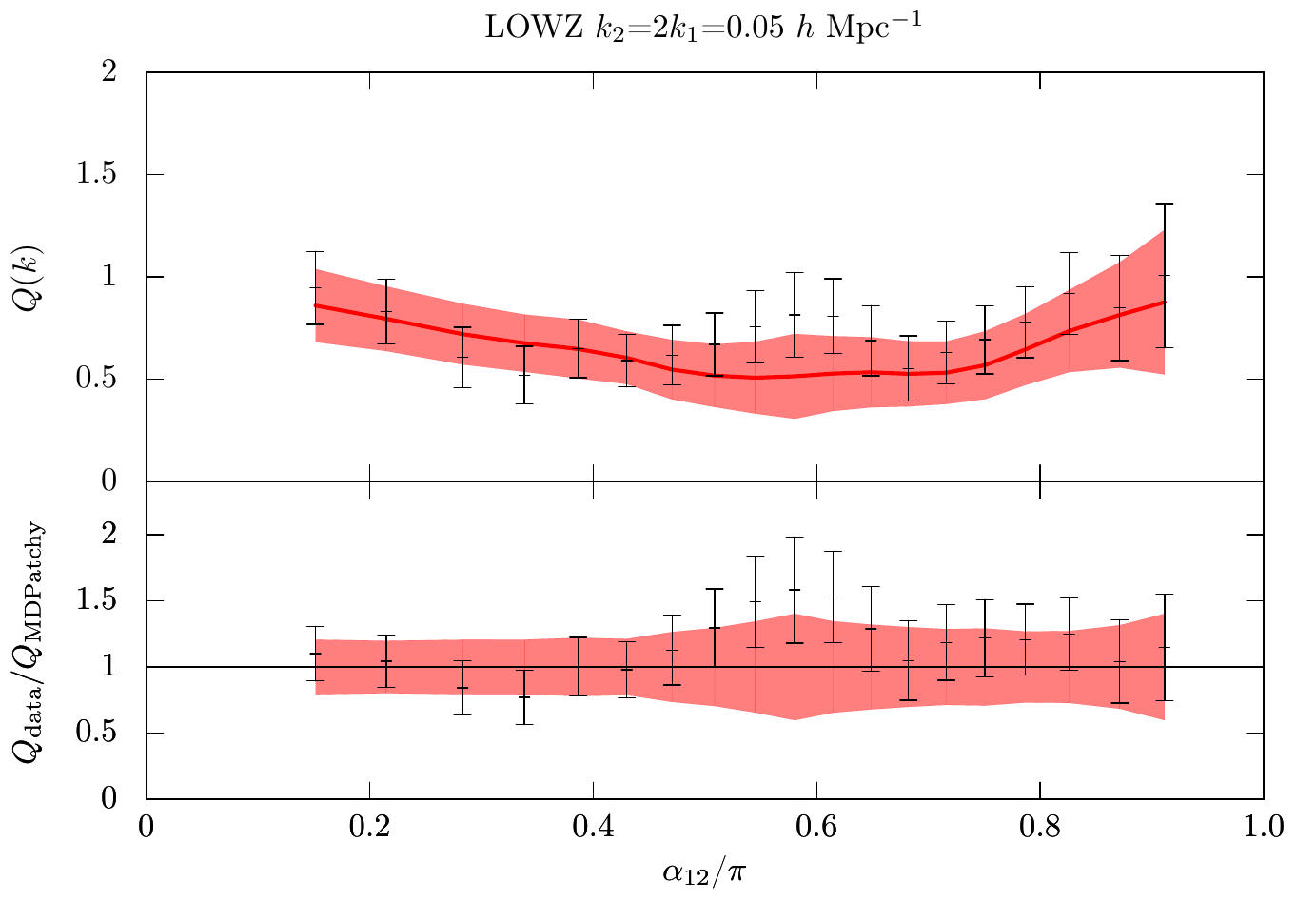}
\hspace{-0.2cm}
\includegraphics[width=6.65cm]{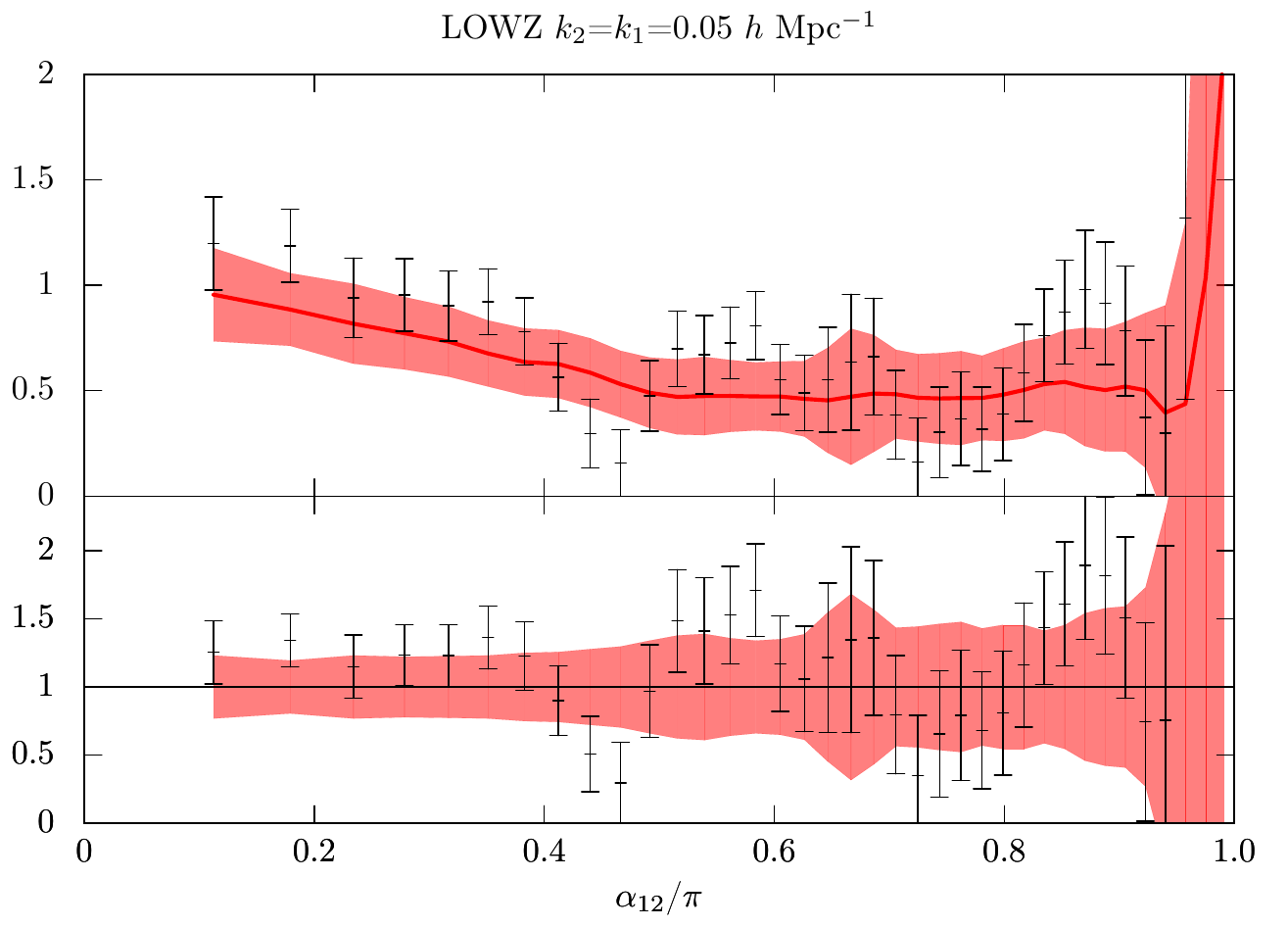}
\hspace{-0.2cm}
\includegraphics[width=6.65cm]{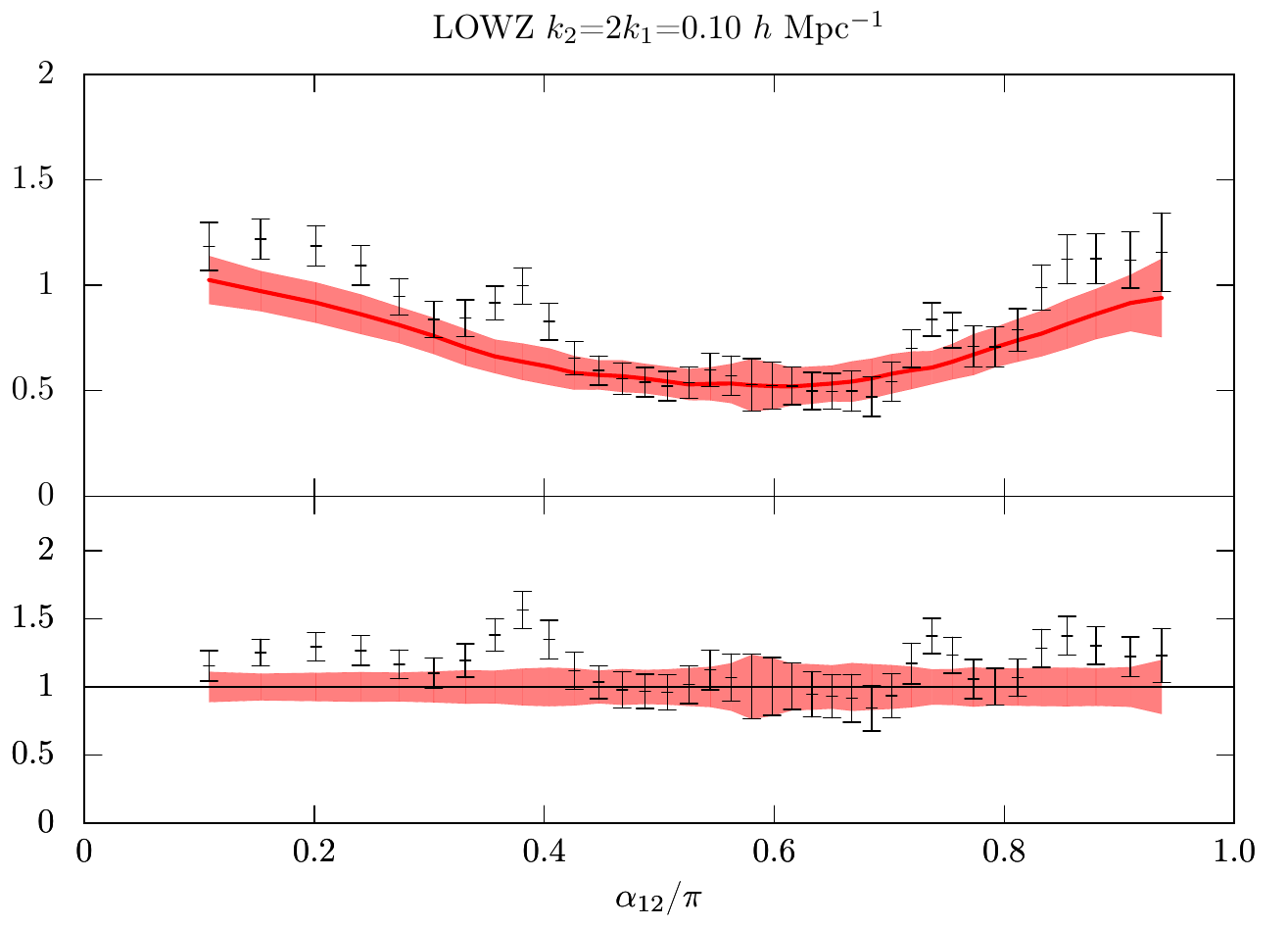}
\\
\hspace{-1.3cm}
\includegraphics[width=6.95cm]{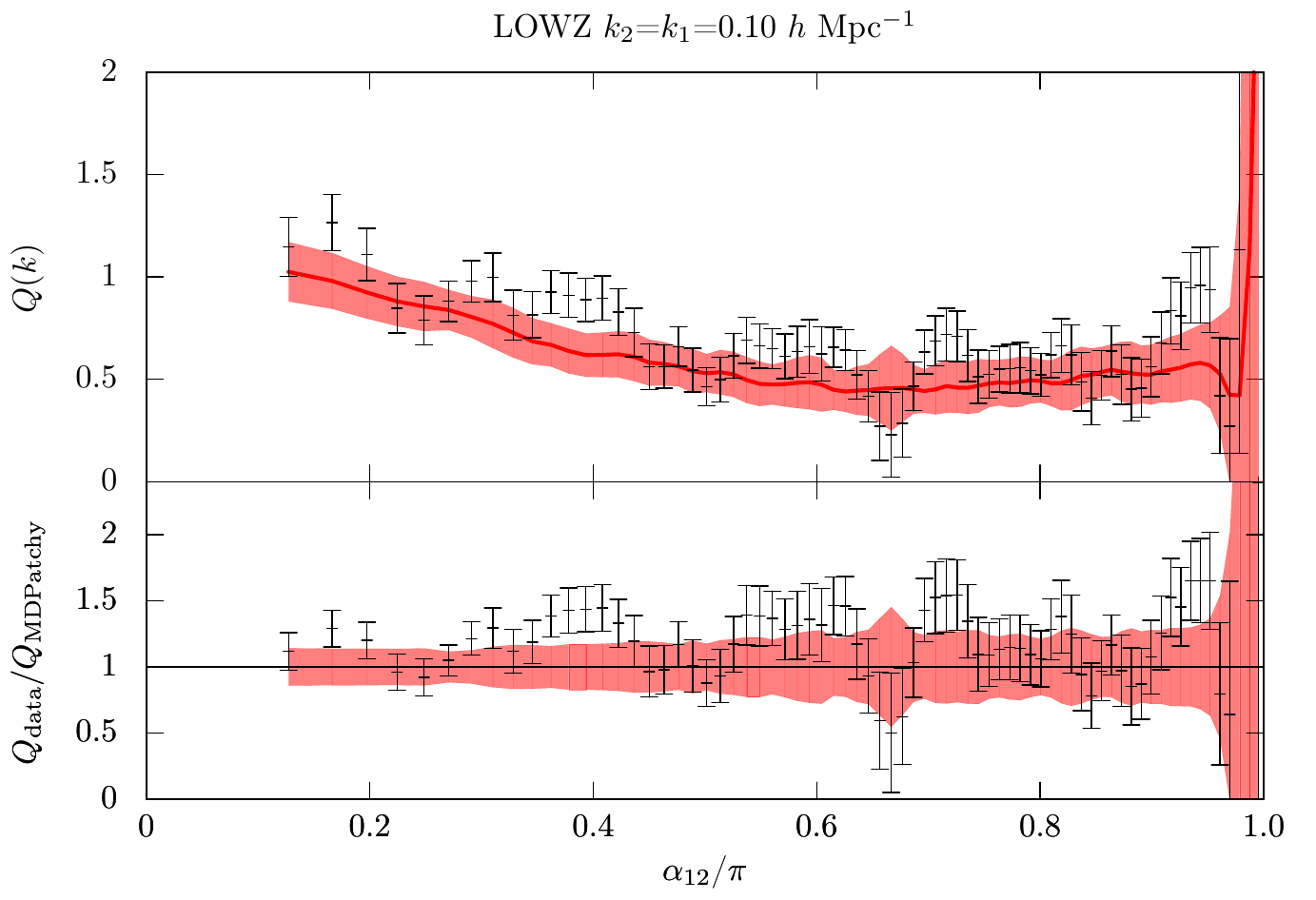}
\hspace{-0.2cm}
\includegraphics[width=6.65cm]{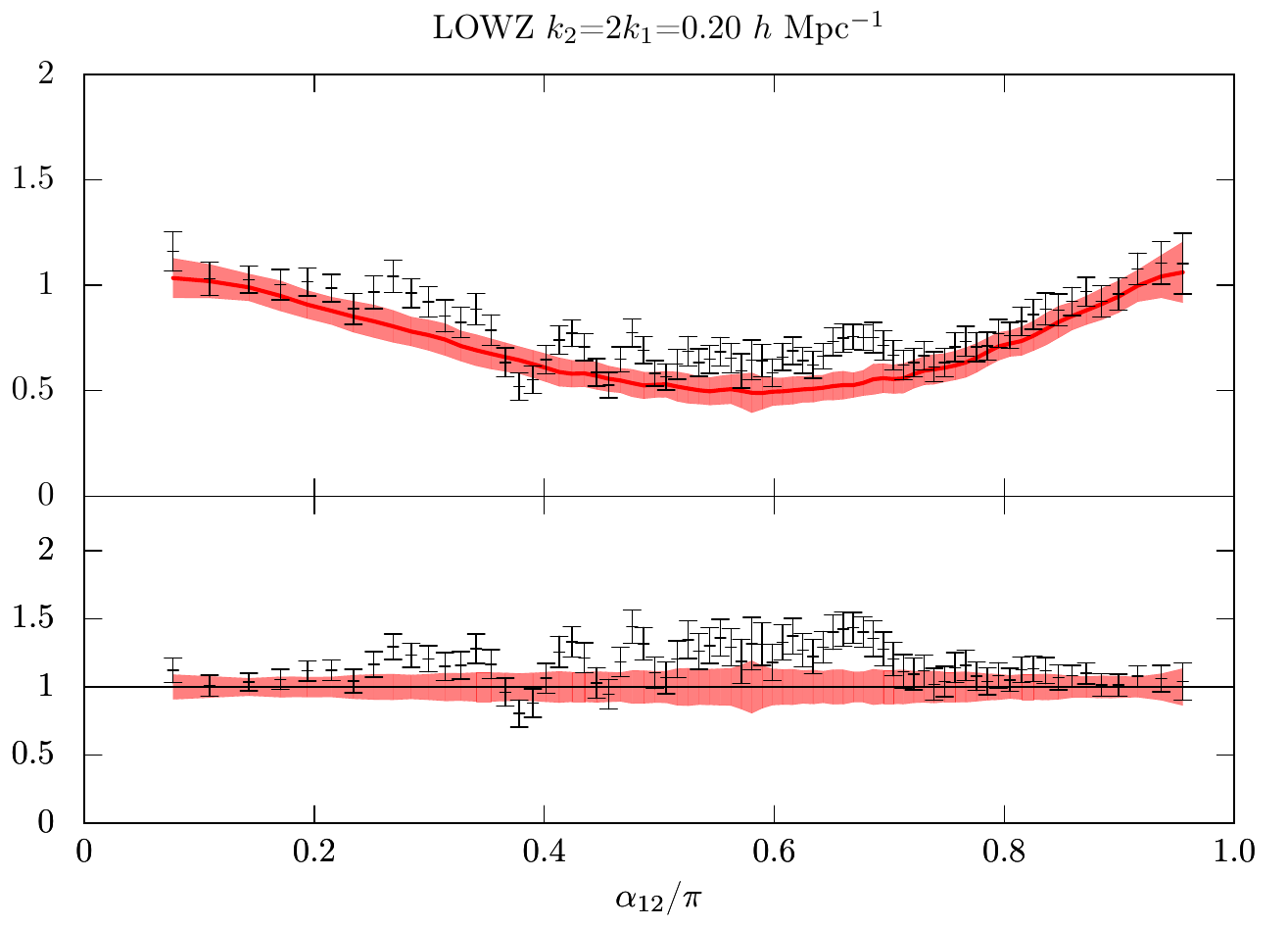}
\hspace{-0.2cm}
\includegraphics[width=6.65cm]{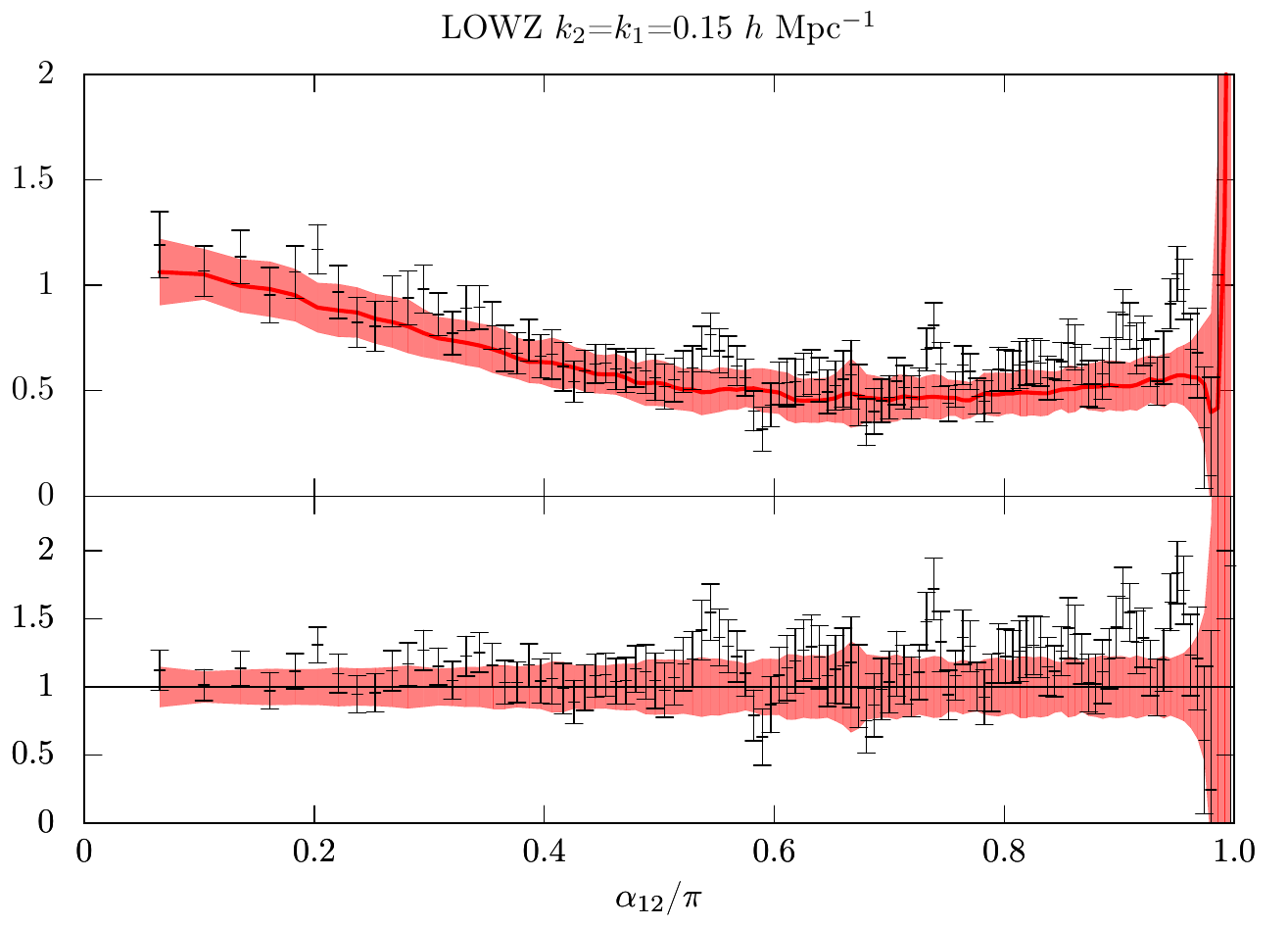}
\end{tabular}
\caption{\label{fig:BSLOWZ} Bispectra and reduced bispectra for LOWZ mocks and observed galaxies for different configurations indicated above each panel. The red solid line corresponds to the mean and the red shaded region to the 1-$\sigma$ contour of 100 \textsc{md patchy} mocks. The black dots correspond to the BOSS DR12 data with the error bars taken from the  \textsc{md patchy} mocks.    }
\end{figure*}

\begin{figure*}
\begin{tabular}{ccc}
\hspace{-1.3cm}
\includegraphics[width=6.8cm]{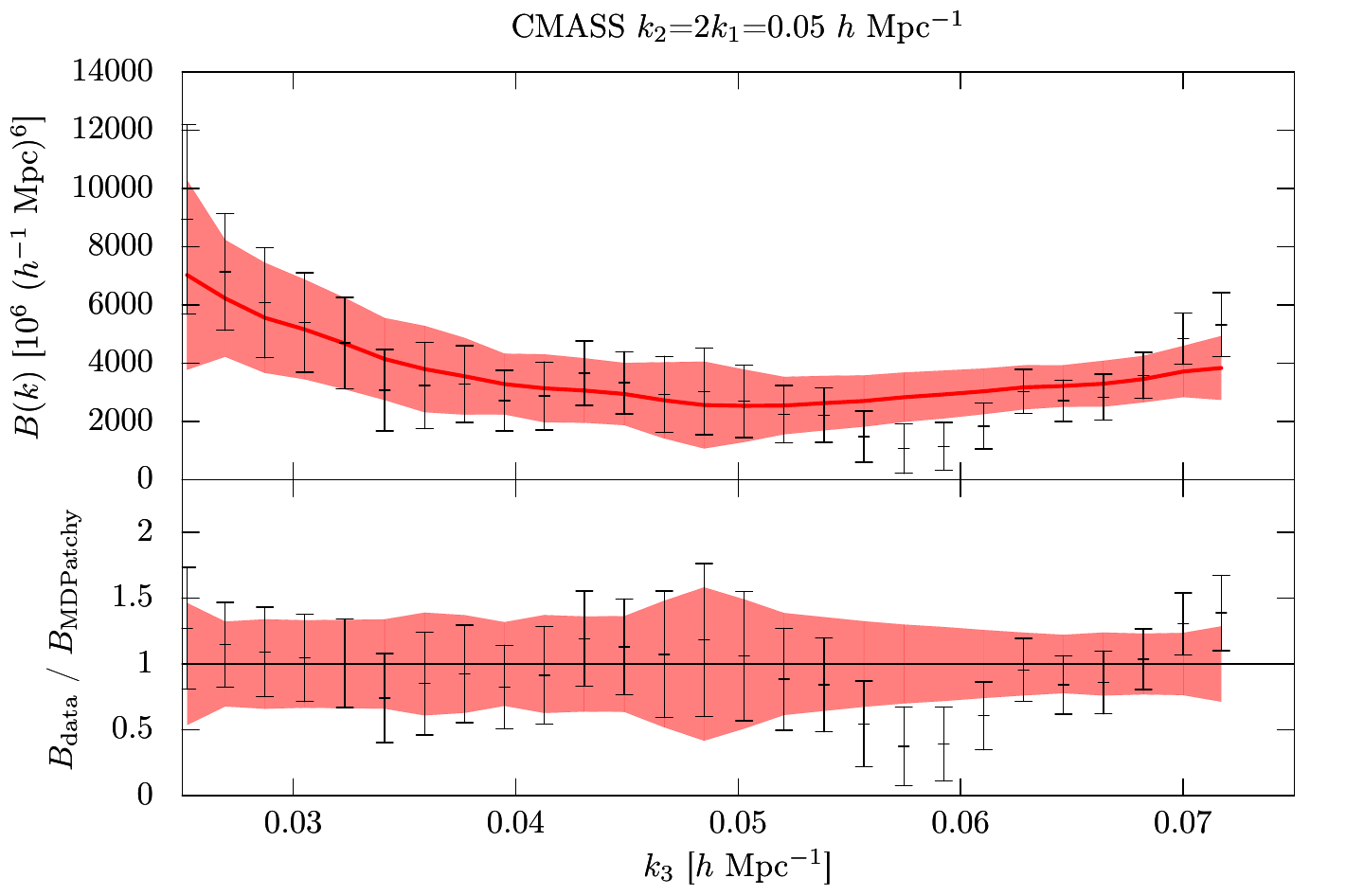}
\hspace{-0.2cm}
\includegraphics[width=6.73cm]{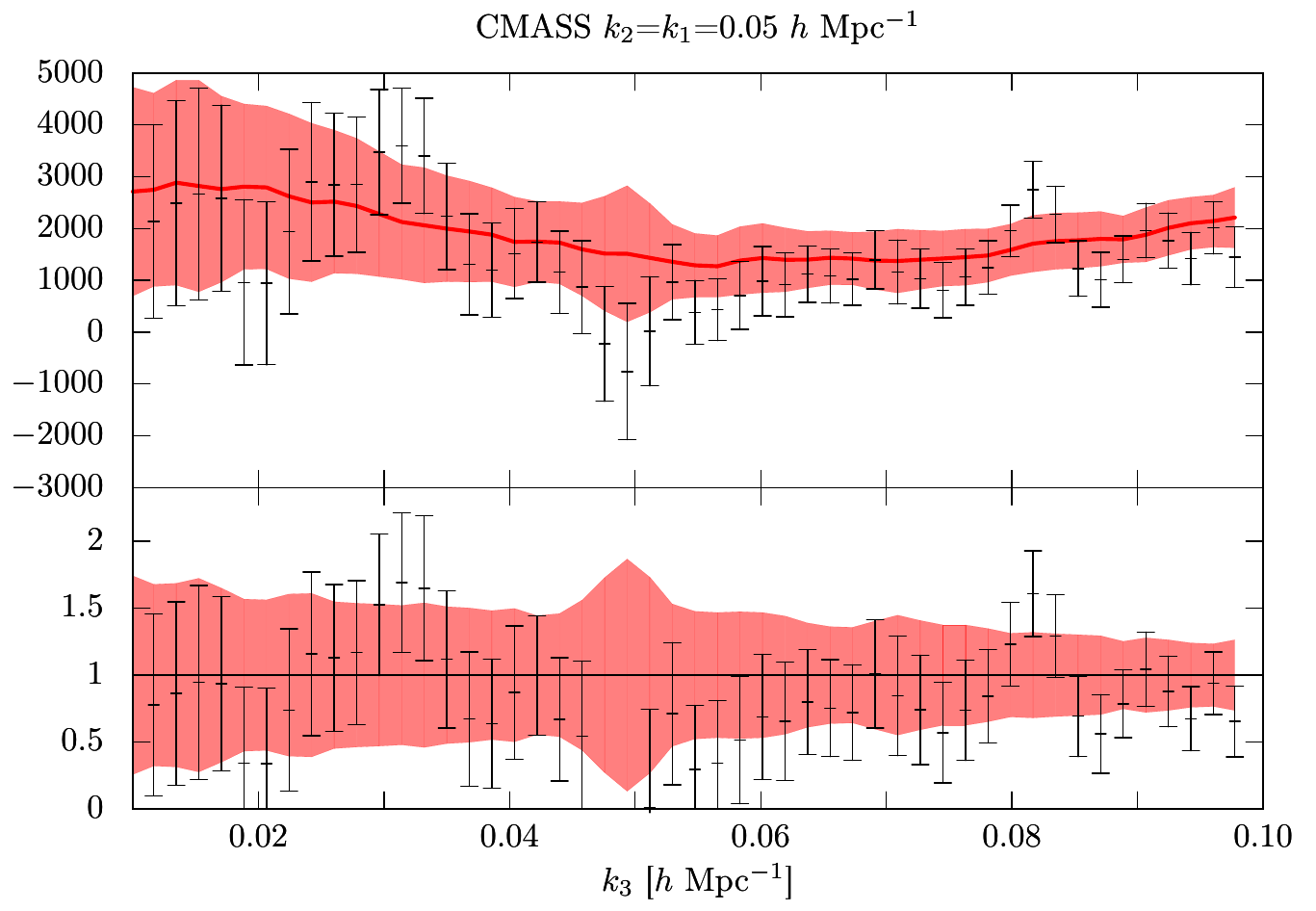}
\hspace{-0.2cm}
\includegraphics[width=6.6cm]{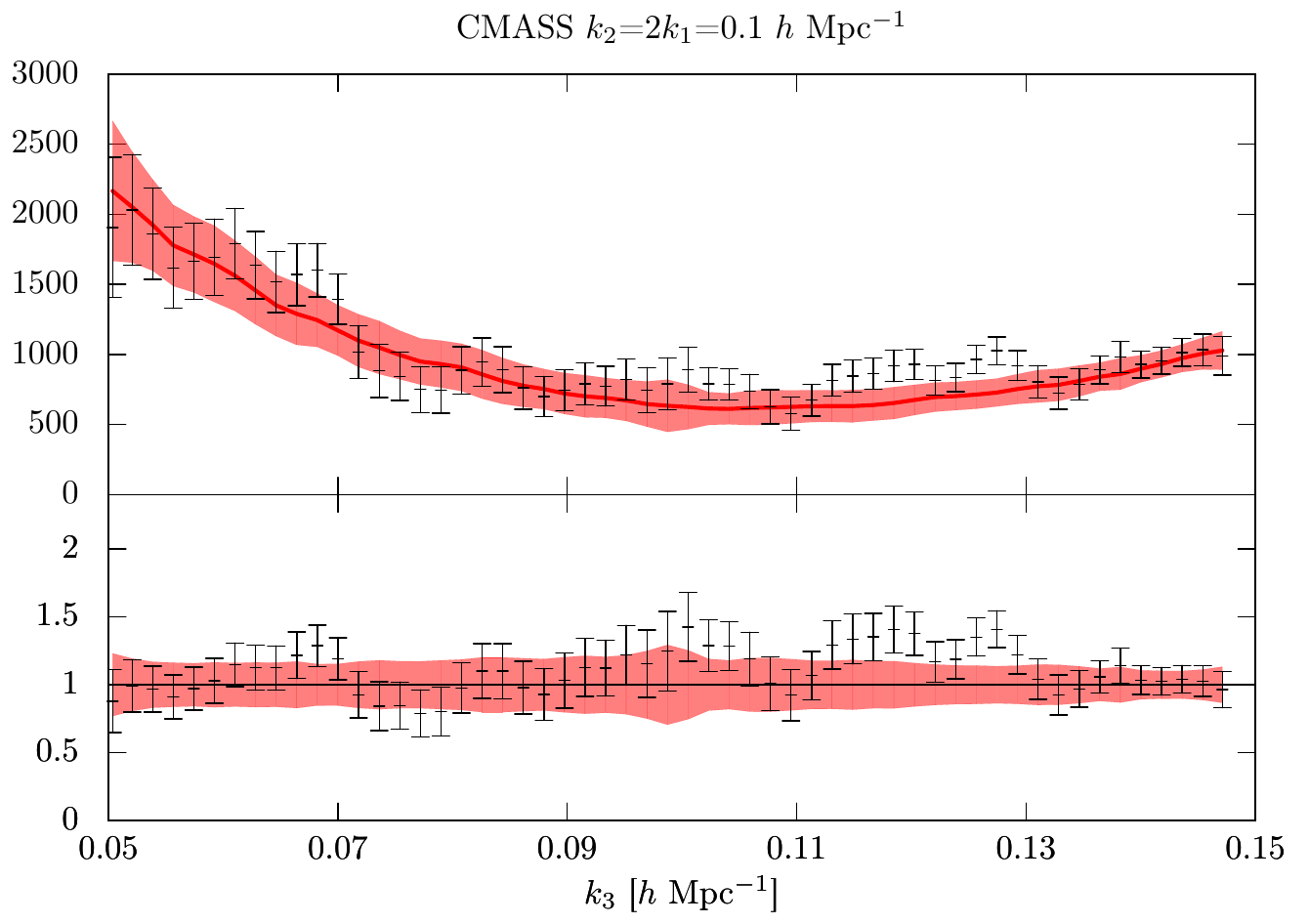}
\\
\hspace{-1.3cm}
\includegraphics[width=6.9cm]{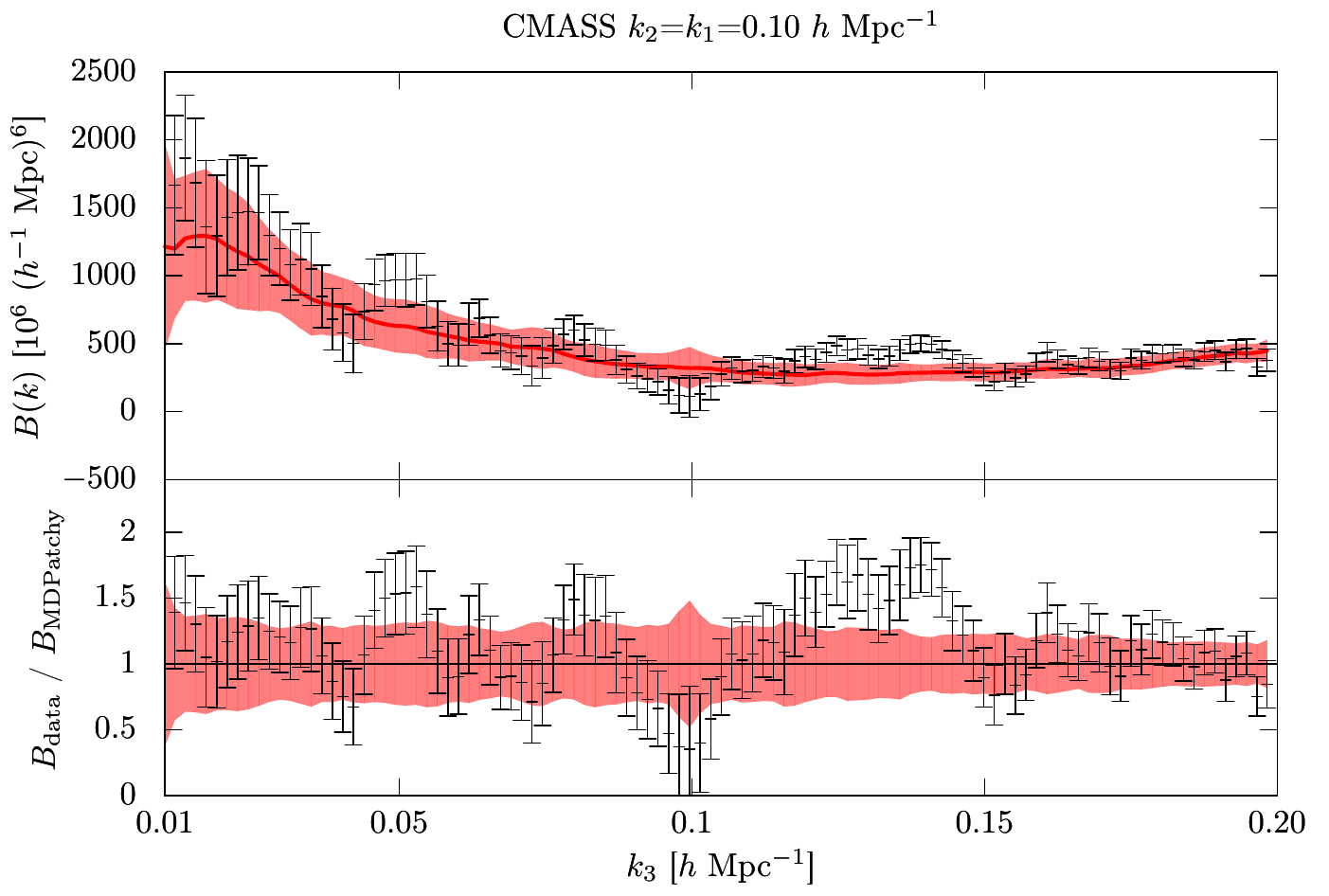}
\hspace{-0.2cm}
\includegraphics[width=6.6cm]{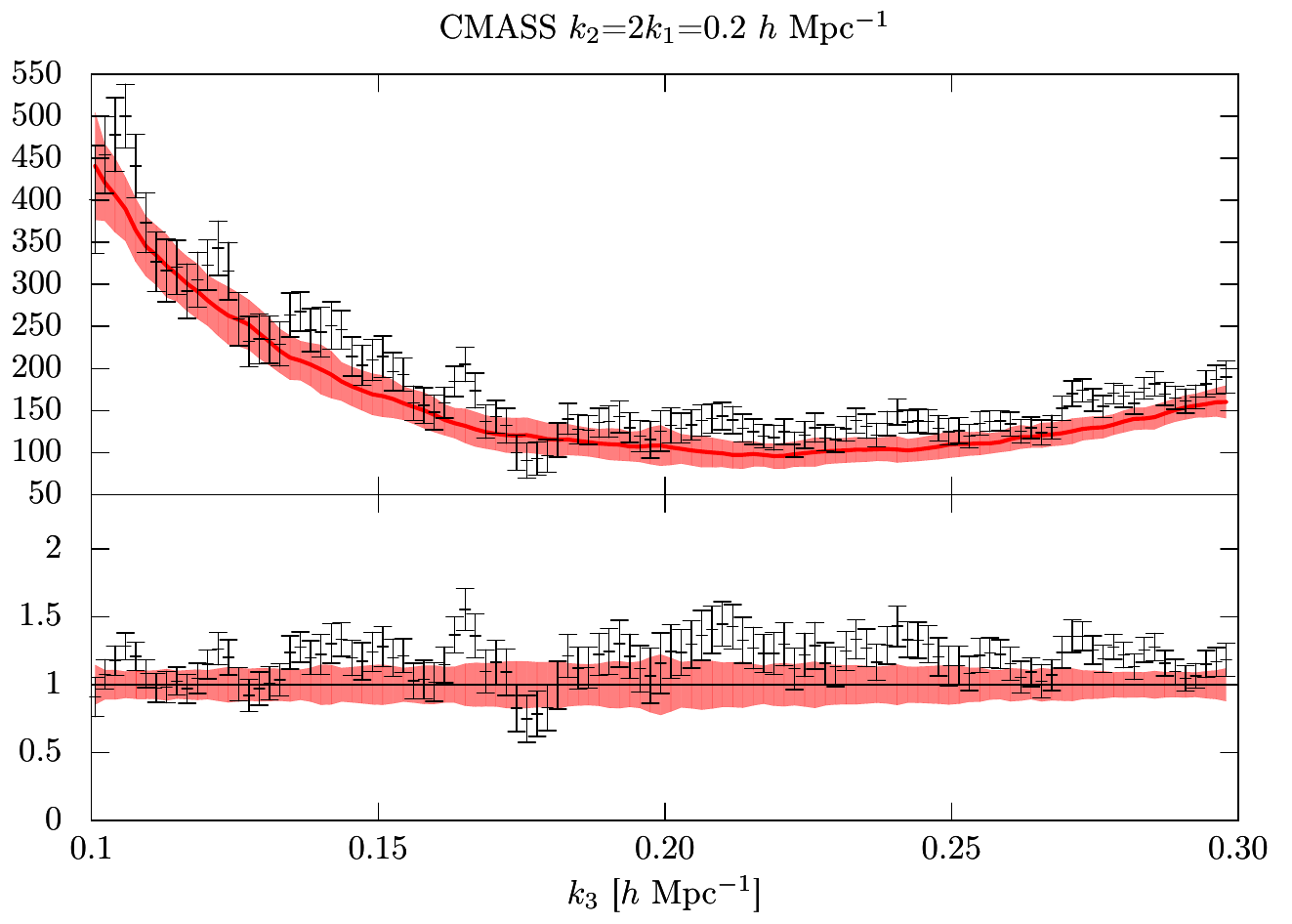}
\hspace{-0.2cm}
\includegraphics[width=6.7cm]{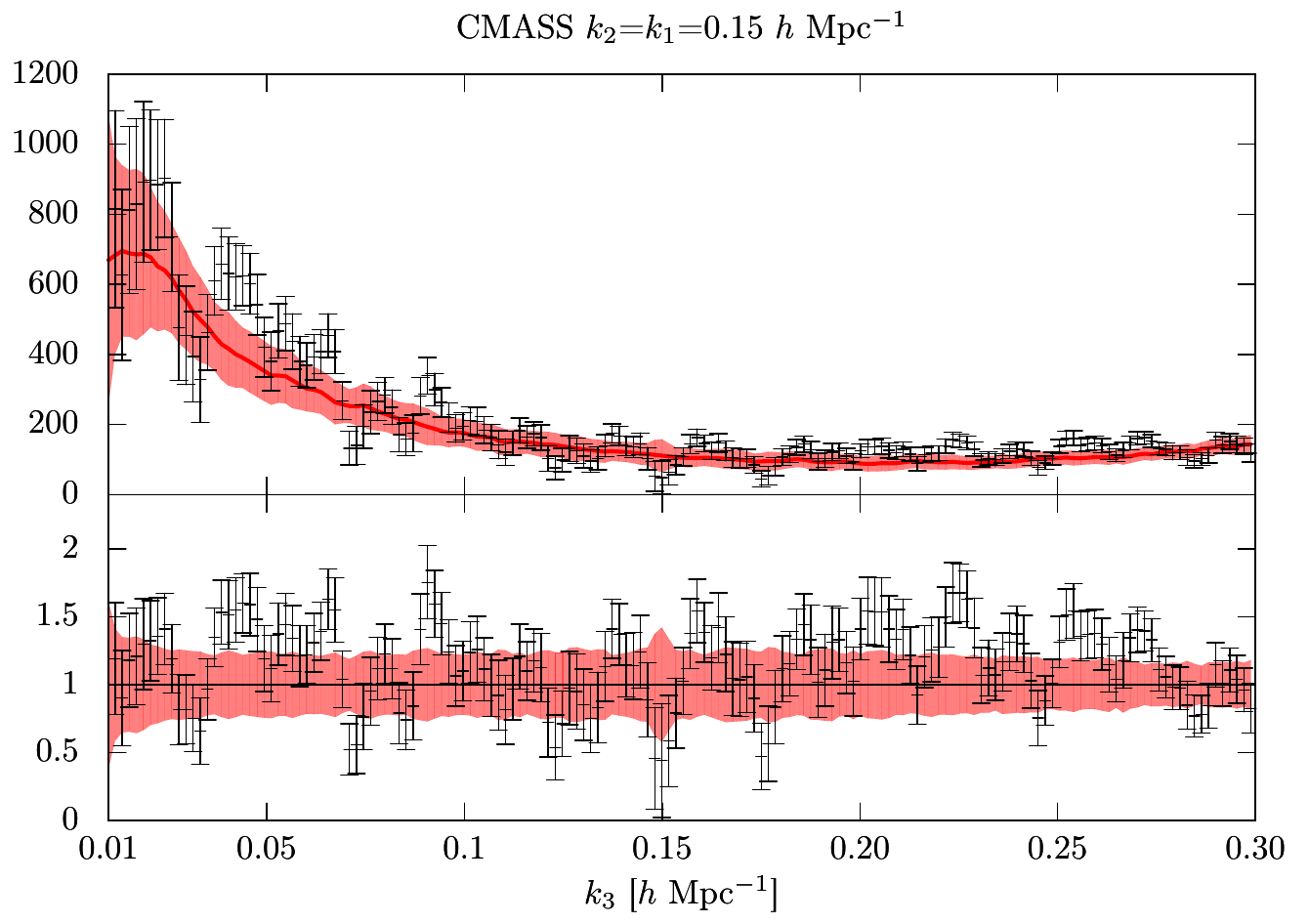}
\\
\hspace{-1.3cm}
\includegraphics[width=6.95cm]{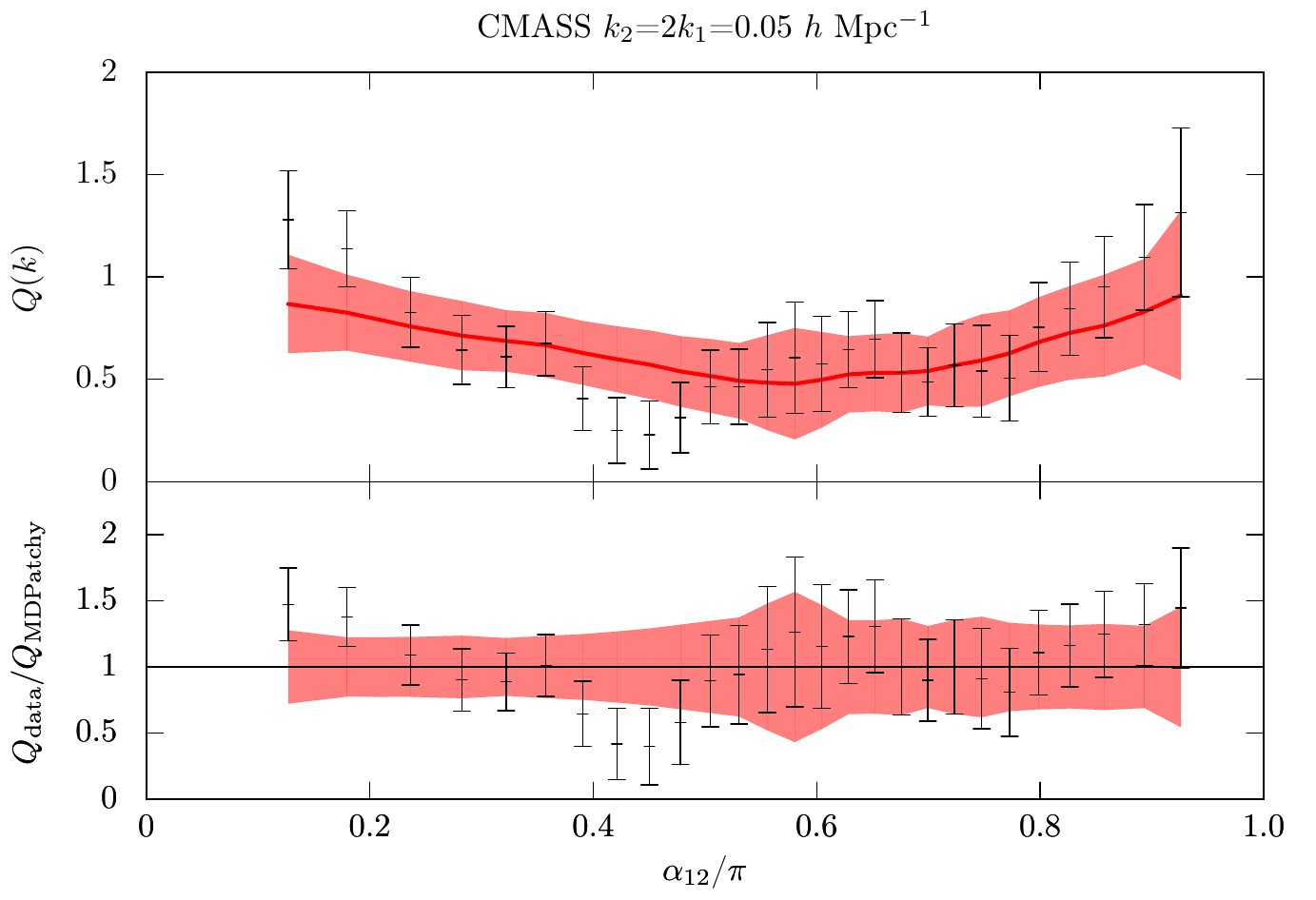}
\hspace{-0.2cm}
\includegraphics[width=6.65cm]{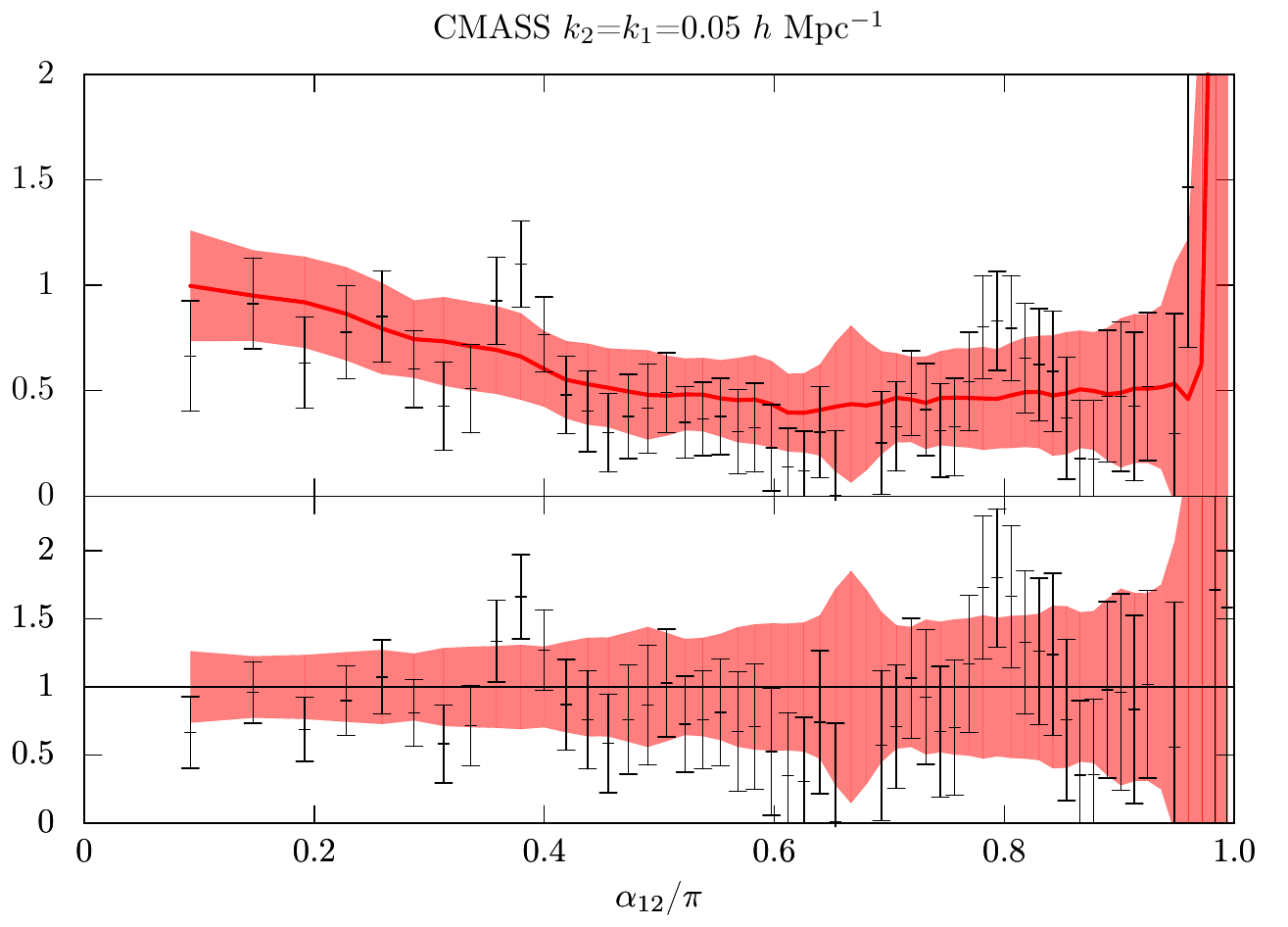}
\hspace{-0.2cm}
\includegraphics[width=6.65cm]{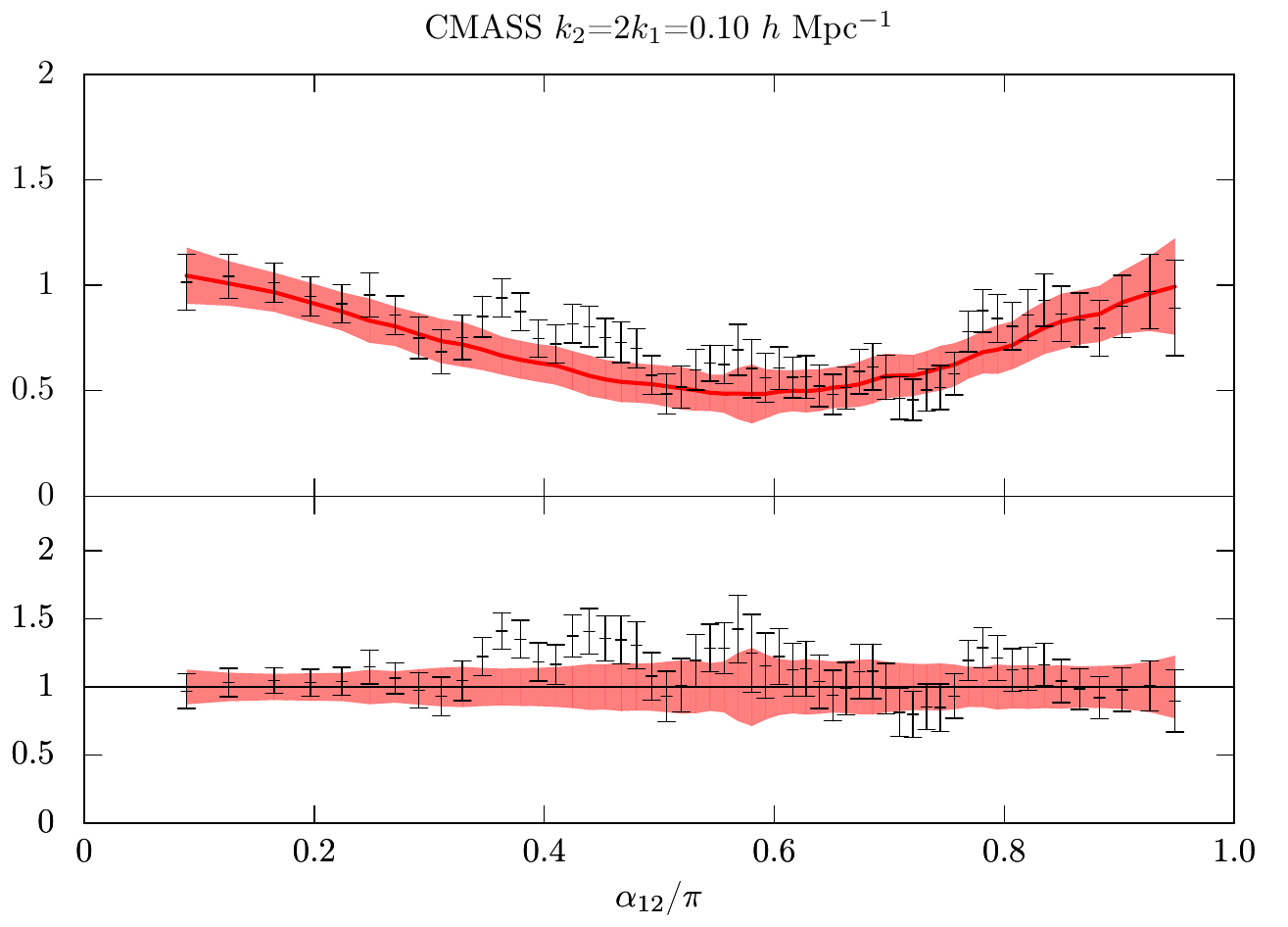}
\\
\hspace{-1.3cm}
\includegraphics[width=6.95cm]{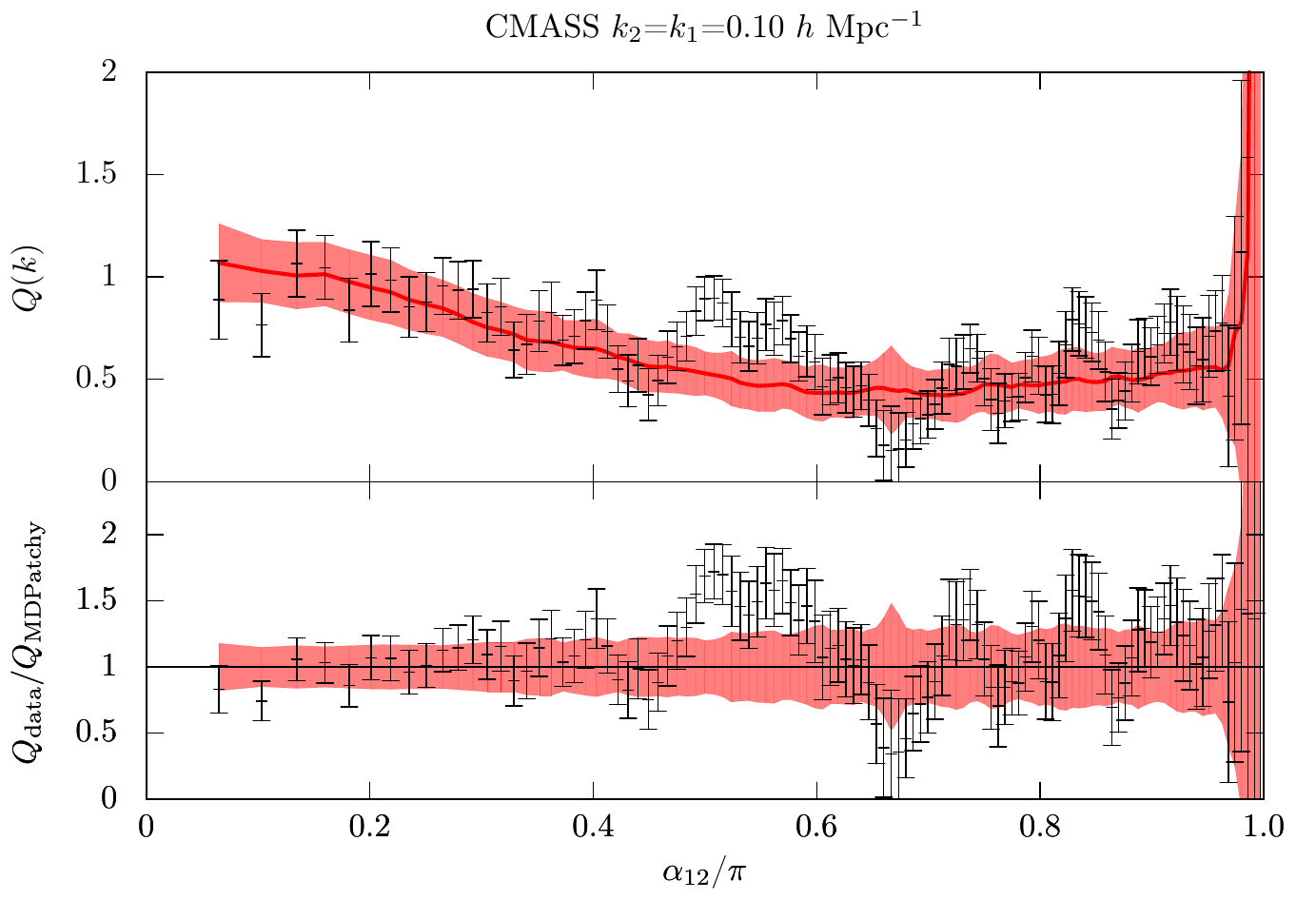}
\hspace{-0.2cm}
\includegraphics[width=6.65cm]{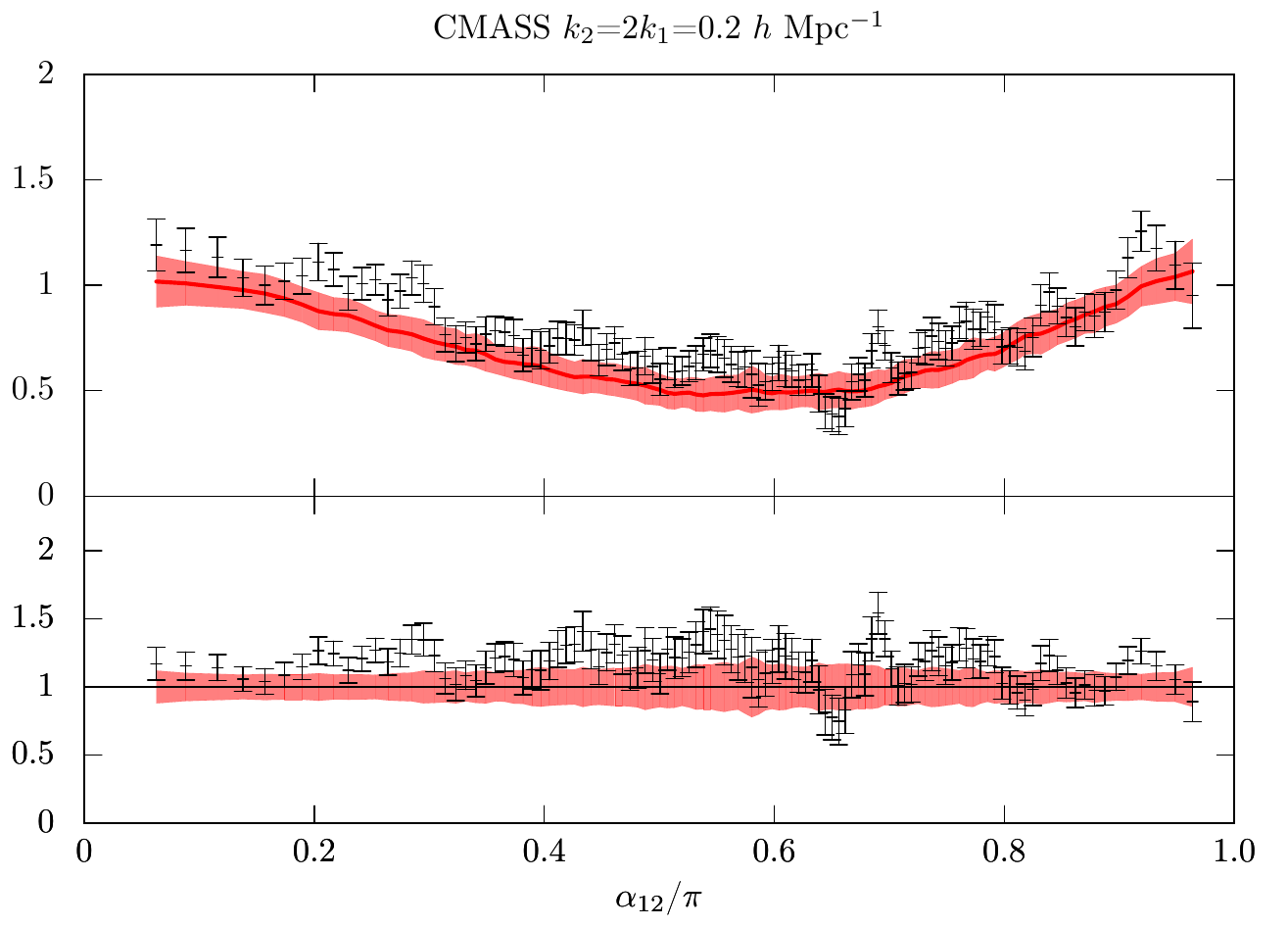}
\hspace{-0.2cm}
\includegraphics[width=6.65cm]{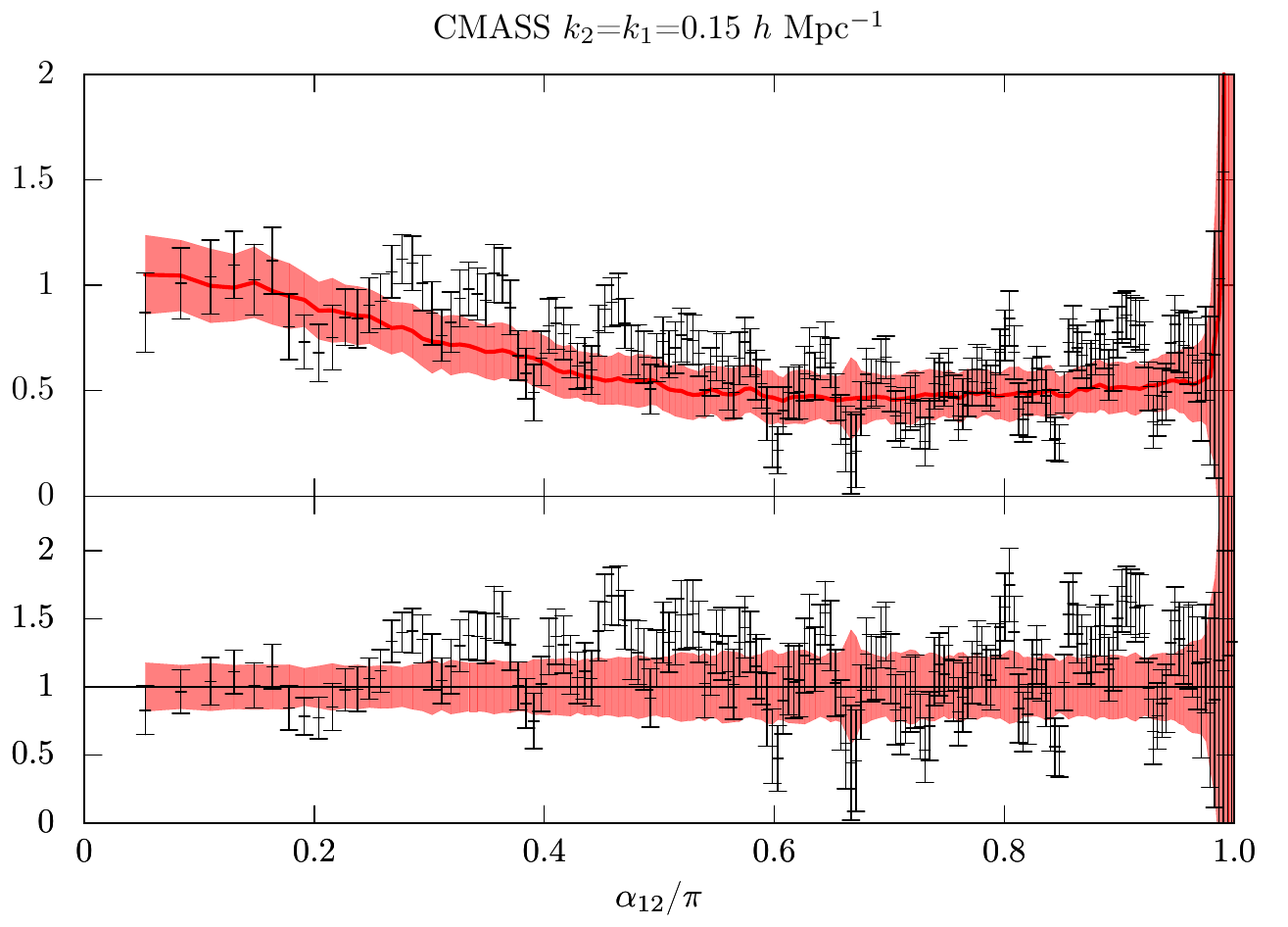}
\end{tabular}
\caption{\label{fig:BSCMASS} Bispectra and reduced bispectra for CMASS mocks and observed galaxies  for different configurations.  The red solid line corresponds to the mean and the red shaded region to the 1-$\sigma$ contour of 100 \textsc{md patchy} mocks. The black dots correspond to the BOSS DR12 data with the error bars taken from the  \textsc{md patchy} mocks.    }
\end{figure*}

\begin{figure*}
\begin{tabular}{cc}
\includegraphics[width=9.cm]{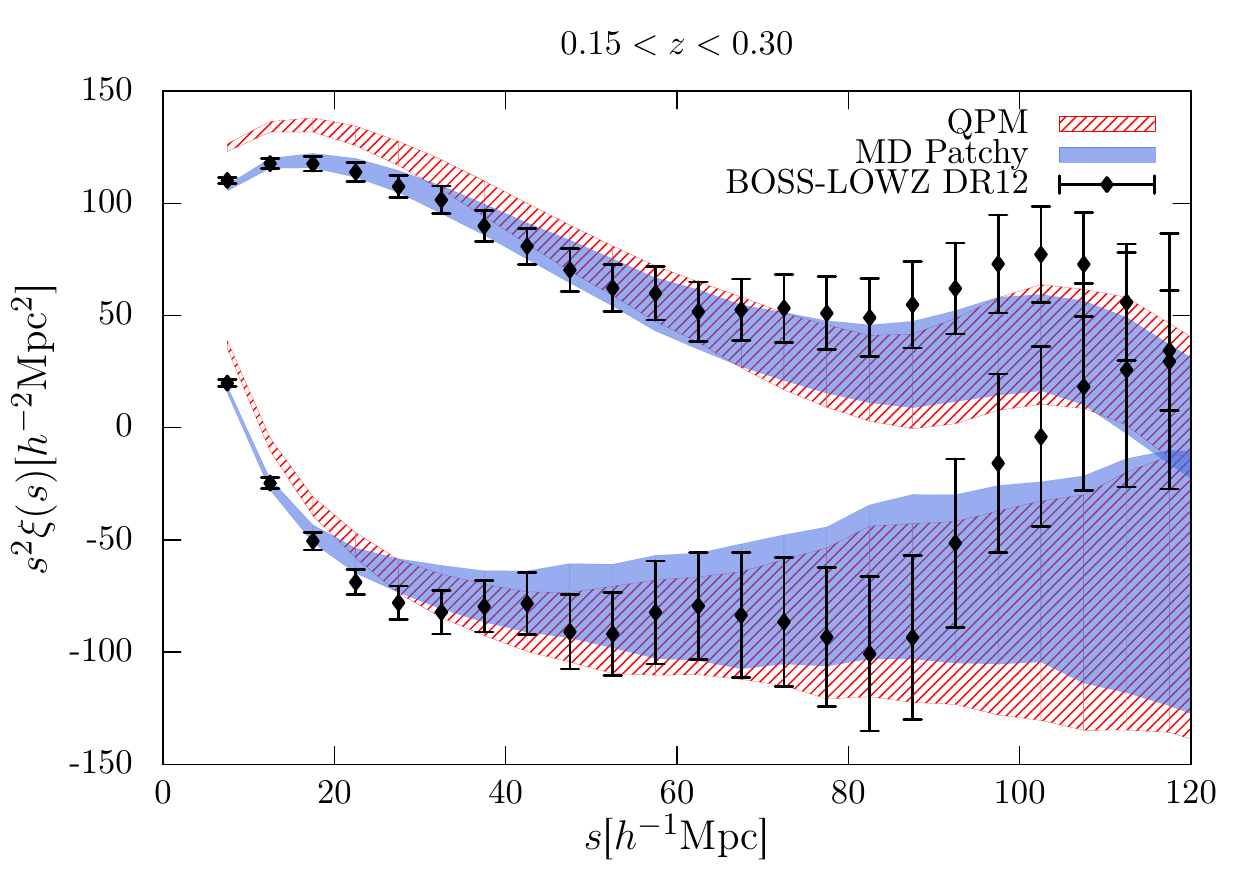}
\includegraphics[width=8.cm]{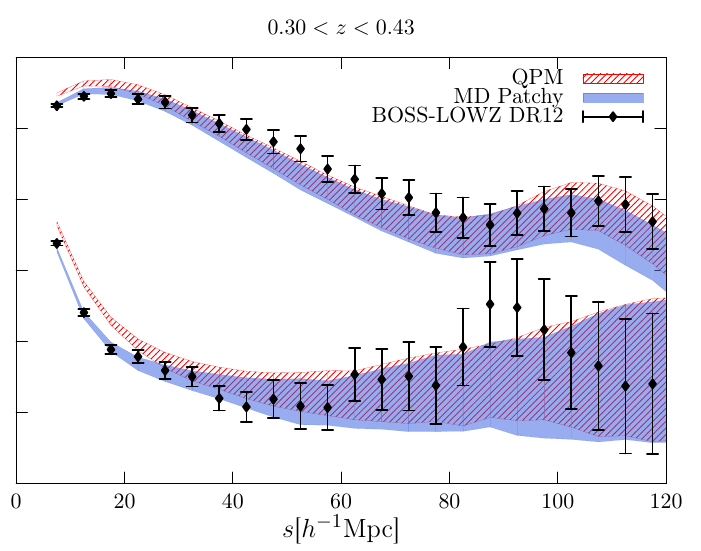}
\\
\includegraphics[width=9.cm]{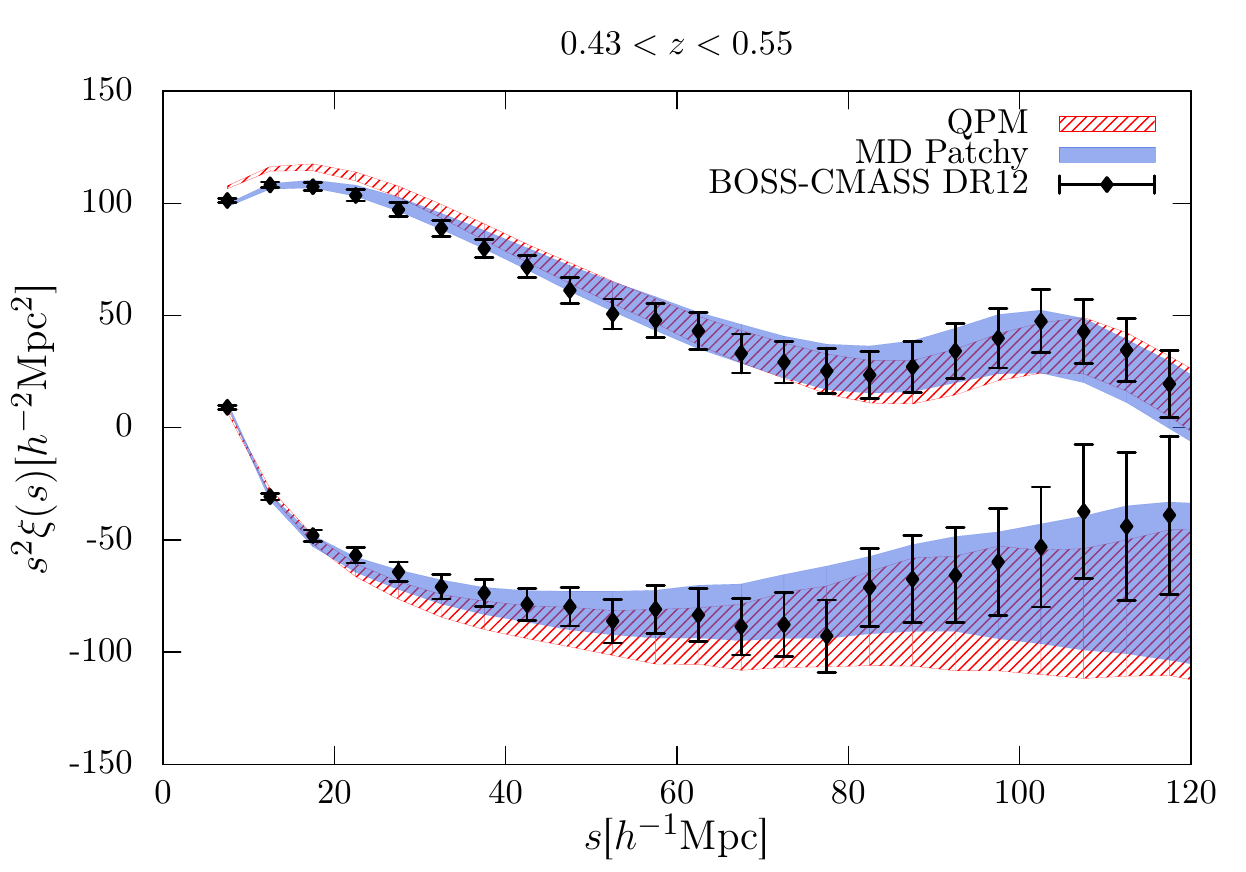}
\includegraphics[width=8.cm]{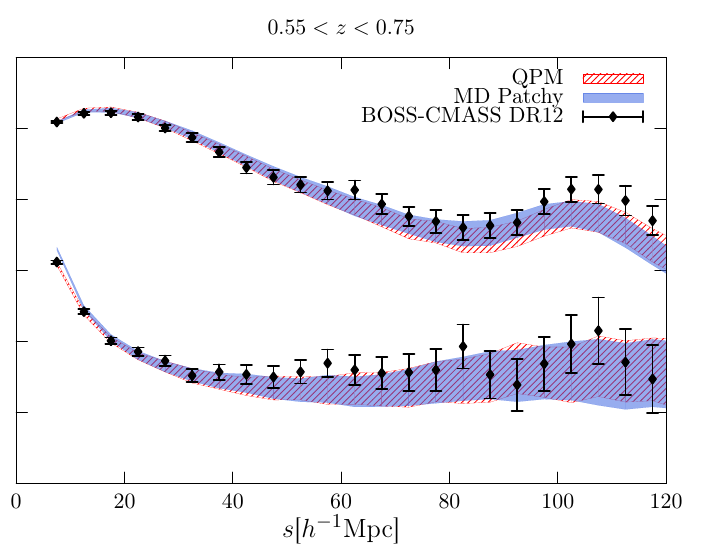}
\end{tabular}
\caption{\label{fig:CFPAQLOWZCMASS}  Monopole and the quadrupole for different redshift bins over the redshift range $0.15<z<0.7$. The black error bars stand for the BOSS DR12 data. The shaded contours represent the 1-$\sigma$ regions according to the \textsc{md patchy} mocks in blue and according to the \textsc{qpm} mocks in red. These measurements are used in the BAO and RSD analysis in Chuang et al. (in prep.).
}
\end{figure*}

\subsection{Monopole and quadrupole in Fourier space}
\label{sec:pk}

\begin{figure}
\hspace{0.cm}
\includegraphics[width=8.cm]{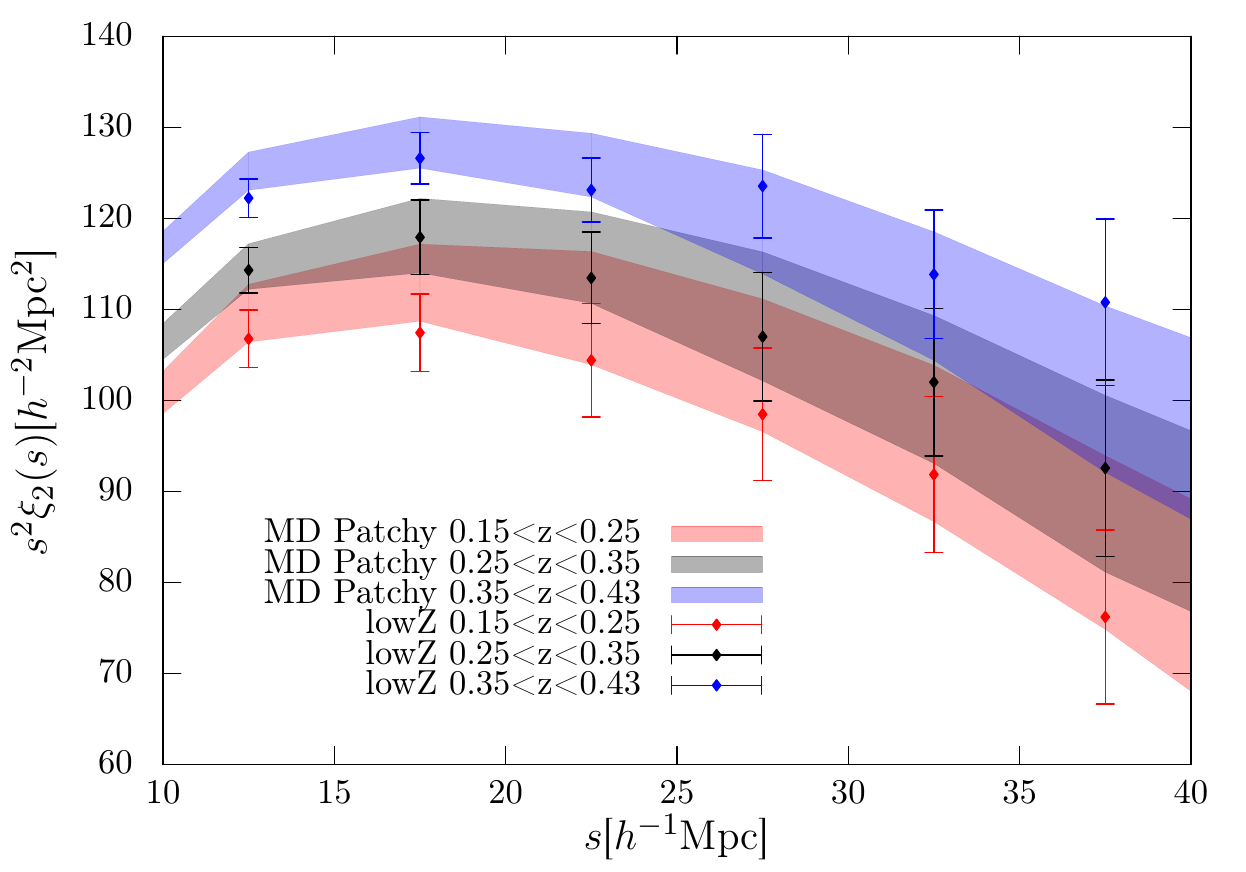}
\caption{\label{fig:LOWZevol}  Monopole showing the evolution for LOWZ. The corresponding redshift bins for the \textsc{patchy} mocks are represented by shaded regions, and the observations by the error bars.  }
\end{figure}
 
The galaxy power spectrum $P$ and the galaxy bispectrum $B$ are the two- and three-point correlation functions in Fourier space. Given the Fourier transform of the galaxy overdensity, $\delta_{\rm g}({\bf x})\equiv\rho_{g}({\bf x})/{\bar \rho}_{\rm g}-1$,
\begin{equation}
\delta_{\rm g}({\bf k})={\int} \dd^3{\bf x}\,\delta_{\rm g}({\bf x})\exp(-i{\bf k}\cdot{\bf x}),
\end{equation}
where $\rho_{\rm g}({\bf x})$ is the number density of objects and $\bar{\rho}_{\rm g}$ its mean value, and the galaxy power spectrum and galaxy bispectrum are defined as,
\begin{eqnarray}
\langle \delta_{\rm g}({\bf k}) \delta_{\rm g}({{\bf k}'})\rangle&\equiv&(2\pi)^3P( k)\delta^D({\bf k}+{\bf k}'),\\
\langle \delta_{\rm g}({{\bf k}_1}) \delta_{\rm g}({{\bf k}_2}) \delta_{\rm g}({{\bf k}_3})\rangle&\equiv&(2\pi)^3B({\bf k}_1, {\bf k}_2)\delta^D({\bf k}_1+{\bf k}_2+{\bf k}_3)\,{,}\nonumber\\
\end{eqnarray}
with $\delta^D$ being the Dirac delta function. Note that the bispectrum is only well defined when the set of $k$-vectors, $k_1$, $k_2$ and $k_3$ close to form a triangle, ${\bf k}_1+{\bf k}_2+{\bf k}_3={\bf 0}$.
It is common to define the reduced bispectrum $Q$ as, 
\begin{equation}
Q(\alpha_{12}|{\bf k}_1,{\bf k}_2)\equiv\frac{B({\bf k}_1,{\bf k}_2)}{P(k_1)P(k_2)+P(k_2)P(k_3)+P(k_1)P(k_3)}.
\end{equation}
where $\alpha_{12}$ is the angle between ${\bf k}_1$ and ${\bf k}_2$.  This quantity is independent of the overall scale $k$ and redshift at large scales and for a power spectrum that follows a power law. Moreover, it presents a characteristic ``U-shape" predicted by  gravitational instability. 
Mode coupling and power law deviations in the actual power spectrum induce a slight scale- and time-dependency in this quantity. However,  in practice it has been observed that at scales of the order of $k\sim0.1\,h\,{\rm Mpc}^{-1}$ the reduced bispectrum does not present a high variation in its amplitude.

The measurement of the bispectrum is performed in the same way as the approach described in \citet[][]{GilMarin2015}. This method consists of generating  $k$-triangles and randomly orientating them in $k$-space. When the number of random triangles is sufficiently large, the mean value of their bispectra tends to the fiducial bispectrum \citep[for details see][]{GilMarin2015}.

\begin{figure*}
\begin{tabular}{ccc}
\hspace{0.5cm}
\includegraphics[width=8.cm]{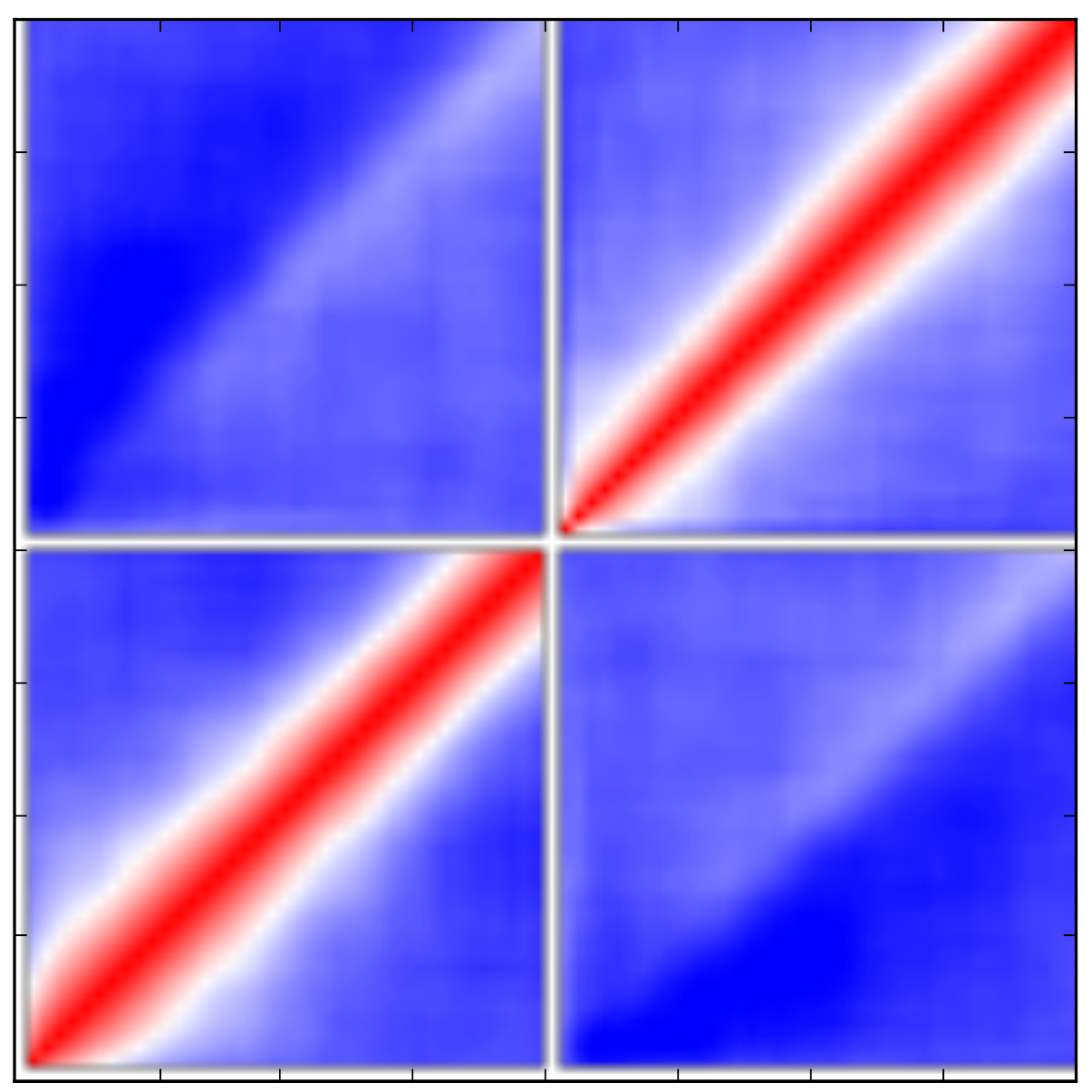}
\put(-255,115){\rotatebox[]{90}{{\large $s \,[h^{-1}\,{\rm Mpc}]$}}}
\put(-242,219){{\large 200}}
\put(-242,193){{\large 150}}
\put(-242,165){{\large 100}}
\put(-237,137){{\large 50}}
\put(-242,109){{\large 200}}
\put(-242,83){{\large 150}}
\put(-242,55){{\large 100}}
\put(-237,30){{\large 50}}
\put(-232,0){{\large 0}}
\put(-205,210){\color{white}$0.15<z<0.30$}
\includegraphics[width=8.cm]{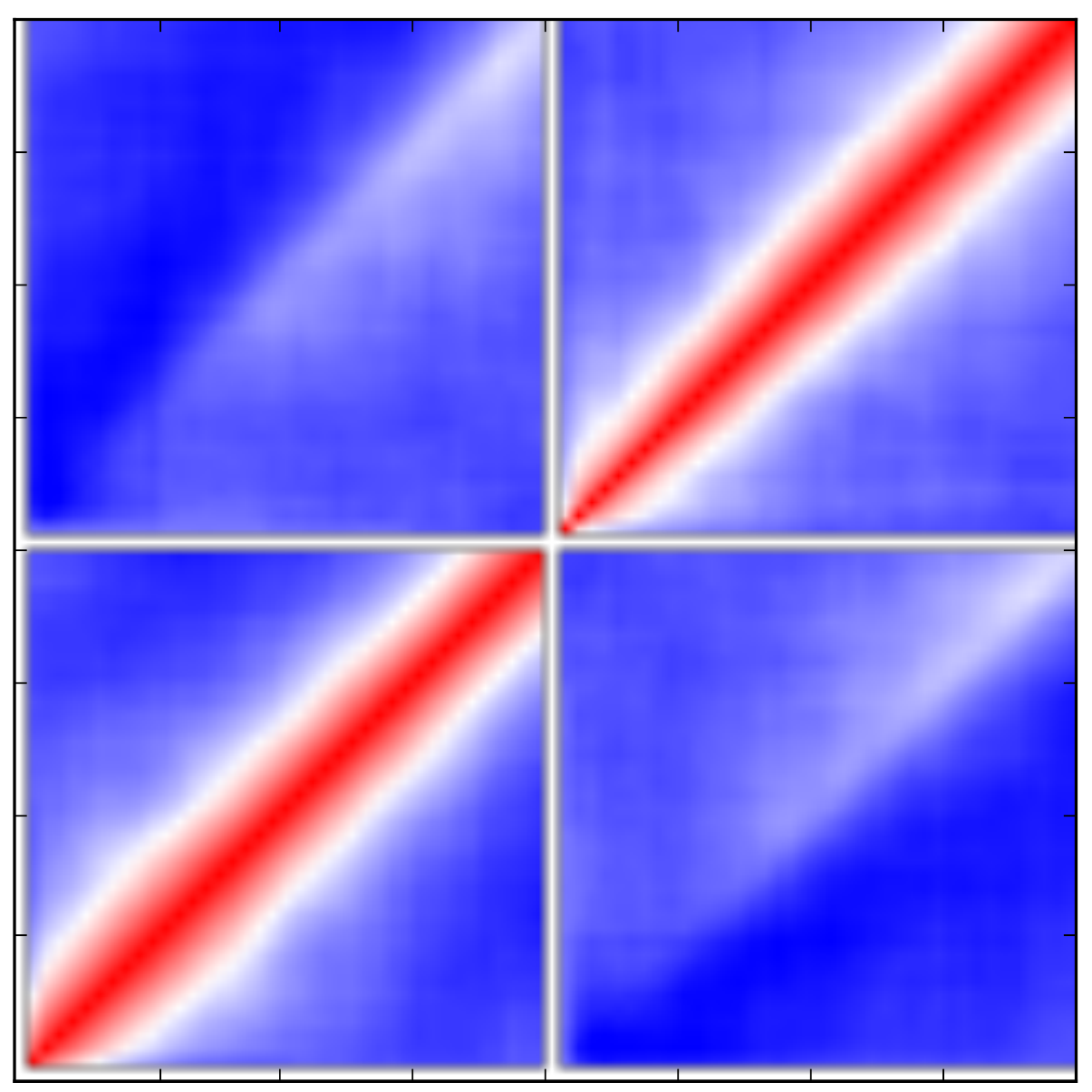}
\put(-205,210){\color{white}$0.30<z<0.43$}
\includegraphics[width=1.6cm]{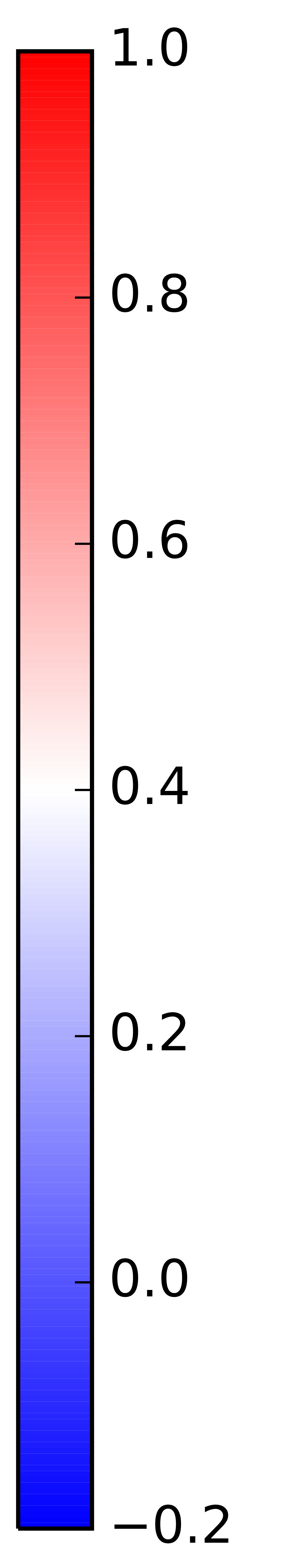}
\\
\hspace{0.5cm}
\includegraphics[width=8.cm]{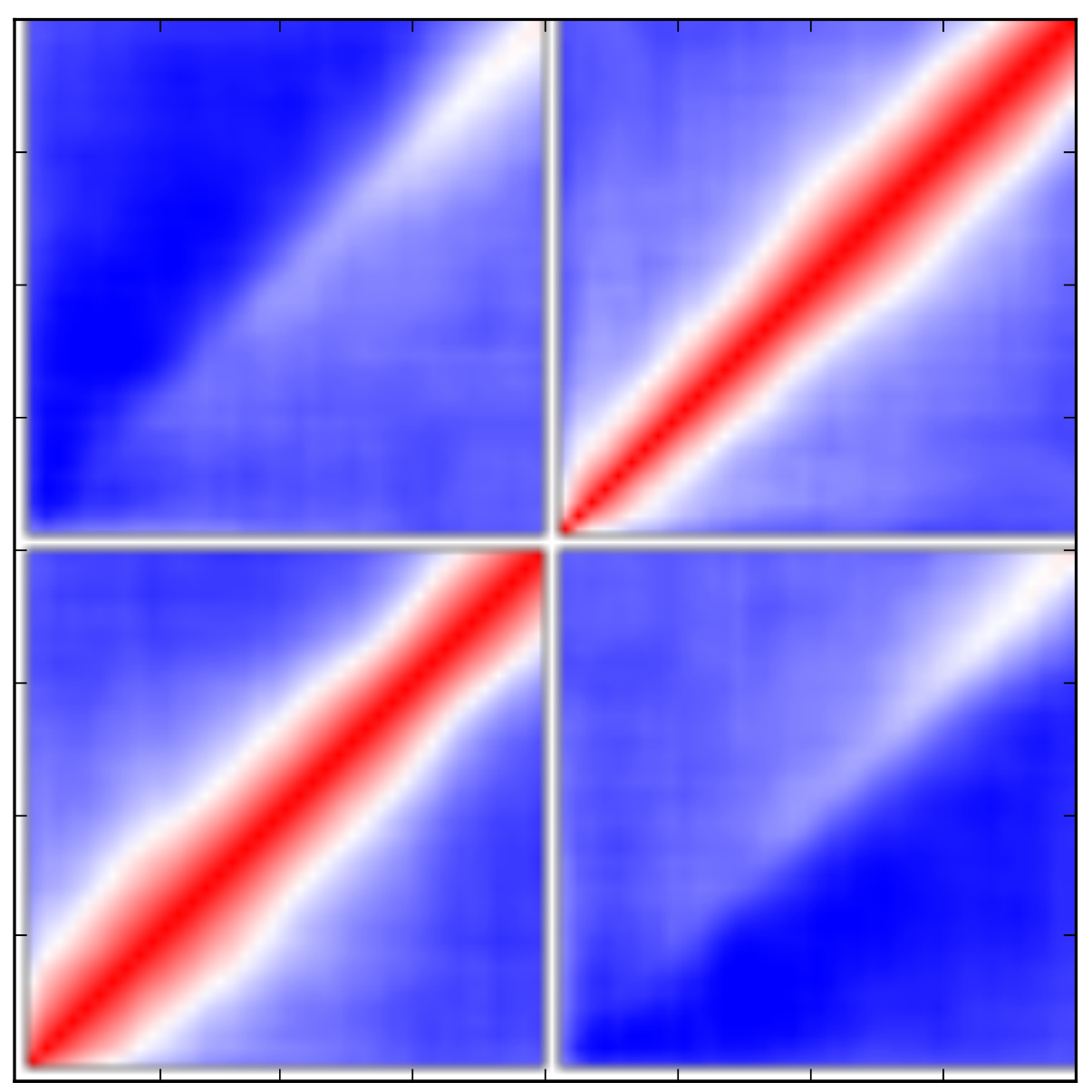}
\put(-242,219){{\large 200}}
\put(-242,193){{\large 150}}
\put(-242,165){{\large 100}}
\put(-237,137){{\large 50}}
\put(-242,109){{\large 200}}
\put(-242,83){{\large 150}}
\put(-242,55){{\large 100}}
\put(-237,30){{\large 50}}
\put(-232,0){{\large 0}}
\put(-255,115){\rotatebox[]{90}{{\large $s \,[h^{-1}\,{\rm Mpc}]$}}}
\put(-10,-8){{\large 200}}
\put(-40,-8){{\large 150}}
\put(-65,-8){{\large 100}}
\put(-90,-8){{\large 50}}
\put(-123,-8){{\large 200}}
\put(-150,-8){{\large 150}}
\put(-177,-8){{\large 100}}
\put(-200,-8){{\large 50}}
\put(-227,-8){{\large 0}}
\put(-140,-20){{\large $s \,[h^{-1}\,{\rm Mpc}]$}}
\put(-205,210){\color{white}$0.43<z<0.55$}
\includegraphics[width=8.cm]{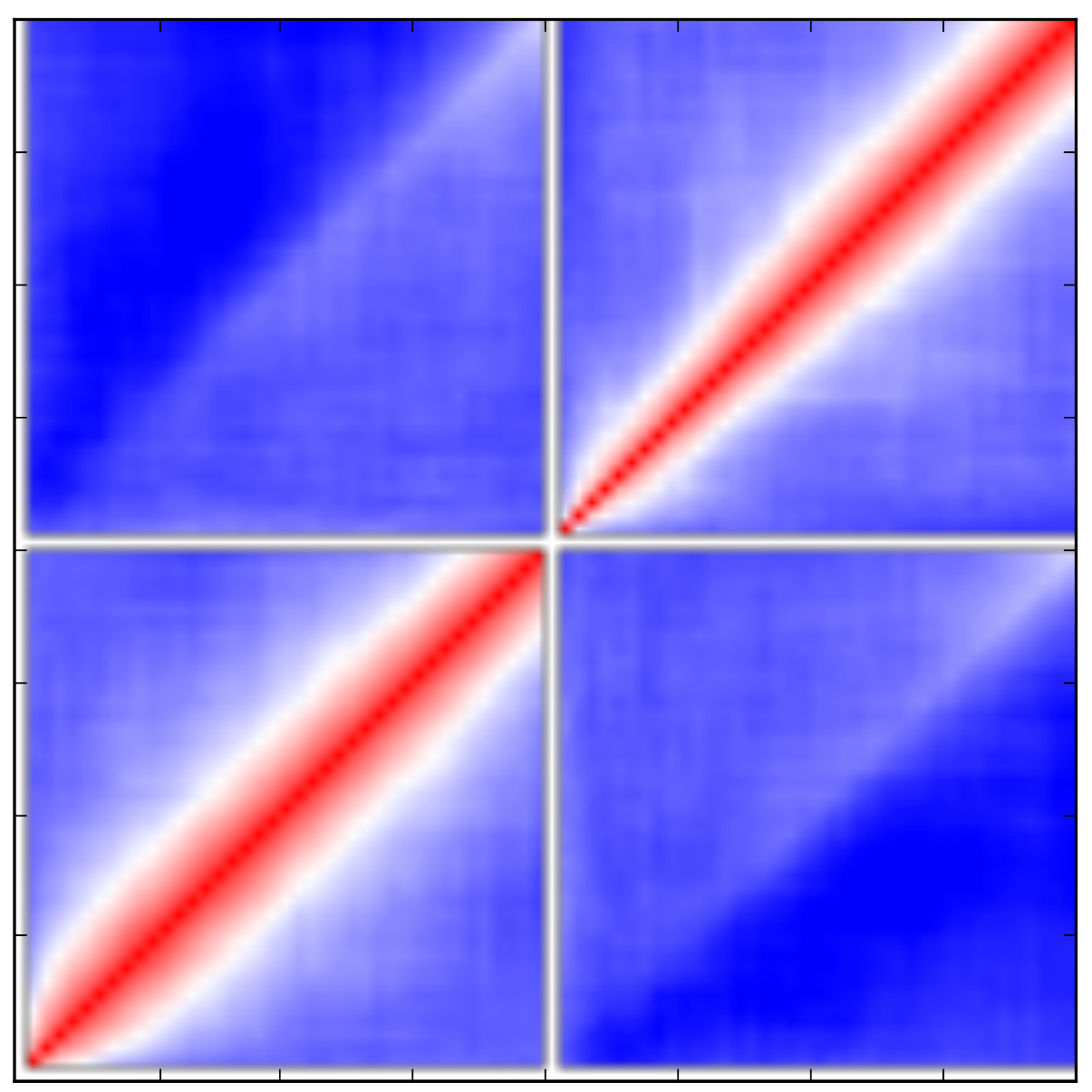}
\put(-10,-8){{\large 200}}
\put(-40,-8){{\large 150}}
\put(-65,-8){{\large 100}}
\put(-90,-8){{\large 50}}
\put(-123,-8){{\large 200}}
\put(-150,-8){{\large 150}}
\put(-177,-8){{\large 100}}
\put(-200,-8){{\large 50}}
\put(-205,210){\color{white}$0.55<z<0.75$}
\put(-140,-20){{\large $s \,[h^{-1}\,{\rm Mpc}]$}}
\includegraphics[width=1.6cm]{covbar}
\end{tabular}
\caption{\label{fig:cov} Cosmic evolution of the correlation matrices for different redshift bins indicated in the legend in bins of 5 $h^{-1}$ Mpc. Lower left block for the monopole, upper right block for the quadrupole, and upper left and lower right blocks for the correlations between the monopole and the quadrupole. See \S \ref{sec:evol} for details of the calculation.  These correlation matrices are used in  the BAO and RSD analysis in Chuang et al. (in prep.).}
\end{figure*}

Discreteness adds a shot noise contribution to the measured power spectrum and bispectrum. In this paper we assume that these contributions are of Poisson type and therefore are given by,

\begin{eqnarray}
P_{\rm sn}(k)&=&\frac{1}{\bar n} \\
B_{\rm sn}({\bf k}_1,{\bf k}_2)&=&\frac{1}{\bar n}\left[ P(k_1)+P(k_2)+P(k_3)\right]+\frac{1}{\bar{n}^2}
\end{eqnarray}
where $k_3=|{\bf k}_1+{\bf k}_2|$ and $\bar{n}$ is the number density of haloes.

For both power spectrum and bispectrum we present the BOSS DR12 data error-bars computed from the dispersion among  2,048 and 100 realizations of \textsc{md patchy} mock catalogues, respectively. 

\begin{figure*}
\begin{tabular}{cc}
\hspace{-0.75cm}
\includegraphics[width=6.7cm]{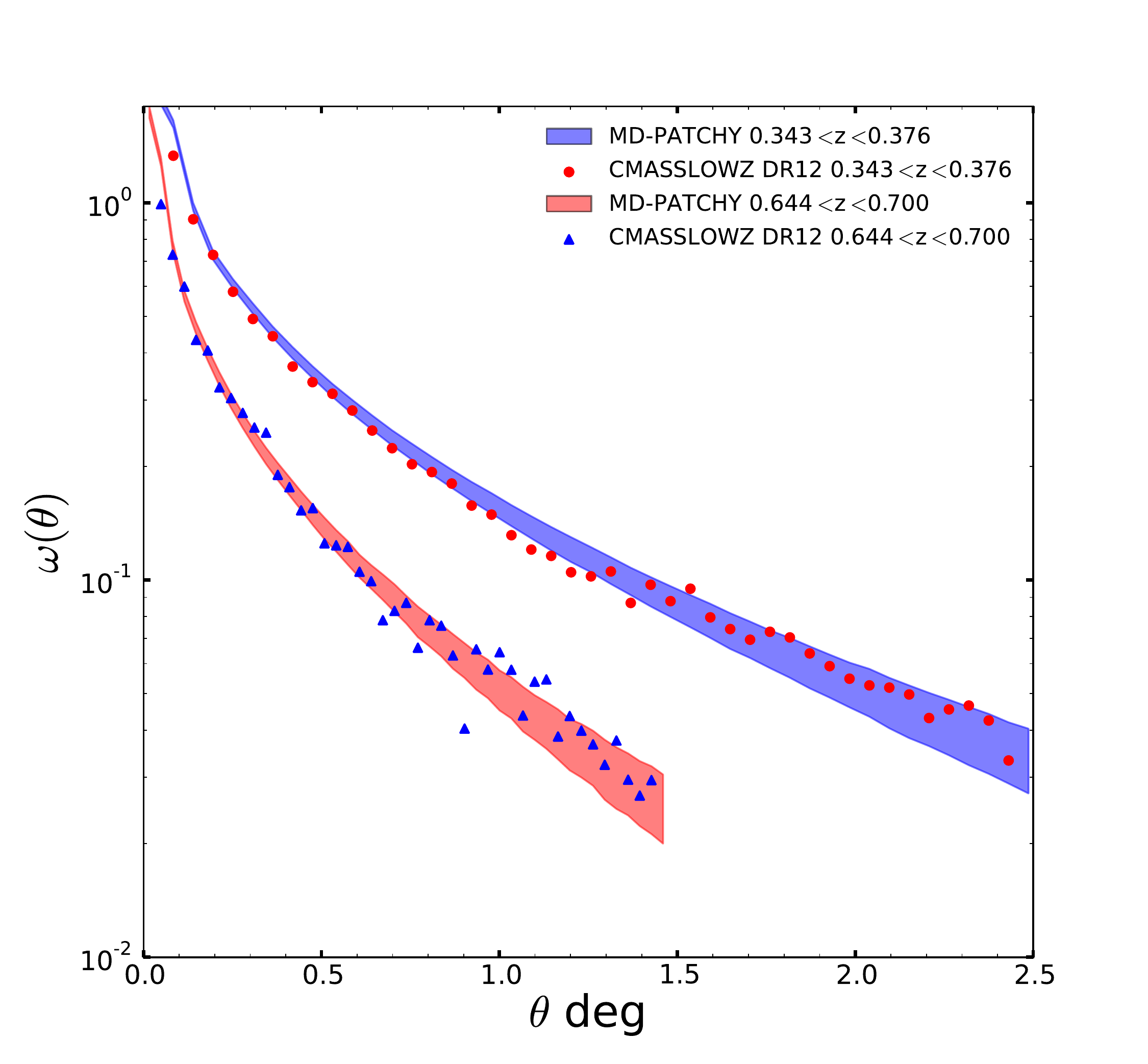}
\includegraphics[width=6.cm]{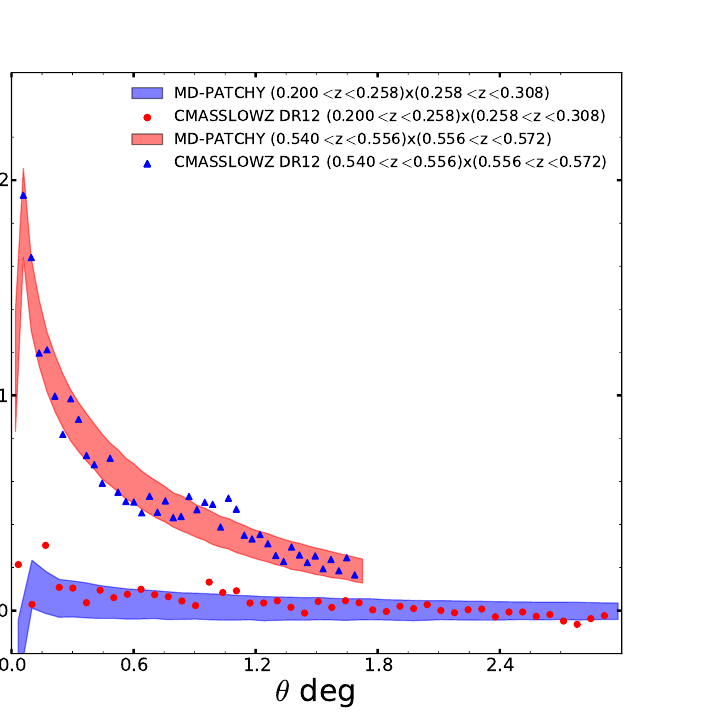}
\includegraphics[width=6.cm]{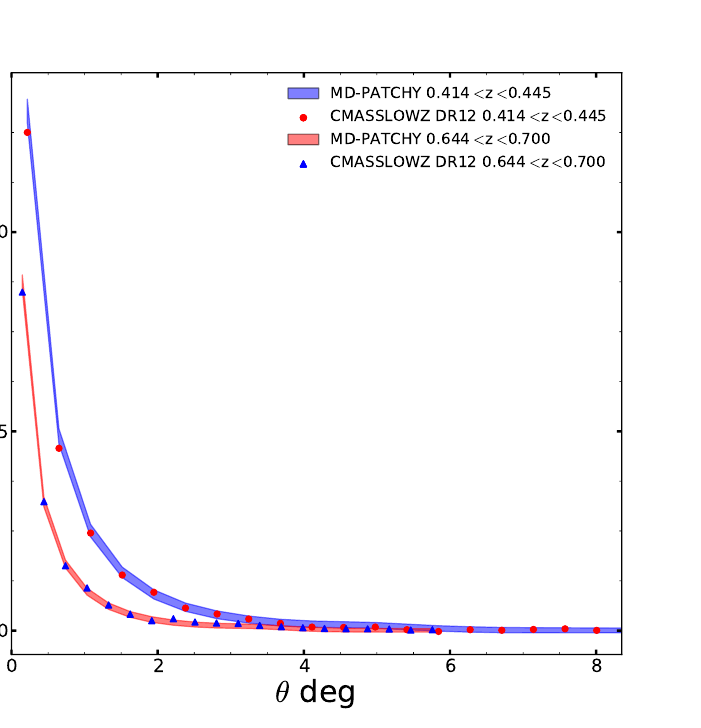}
\end{tabular}
\caption{\label{fig:ang}  Angular correlation functions based on the combined sample. Left panel: Angular auto-correlation function on small scales for two different tomographic bins (see key for redshift ranges), where colour bands are the mean and 1-$\sigma$ region of \textsc{md patchy} and symbols correspond to the measurements on the DR12 combined sample. Central panel: angular cross-correlation function on small scales  between different tomographic bins, following the same key as the left-hand panel. Right-hand panel: large-scale angular auto-correlation function for two different redshift bins. These measurements are used in the tomographic analysis of galaxy clustering in Salazar-Albornoz et al.~(in prep.).    }
\end{figure*}

\begin{figure*}
\begin{tabular}{cc}
\hspace{-0.75cm}
\includegraphics[width=7.cm]{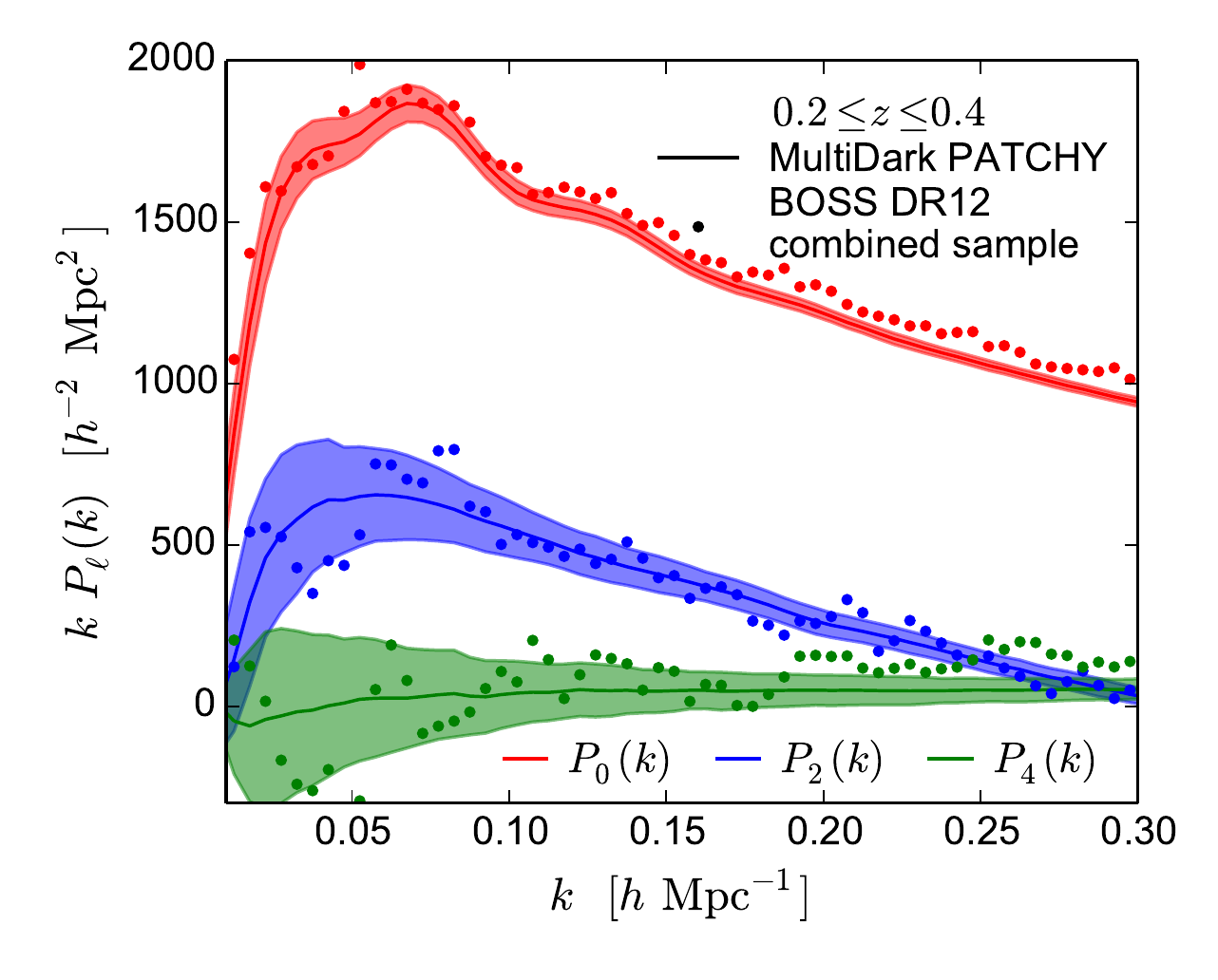}
\includegraphics[width=5.7cm]{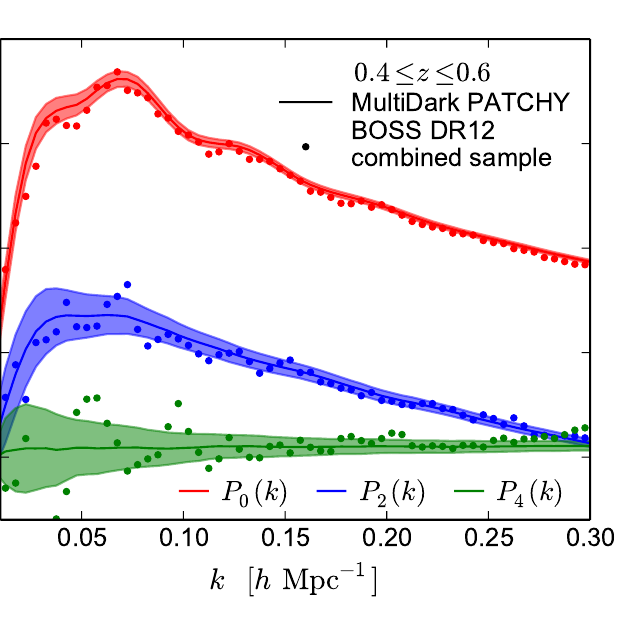}
\includegraphics[width=5.7cm]{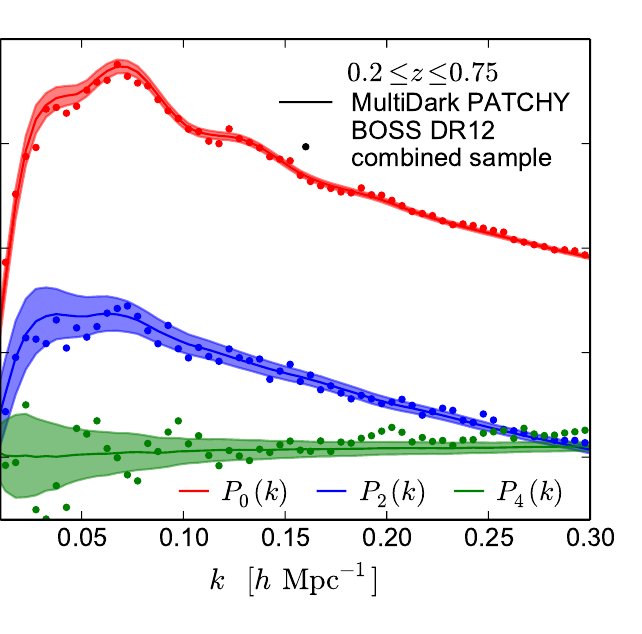}
\end{tabular}
\caption{\label{fig:mult} Multipole moments based on the combined sample: monopole $P_0$, quadrupole $P_2$ and hexadecapole $P_4$ for different redshift bins (see key for redshift ranges), where colour bands are the mean of \textsc{md patchy} and symbols correspond to the measurements on the DR12 combined sample. These measurements are used in the wedges analysis of galaxy clustering in Grieb et al.~(in prep.).    }
\end{figure*}

The Fourier space analysis has been used to improve  the modelling of the RSD  in the galaxy mock catalogues. We have assigned higher peculiar random motions to about 10\% of the galaxies to fit the quadrupole of the data with a specific value for each of the 10 redshift bins.
The resulting monopoles and the quadrupoles show a good agreement with the observations  over the range relevant to BAOs and RSDs up to at least $k\simeq0.3\,h^{-1}$ Mpc for both LOWZ and CMASS (see Fig.~\ref{fig:MQFourier}). 
{ This agreement is further supported after BAO reconstruction, as can be seen in Fig.~\ref{fig:baorec}.}
Only towards the very large scales ($k\lsim0.02\,h^{-1}$ Mpc) we can find that the observed monopole tends to be larger than the mock catalogues (both \textsc{md patchy} and \textsc{qpm}). This hints towards the  discrepancy in the monopole found in configuration space (see the previous section).
Although the \textsc{patchy} method can potentially yield accurate two-point statistics up to $k\sim1\,h^{-1}$ Mpc \citep[see][]{KitauraPatchy,ChuangComp15}, we have restricted the study to lower $k$s, as the analysis of BAOs and RSDs will not be done beyond $k=0.3\,h^{-1}$ Mpc, and the computation of power spectra for thousands of mocks with large grids becomes very expensive. 

This fitting procedure had, however, as a consequence that the three-point correlation function is slightly less precise at angles close to $\theta\sim0$ and $\theta\sim\pi$, as can be seen in Fig~\ref{fig:3pt}, which prior to this operation was fully compatible with the reference catalogue. In fact the reference {BigMultiDark} catalogue used in this study showed a highly discrepant quadrupole, as compared to the observations. This has been deeply analysed and a better agreement has been found based on an improved HAM procedure applied to the {BigMultiDark} simulation  \citep[see][companion paper]{2015arXiv150906404R}, which however was not available at the moment of the generation of the \textsc{md patchy} mocks. 
The HOD model adopted in the \textsc{qpm} mock catalogues assumed about 10\% satellite galaxies. This yields a compatible quadrupole for the CMASS galaxies. However, as these catalogues were not iteratively calibrated for different redshift slices, their agreement with the LOWZ galaxies is less accurate.

A detailed analysis of the bispectra is presented in Figs.~\ref{fig:BSLOWZ} and \ref{fig:BSCMASS} demonstrating a reasonable agreement between the mocks and the observations for different configurations of triangles across a wide range of scales, given the high uncertainties introduced by the mask, selection function, and cosmic variance.

\subsection{Cosmic evolution}
\label{sec:evol}

The cosmic evolution modelled in the \textsc{md patchy} mocks was achieved by fitting the clustering of 10 redshift bins for the full redshift range spanning about 5 Gyr. This implied running structure formation with {ALPT} for each redshift, i.e., modelling the growth of structures and the growth rate, and additionally fitting the galaxy bias evolution and the nonlinear RSDs.
The evolution of clustering for both sets of mocks in the full redshift range is shown in Fig.~\ref{fig:CFPAQLOWZCMASS}.
While the correlation function for CMASS galaxies does not show strong differences along the CMASS redshift range, this evolution is very apparent for the LOWZ sample.  Fig.~\ref{fig:LOWZevol} shows the comparison between the mocks and the observations for different LOWZ in more detail.  The \textsc{qpm} mocks do not include a detailed  cosmic evolution within LOWZ or CMASS being based on mean redshifts for each case. This explains why these mocks lose  accuracy in the two-point statistics towards low redshifts.

We investigate now the cosmic evolution of the covariance matrices derived from the \textsc{md patchy} mocks\footnote{Covariance matrices for the different catalogues (LOWZ, CMASS, and combined sample) will be made publicly available with the publication of the galaxy catalogue.} computed as in \citet[][]{AAB14}:
\be
{\rm cov}{[i,j]}=\frac{\sum_l(\xi^l_i-\langle\xi^l_{i}\rangle)(\xi^l_j-\langle\xi^l_{j}\rangle)}{N_{\rm s}-1}\,,
\ee
with bins $i$ and $j$, mock sample $l$, and  $N_{\rm s}$ being the number of simulations.

The correlation matrices for different redshift bins shown in Fig.~\ref{fig:cov} were constructed upon the covariance matrices following
\be
{\cal C}{[i,j]}=\frac{{\rm cov}{[i,j]}}{\sqrt{{\rm cov}{[i,i]}}\sqrt{{\rm cov}{[j,j]}}}\,.
\ee
We find that the correlation matrices vary in subsequent redshift bins.
First, the correlation matrices are increasingly correlated close to the diagonal for both the monopole and the quadrupole towards lower redshifts, as expected from gravitational evolution coupling different scales. This is seen in Fig.~\ref{fig:cov} as  the diagonal red band becomes broader especially comparing the highest redshift bin with the lower ones. Secondly, we find that moderate off-diagonal correlations present at higher redshifts disappear towards lower redshifts. And thirdly, we can see that the correlation between the monopole and the quadrupole at large scales becomes maximal in the redshift bin $0.43<z<0.55$, as can be seen in the white region in the lower right and upper left blocks. This ``triangular'' correlation is expected from linear theory \citep[see Eqs.~7 and 9 in][]{Chuang13a}. 

Further calculations of the correlation functions including \textsc{qpm} mocks are shown in  companion publications \citep[][companion papers]{2015arXiv150906386G,gilBAO15}.

{ Additionally we show in Fig.~\ref{fig:ang} the angular correlation function and in Fig.~\ref{fig:mult} the multipole moments (including the hexadecapole)  for different redshift bins based on the combined sample showing  good agreement between the \textsc{md patchy} mocks and the data. }

\section{Future work}
\label{sec:future}

We have taken advantage in this survey of the characteristic bias of LRGs, 
being massive objects residing in high density regions. This work confirms that threshold bias is an essential ingredient to explain the clustering of LRGs. This  facilitates our analysis, 
since the low density filamentary network did not need to be accurately described, and it has permitted us to rely on low resolution (augmented Lagrangian) PT based methods. This  will no longer apply for upcoming surveys based on emission line galaxies residing in the whole cosmic web.  One could improve the methodology presented in this work by substituting the structure formation model based on PT with a more accurate one (e.g.~\textsc{cola}). Whether this is necessary, or whether more efficient alternative approaches are sufficient (e.g.~{ALPT} with \textsc{muscle} corrections), will be investigated in future works.

Nonlocal bias was  only considered in the mass assignment step, but neglected in the generation of the full galaxy population. This may become  important  to model for emission line galaxies, and needs a deeper analysis.

The approximate  ``halo exclusion'' modelling  is mainly responsible for the deviation  in the clustering of the most massive objects, and could be improved by taking their full distribution of relative distances, instead of taking a sharp minimum separation for each mass bin, as  is done here.

Another aspect which still needs to be improved in the catalogues is the clustering on sub-Mpc scales. We have randomly assigned  positions of dark matter particles to the mock galaxies without considering that some of them are satellites of central galaxies. This implies that these mocks are not appropriate for  fibre-collision analysis.  For the time being we will leave the mock catalogues as they are, since most of the studies are not affected by this.  Nevertheless, we would like to stress that this aspect can  easily be corrected by assigning to a fraction of the mock galaxies close positions to the major most massive ones in the neighbourhood, without the need of redoing the catalogues. The \textsc{qpm} mocks better model fibre collisions, as the HOD adopted in this work  successfully reproduced the fraction of close satellites and central galaxies \citep[][companion paper]{2015arXiv150906386G}. 

 Also the photometric calibration systematics, presumably responsible for the excess of power in the data towards large scales, require further investigation.

We have considered one fiducial cosmology. It would be, however, interesting to provide sets of mock catalogues running over different combinations of cosmological parameters.

Let us finally mention that we have ignored in this study super-survey modes, which may be especially relevant for the analysis of the power spectrum at very large scales \citep[][]{Takada13,Lia14,Li14,Carron15}. 

We aim at addressing all these issues in future works.

\section{Summary and conclusions}
\label{sec:conc}

We have presented  12,288 mock galaxy catalogues for the BOSS DR12, including all relevant physical and observational effects, to enable a robust analysis of BAOs and RSDs. 

The main features of these mock catalogues are as follows:
\begin{itemize}

\item large number of catalogues: 2,048 for each LOWZ, CMASS, and combined LOWZ+CMASS and northern and southern galactic cap,

\item accurate structure formation model on scales of a few Mpc,

\item accurate galaxy bias model including nonlinear, stochastic,  threshold bias, and a nonlocal  bias dependence on the tidal field tensor and the  exclusion effect separation of massive objects,

\item modelling redshift evolution of galaxy bias, growth of structures, growth rate, and nonlinear RSDs, 

\item and additional survey features, such as geometry, sector completeness, veto masks and radial selection functions.

\end{itemize}

{  The same degree of accuracy is achieved for the BOSS DR11 \textsc{md patchy} mocks, for which only 6,000 lightcone mock catalogues were produced (1,000 for each LOWZ, CMASS, and combined LOWZ+CMASS and northern and southern galactic cap). 

The \textsc{md patchy} mocks have shown a better match to the data than the \textsc{qpm} mocks in terms of two- and three-point statistics. Investigating the origin for these differences can be interesting as the physical models, and in particular the galaxy bias, adopted in each method are quite different.}

We note that neglecting the stochastic bias considered in the \textsc{md patchy} mocks, modelling the deviation from  Poisson shot noise (predominantly over-dispersion), could  underestimate  the clustering uncertainties.

The  mock catalogues have enabled a robust analysis of the BOSS data yielding the necessary error estimates and the validation of the analysis methods. 
In particular the studies include the following:
\begin{itemize}
\item a full clustering analysis  (S{\'a}nchez et al.~in prep., Grieb et al.~in prep.: see Fig.~\ref{fig:mult}),
\item  a tomographic analysis of the large-scale angular galaxy clustering, where full light-cone effects (e.g. growth, bias and velocity field evolution) are essential (Salazar-Albornoz et al.~in prep.: see Fig.~\ref{fig:ang}), 
\item a  study of the BAOs reconstructions (see Vargas-Magana et al.~in prep., and Fig.~\ref{fig:baorec} showing the performance on  the \textsc{md patchy} mocks),
\item and a  RSD analysis \citep[][companion paper;  Beutler et al in prep.]{2015arXiv150906386G}.
\end{itemize}

We have demonstrated that the \textsc{md patchy} BOSS DR12 mock galaxies match, in general within 1-$\sigma$, the clustering properties of the BOSS LRGs for the monopole, quadrupole, and hexadecapole of the two-point correlation function both in configuration and Fourier space. { In particular we achieve a high accuracy in the modelling of the monopole up to $k\sim0.3\,h\,{\rm Mpc}^{-1}$.} We have furthermore shown that we also obtain three-point statistics with the same level of accuracy as $N$-body based catalogues at scales larger than a few Mpc, which are close to the observations.

The good agreement between the models and the observations demonstrates the level of accuracy reached in cosmology, our understanding of structure formation, galaxy bias, and observational systematics.

All the mock galaxy catalogues and the corresponding covariance matrices will be made publicly available together with the release of the BOSS DR12 galaxy catalogue.

\section*{acknowledgements}

FSK, SRT, CC, CZ, FP, AK, and CGS acknowledge support from the Spanish MICINNs Consolider-Ingenio 2010 Programme under grant MultiDark CSD2009-00064, MINECO Centro de Excelencia Severo Ochoa Programme under grant SEV- 2012-0249, and grant AYA2014-60641-C2-1-P.  FSK and CZ also want to thank the Instituto de F{\'i}sica Te{\'o}rica UAM/CSIC for the hospitality and support during several visits, where part of this work was completed. FP wishes to thank the Lawrence Berkeley National Laboratory for the hospitality during the development of this work, he also acknowledges the Spanish MEC “Salvador de Madariaga” programme, Ref.~PRX14/00444. 
HGM is grateful for support from the UK Science and Technology Facilities Council through the grant
ST/I001204/1. HG acknowledges the support of the 100 Talents Programme of the Chinese Academy of Sciences. GY wishes to thank MINECO  (Spain) for financial support under project grants AYA2012-31101 and FPA2012-34694.  He also thanks the Red Espa\~nola de Supercomputaci\'on  for granting computing time in the Marenostrum supercomputer, in which part of this work has been done.
AGS, SSA and JNG acknowledge support from the Transregional Collaborative Research Centre TR33 ‘The Dark Universe’ of the German Research Foundation (DFG).
AJC is supported by supported by the European Research Council under the European Community's Seventh Framework Programme FP7-IDEAS-Phys.LSS 240117. Funding for this work was partially provided by the Spanish MINECO under project MDM-2014-0369 of ICCUB (Unidad de Excelencia 'Mar{\'\i}a de Maeztu').

The massive production of all \textsc{MultiDark patchy} BOSS DR12 mocks has been performed at the BSC Marenostrum supercomputer, the Hydra cluster at the Instituto de F{\'i}sica Te{\'o}rica UAM/CSIC and NERSC at the Lawrence Berkeley National Laboratory.

 The {BigMultiDark} simulations have been performed on the SuperMUC supercomputer at the Leibniz-Rechenzentrum (LRZ) in Munich, using the computing resources awarded to the PRACE project number 2012060963. We want to thank V.~Springel for providing us with the optimized version of GADGET-2. 

Numerical computations for the power spectrum multipoles and bispectrum were performed on the Sciama High Performance Compute (HPC) cluster which is supported by the ICG, SEPNet and the University of Portsmouth. 

This research also used resources of the National Energy Research Scientific Computing Center, a DOE Office of Science User Facility supported by the Office of Science of the U.S. Department of Energy under Contract No. DE-AC02-05CH11231.

Funding for SDSS-III has been provided by the Alfred P. Sloan Foundation, the Participating Institutions, the National Science Foundation, and the U.S. Department of Energy Office of Science. The SDSS-III web site is http://www.sdss3.org/.

SDSS-III is managed by the Astrophysical Research Consortium for the
Participating Institutions of the SDSS-III Collaboration including the
University of Arizona,
the Brazilian Participation Group,
Brookhaven National Laboratory,
University of Cambridge,
Carnegie Mellon University,
University of Florida,
the French Participation Group,
the German Participation Group,
Harvard University,
the Instituto de Astrofisica de Canarias,
the Michigan State/Notre Dame/JINA Participation Group,
Johns Hopkins University,
Lawrence Berkeley National Laboratory,
Max Planck Institute for Astrophysics,
Max Planck Institute for Extraterrestrial Physics,
New Mexico State University,
New York University,
Ohio State University,
Pennsylvania State University,
University of Portsmouth,
Princeton University,
the Spanish Participation Group,
University of Tokyo,
University of Utah,
Vanderbilt University,
University of Virginia,
University of Washington,
and Yale University.



\bibliographystyle{mnras}


\bibliography{lit}

\vspace{0.5cm}

{\hspace{-0.65cm}$^1$ Leibniz-Institut f{\"u}r Astrophysik Potsdam (AIP), An der Sternwarte 16, D-14482 Potsdam, Germany\\
$^2$ Instituto de F\'{\i}sica Te\'orica, (UAM/CSIC), Universidad Aut\'onoma de Madrid,  Cantoblanco, E-28049 Madrid, Spain \\
$^3$ Campus of International Excellence UAM+CSIC, Cantoblanco, E-28049 Madrid, Spain \\
$^4$ Departamento de F\'{\i}sica Te\'orica, Universidad Aut\'onoma de Madrid, Cantoblanco, 28049, Madrid, Spain \\
$^5$ Tsinghua Center for Astrophysics, Department of Physics, Tsinghua University, Haidian District, Beijing 100084, China\\
$^6$ Instituto de Astrof\'{\i}sica de Andaluc\'{\i}a (CSIC), Glorieta de la Astronom\'{\i}a, E-18080 Granada, Spain \\
$^{7}$ Institute of Cosmology \& Gravitation, University of Portsmouth, Dennis Sciama Building, Portsmouth PO1 3FX, UK\\
$^{8}$ Key Laboratory for Research in Galaxies and Cosmology of Chinese Academy of Sciences, Shanghai Astronomical Observatory, \\Shanghai 200030, China\\
$^{9}$ Department of Physics and Astronomy, University of Utah, UT 84112, USA\\
$^{10}$ Astronomy Department, New Mexico State University, Las Cruces, NM 88003, USA\\
$^{11}$ Severo Ochoa Associate Researcher at the Instituto de F{\'i}sica Teorica (UAM/CSIC), E-28049, Madrid, Spain\\
$^{12}$ Instituto de Astrof{\'i}sica de Canarias (IAC), C/V{\'i}a L{\'a}ctea, s/n, E-38200, La Laguna, Tenerife, Spain. \\
$^{13}$ Dpto. Astrof{\'\i}sica, Universidad de La Laguna (ULL), E-38206 La Laguna, Tenerife, Spain.\\
$^{14}$Center for Cosmology and Particle Physics, New York University, 4 Washington Place, New York, NY 10003, USA\\
$^{15}$ Harvard-Smithsonian Center for Astrophysics, 60 Garden St., Cambridge, MA 02138, USA\\
$^{16}$ Lawrence Berkeley National Laboratory, 1 Cyclotron Road, Berkeley, CA 94720, USA\\
$^{17}$ Departments of Physics and Astronomy, University of California, Berkeley, CA 94720, USA\\
$^{18}$ Max-Planck-Institut f\"ur extraterrestrische Physik, Postfach 1312, Giessenbachstr., D-85741 Garching, Germany\\
$^{19}$ Universit\"ats-Sternwarte M\"unchen, Scheinerstrasse 1, D-81679 Munich, Germany\\
$^{20}$ Instituto de F{\'i}sica, Universidad Nacional Aut{\'o}noma de M{\'e}xico, Apdo. Postal 20-364, 01000 M{\'e}xico,  D.F., M{\'e}xico\\
$^{21}$ Institut de Ci{\`e}ncies del Cosmos (ICCUB), Universitat de Barcelona (IEEC-UB), Mart{\'\i} i Franqu{\`e}s 1, E08028 Barcelona, Spain\\
$^{22}$ Department of Physics and Astronomy, The Johns Hopkins University, Baltimore, MD 21218, USA\\
$^{23}$ Center for Cosmology and AstroParticle Physics, The Ohio State University, Columbus, OH 43210, USA}


\end{document}